\documentclass[%
 reprint,
superscriptaddress,
nofootinbib,
 amsmath,amssymb,
 aps,
 pra,
]{revtex4-2}
\usepackage{graphicx} 
\usepackage{pgfplotstable}
\usepackage{booktabs}
\usepackage{siunitx}
\usepackage{float}
\usepackage{makecell}

\usepackage[hyperfootnotes=false,colorlinks=true]{hyperref}
\hypersetup{
    linkcolor=black,           
    citecolor=blue,        
    urlcolor=blue,         
    pdfborder={0 0 1}      
}

\newcommand{\ddn}{$d$($d$,$n$)$^3$He}
\newcommand{\ddp}{$d$($d$,$p$)$t$}
\newcommand{\dpg}{$d$($p$,$\gamma$)$^3$He}
\newcommand{\npg}{$p$($n$,$\gamma$)$d$}

\begin{document}

\title{A data-driven prediction for the primordial deuterium abundance}
\author{Timothy Launders}
\email{tlaunder@bu.edu}
\thanks{ORCID: \href{https://orcid.org/0009-0006-7718-4671}{0009-0006-7718-4671}}
\affiliation{Physics Department, Boston University, Boston, MA 02215, USA}
\author{Cara Giovanetti}
\email{cgiovanetti@lbl.gov}
\thanks{ORCID: \href{https://orcid.org/0000-0003-1611-3379}{0000-0003-1611-3379}}
\affiliation{Theory Group, Lawrence Berkeley National Laboratory, Berkeley, CA 94720, USA}
\affiliation{Leinweber Institute for Theoretical Physics, University of California, Berkeley, CA 94720, USA}
\author{Hongwan Liu}
\email{hongwan@bu.edu}
\thanks{ORCID: \href{https://orcid.org/0000-0003-2486-0681}{0000-0003-2486-0681}}
\affiliation{Physics Department, Boston University, Boston, MA 02215, USA}

\date{\today}

\begin{abstract}
	We predict the primordial deuterium abundance using a novel, fully data-driven approach, where we use Gaussian process regression to fit experimental nuclear reaction data for \ddn,~\ddp,~and \dpg, three reactions to which the primordial deuterium abundance is most sensitive.  Using the \textit{Planck} determination of the baryon density, we predict $10^5\times\mathrm{D/H} = 2.442\pm0.040$ in standard Big Bang Nucleosynthesis, $1.70\sigma$ below the Cooke \textit{et al.}\ measurement. Our result is consistent with predictions relying on first principles calculations of the deuterium burning cross sections. With the inferred baryon density from a combined fit to \textit{Planck}, ACT DR6, and SPT-3G D1, this discrepancy worsens to $1.98\sigma$. We validate our approach and confirm that Gaussian processes make unbiased D/H predictions with appropriately-sized uncertainties. We repeat our validation tests for low-degree polynomial fits, a technique used in previous analyses, and find that they systematically over-predict D/H. Our results highlight the need for improved measurements of the \ddn~and \ddp~S-factors at energies between $0.1$ and $\qty{0.6}{MeV}$.
\end{abstract}

\maketitle

Following the Big Bang, the universe existed in a hot, dense state filled with Standard Model particles in thermal equilibrium. As the universe expanded, the baryons reached a temperature of $\sim\qty{d9}{\kelvin}$, at which point nuclear processes fused protons and neutrons together to form heavier nuclei, including deuterium, helium-4, and lithium-7. These reactions quickly shut off once the baryon temperatures drops below $\sim\qty{d8}{\kelvin}$, ending this process of Big Bang Nucleosynthesis (BBN). The relative abundances of nuclei produced are highly sensitive to the physics of the primordial universe during this period. 

When combined with theoretical predictions, measurements of two abundances---primordial helium-4 and deuterium---make BBN a powerful test of cosmology. The recent sub-percent-level observational uncertainty of the helium-4 abundance~\cite{Aver2026} cements BBN's status as a precision tool in cosmology, enabling leading constraints on the number of early-universe relativistic species~\cite{Yeh2026} and surpassing the precision of similar constraints from the Cosmic Microwave Background (CMB)~\cite{Planck2018,ACT2025b, SPT2025}. Measurements of the primordial deuterium abundance relative to hydrogen D/H have also reached the percent level with the result of Cooke \textit{et al.}~\cite{Cooke2018}:
\begin{equation}
	10^{5} \times \mathrm{D/H} = 2.527 \pm 0.030,
\end{equation}
providing competitive constraints on the baryon density, $\omega_b$.

Prediction uncertainties for the helium-4 abundance are currently a factor of 6.5 smaller than measurement uncertainty~\cite{Aver2026},
and so the main focus of BBN theory should be on D/H, where measurement and theory uncertainties are of the same order. Calculating D/H involves integrating the rate of change of light nuclei abundances due to nuclear reactions as a function of time. These reactions (\textit{e.g.}\ \ddp) are characterized by the S-factor, which describes how nuclear interactions scale with energy $E$:
\begin{equation}
	S(E) = \sigma(E)E e^{2\pi\eta}.
\end{equation}
Above, $\sigma(E)$ is the cross section, $\eta = \alpha Z_0 Z_1\sqrt{{\mu_{12} c^2}/{2E}}$ is the Sommerfeld parameter, $\mu_{12}$ is the reduced mass of the two-body system, and $Z_0$ and $Z_1$ are the charges of the two nuclei~\cite{Iliadis2015}. $\alpha$ is the fine structure constant, and $c$ is the speed of light.  Thermal averages of S-factors over the equilibrium distribution of nuclei give reaction rates. To compute these rates, one must therefore perform a fit to S-factor data. Uncertainties in S-factor data, propagated to uncertainties in reaction rates, are responsible for the largest uncertainties in theoretical predictions for D/H. 

Repeating this procedure for all nuclear reactions active during BBN defines a reaction network.  Two widely-used reaction networks are those found in the PArthENoPE~\cite{Gariazzo2022} and PRIMAT~\cite{Pitrou2018} BBN codes. The PArthENoPE network uses low-degree polynomials to fit S-factor data~\cite{Serpico2004, Pisanti2021}, whereas the PRIMAT network shifts first-principles S-factor calculations by an overall scaling to fit experimental data~\cite{Inesta2017, Moscoso2021}. In both approaches, uncertainties in D/H are parametrized by rescaling each relevant rate by a single parameter. Using the public BBN code LINX~\cite{Giovanetti2025} and marginalizing over the \textit{Planck} 2018 TT,TE,EE+lowE+lensing $\Lambda$CDM CMB value for the baryon density~\cite{Planck2018},
\begin{equation}
	\omega_b = 0.02237 \pm 0.00015,
\end{equation}
the prediction with the PArthENoPE network is 
\begin{equation}
	10^5\times\mathrm{D/H} = 2.512 \pm 0.042,
	\label{eqn:parth_DH}
\end{equation}
in agreement with the Cooke \textit{et al.}\ measurement at $0.3\sigma$.\footnote{The uncertainty is Eq.~\eqref{eqn:parth_DH} is smaller than that reported in Ref.~\cite{Pisanti2021} due to differences in error propagation.  In this result, reaction rate uncertainties are marginalized.} Repeating the same calculation for the PRIMAT network predicts
\begin{equation}
	10^5\times\mathrm{D/H} = 2.444 \pm 0.037,
	\label{eqn:primat_DH}
\end{equation}
a $1.7\sigma$ disagreement with Cooke \textit{et al}.

In this paper, we propose and rigorously test a new method to predict D/H using Gaussian processes (GPs). GPs can be understood as probability distributions of functions consistent with data~\cite{Rasmussen2006}. Samples drawn from GPs are not restricted to a particular functional form, enabling a flexible parametrization of reaction rates that are consistent with the experimental S-factor data. 

We estimate the uncertainty of our D/H prediction by sampling from the GPs. Through sampling, we explore the space of possible reaction rates more fully than the traditional mean rate rescaling. Additionally, we optimize the GPs with a leave-dataset-out procedure that weighs each nuclear reaction dataset equally and mitigates overfitting to any individual experiment. 

To test the robustness of our predictions, we perform a Monte Carlo (MC) analysis with mock data. We find that our GP method make unbiased predictions at current experimental uncertainties, and contain the true value within $1\sigma$ ($2 \sigma$) for 66\%--69\% (95\%--96\%) of realizations, demonstrating accurate coverage. Having validated the GP method, we obtain
\begin{equation}
	10^5\times\mathrm{D/H} = 2.442 \pm 0.040 \,,
\end{equation}
marginalizing over the \textit{Planck} posterior for $\omega_b$.
Finally, we repeat our MC analysis for the low-degree polynomial method, finding that it generally over-predicts D/H.

\section{Results}
\label{sec:results}
\subsection{BBN Prediction Pipeline}

We perform GP regression on experimental data for the nuclear reactions \ddn,~\ddp,~and \dpg~in this work.  These reactions have the greatest influence on D/H (see \textit{e.g.}\ Refs.~\cite{Cyburt2004, Coc2010, Giovanetti2025}).

To illustrate our pipeline from experimental data to reaction rate samples, we demonstrate GP regression on \ddp~S-factor data, which we take from Refs.~\cite{Krauss1987, Brown1990, Greife1995, Leonard2006}. Each data point $i$ in dataset $k$ has statistical noise $\sigma_{ik}$ and correlated normalization uncertainty $\epsilon_k$ from systematics. Details on the GP fit to the other two rates, choice of datasets and associated uncertainties are discussed in Methods.

GP regression starts by assuming a joint multivariate Gaussian prior distribution on the S-factor $\boldsymbol{S}$ at a vector of $n$ energies $\boldsymbol{E}$, where there are experimental data, and $m$ energies $\boldsymbol{E}_*$ where we predict the S-factor. The experimentally observed $\boldsymbol{S}(\boldsymbol{E})$ update this prior, yielding a posterior that is the conditional distribution of $\boldsymbol{S}(\boldsymbol{E}_*)$ given $\boldsymbol{S}(\boldsymbol{E})$. The posterior is another Gaussian that can be derived analytically, yielding a principled and convenient way to draw samples consistent with the data.

The partitioned prior distribution is 
\begin{equation}
	\begin{bmatrix}
   		\boldsymbol{S} \\ \boldsymbol{S}_*
	\end{bmatrix} \sim \mathcal{N} \left( 
	\begin{bmatrix}
    		\boldsymbol{\mu}(\boldsymbol{E}) \\ \boldsymbol{\mu}(\boldsymbol{E}_*)
	\end{bmatrix} , \begin{bmatrix}
    		k(\boldsymbol{E}, \boldsymbol{E}) + \Sigma& k(\boldsymbol{E}, \boldsymbol{E}_*) \\ k(\boldsymbol{E}_*, \boldsymbol{E}) & k(\boldsymbol{E}_*, \boldsymbol{E}_*)
	\end{bmatrix} \right).
	\label{eqn:gp_partitioned}
\end{equation}
The covariance matrix for experimental data $\Sigma$ contains total uncertainties on the diagonal and correlated normalization uncertainties within each experiment as off-diagonal elements. We add $\Sigma$ to the kernel in the top-left block to capture all reported experimental uncertainties within the GP. 

$\boldsymbol{\mu}$ is the prior mean, set to zero for our fiducial analysis. This is a standard choice in the literature, though our D/H prediction has little sensitivity to setting the prior mean to theoretical calculations (see Methods).

The kernel function $k(E_i,E_j)$ encodes correlations between S-factor values at different energies. Different kernels enforce different assumptions regarding the S-factor; our choice is discussed in detail in Methods, but we observe little sensitivity in our result to a variety of kernels. The kernel includes hyperparameters, such as the characteristic length scale of correlations, selected using a form of cross-validation where each dataset receives equal weight during optimization (see Methods). 

Introducing shorthand $K_{11}$ for the $n\times n$ upper-left block of the full covariance, $K_{22}$ for the $m \times m$ lower-right block, and so forth, the posterior distribution of $\boldsymbol{S}_*$ conditioned on $\boldsymbol{S}$ is given by $\mathcal{N}(\boldsymbol{\mu}_{\boldsymbol{S}_*|\boldsymbol{S}}, \Sigma_{\boldsymbol{S}_*|\boldsymbol{S}})$, where~\cite{Rasmussen2006}:
\begin{equation}
	\begin{gathered}
		\boldsymbol{\mu}_{\boldsymbol{S}_* | \boldsymbol{S}} = \boldsymbol{\mu}(\boldsymbol{E_*}) + K_{21} K_{11}^{-1}(\boldsymbol{S} - \boldsymbol{\mu}(\boldsymbol{E})), \\
		\quad \Sigma_{\boldsymbol{S}_* | \boldsymbol{S}} = K_{22} - K_{21} K_{11}^{-1} K_{12}.
	\end{gathered}
	\label{eqn:gp_conditional}
\end{equation}

\begin{figure}
    	\centering
    	\begin{minipage}{.48\textwidth}
        		\centering
        		\includegraphics[width=\linewidth]{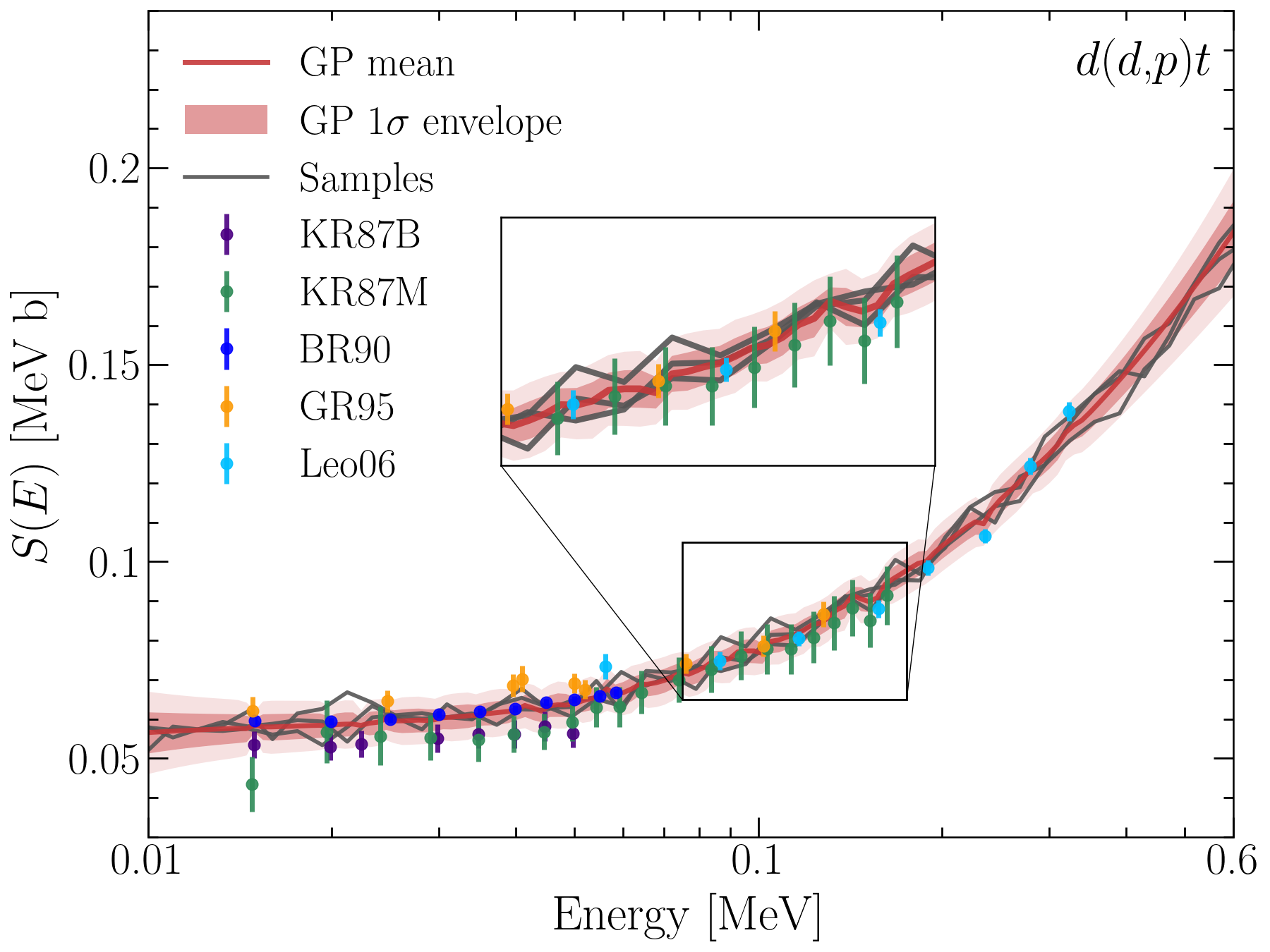}
    	\end{minipage}
    	\vspace{0mm}
    	\begin{minipage}{.48\textwidth}
        		\centering
        		\includegraphics[width=\linewidth]{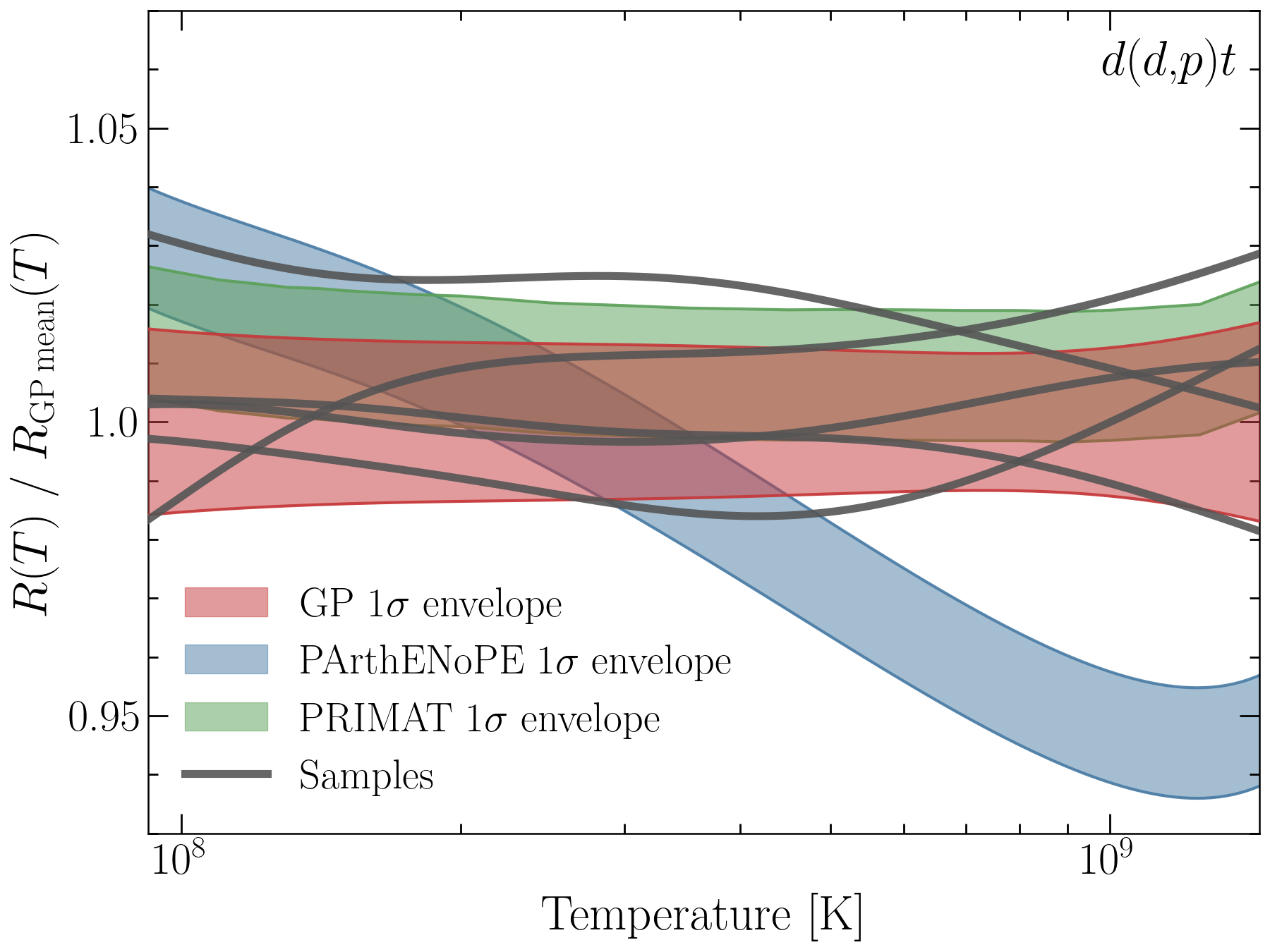}
    	\end{minipage}
	\caption{Gaussian process regression on \ddp~S-factor data. Samples drawn from the GP posterior are shown in grey. \textit{Top}: GP S-factor mean, $1\sigma$, and $2\sigma$ envelopes plotted along with data used for regression~\cite{Krauss1987, Brown1990, Greife1995, Leonard2006} shown in red. Three sample draws are overlayed (grey). \textit{Bottom}: GP $1\sigma$ reaction rate envelope (red) and five S-factor samples thermally averaged with Eq.~\eqref{eqn:thermal_average} in the relevant BBN temperature range (grey). The $1\sigma$ rate envelopes derived from PArthENoPE~\cite{Gariazzo2022} (blue) and PRIMAT~\cite{Pitrou2018} (green) are shown for comparison. All rates are normalized to the GP mean rate.}
	\label{fig:ddtp_samples}
\end{figure}

The top panel of Figure~\ref{fig:ddtp_samples} shows the S-factor posterior for \ddp.~The posterior is consistent with each experiment despite some scatter between them---our cross-validation approach avoids overfitting to data with very small uncertainties, which is important when combining disparate datasets. Thus, the relatively small uncertainties of Refs.~\cite{Brown1990} (dark blue) and~\cite{Leonard2006} (light blue) do not dominate the posterior. At high energies, where there are only data from Ref.~\cite{Leonard2006}, the posterior does not compress to match their uncertainties; these data are correlated and the GP has established a wider posterior where they overlap with others.

S-factor samples appear jagged, a consequence of the particular kernel chosen for this analysis. The kernel must allow for some small-scale fluctuations samples to be consistent with and capture variations in the data. This does not negatively impact D/H predictions since samples are integrated during thermal averaging to get the reaction rate $R(T)$:
\begin{equation}
	\begin{split}
		R(T) = N_A \langle \sigma v \rangle = \sqrt{\frac{8}{\pi \mu_{12}}} \frac{N_A}{(k_B T)^{3/2}} \\
		\times \int_{0}^{\infty} dE \, e^{-2 \pi \eta} S(E) e^{-E/k_B T},
	\end{split}
	\label{eqn:thermal_average}
\end{equation}
where $k_B$ is the Boltzmann constant and $N_A$ is Avogadro's number. This is the quantity that enters directly into the BBN prediction.  Compared with traditional mean rate rescaling, our samples---shown normalized to the posterior mean in the bottom panel of Figure~\ref{fig:ddtp_samples}---more completely span functional forms that are consistent with the data. Our rate is comparable to the PRIMAT rate~\cite{Pitrou2018} but quite different from the PArthENoPE rate~\cite{Pisanti2021}. We do not agree as closely with the PRIMAT rate for \ddn~(see Methods). 

After performing GP regression on \ddp,~\ddn, and \dpg~data, we are equipped to calculate D/H. We use the the recent measurement of Ref.~\cite{Chen2026} for \npg,  and Ref.~\cite{Czarnecki2018} for the neutron lifetime.  We use the rates from PRIMAT for the remaining 9 reactions required to predict the primordial deuterium abundance; switching to the PArthENoPE network instead for the other reactions reduces the predicted deuterium abundance by $\lesssim0.17\%$, which is negligible compared to the current experimental uncertainties. Rate uncertainties are propagated in the same manner as described in Ref.~\cite{Giovanetti2025}. 

\subsection{Monte Carlo Validation Tests}
\label{sec:MC}

\begin{figure*}[t]
	\centering
	\begin{minipage}{0.48\textwidth}
		\centering
		\includegraphics[width=\linewidth]{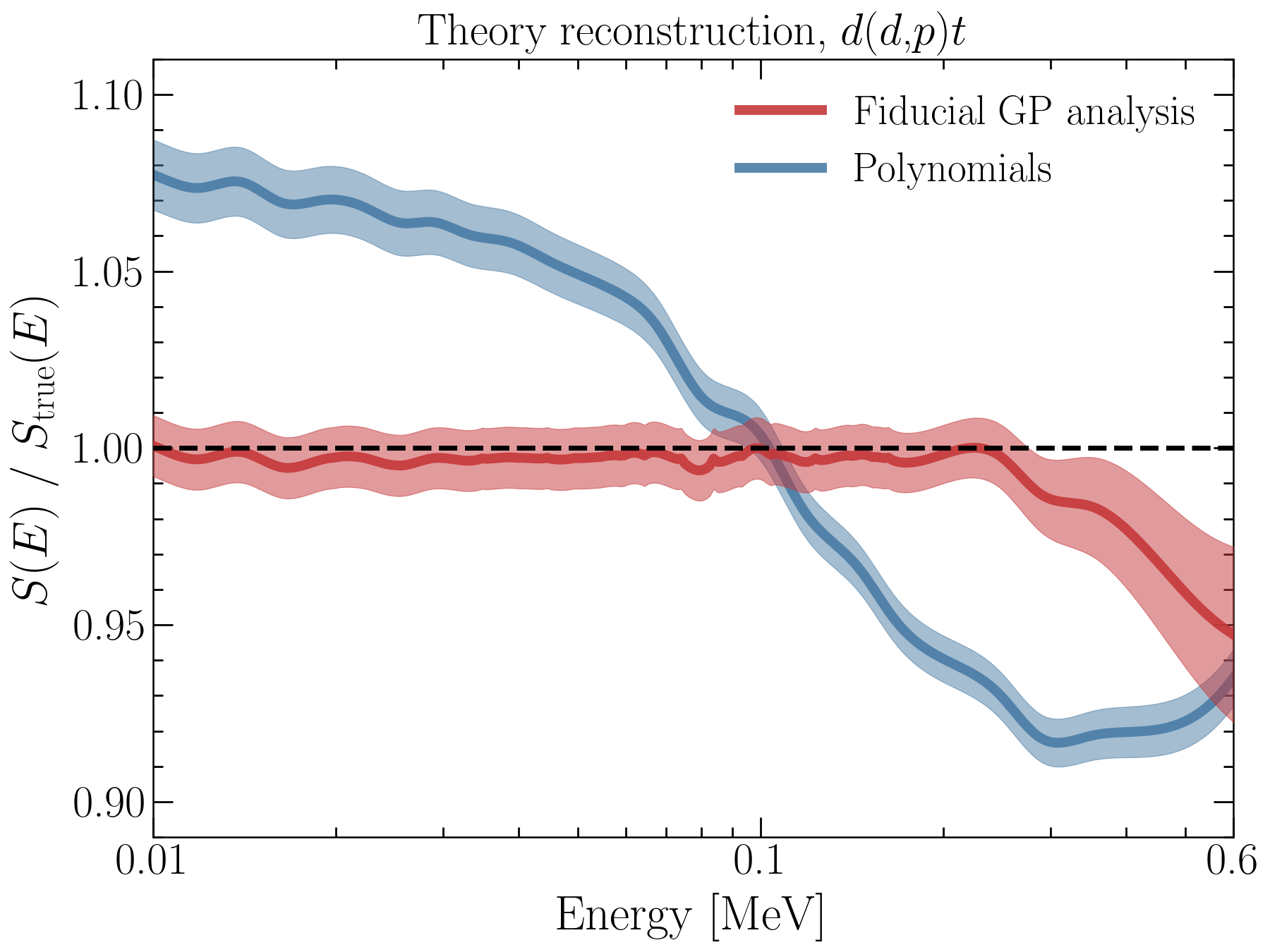}
	\end{minipage}
	\begin{minipage}{0.48\textwidth}
		\centering
		\includegraphics[width=\linewidth]{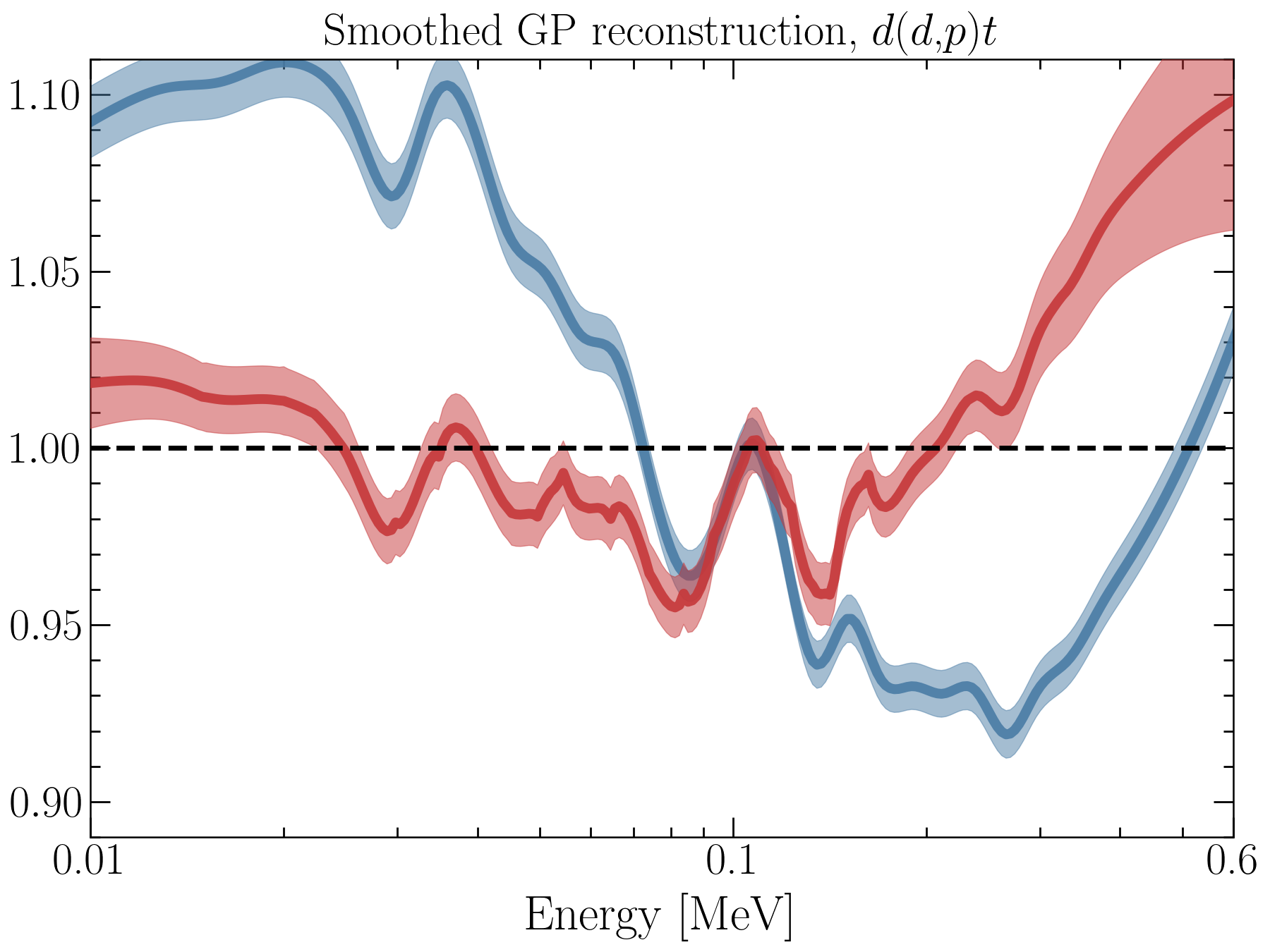}
	\end{minipage}
	\begin{minipage}{0.48\textwidth}
		\vspace{2mm}
		\centering
		\includegraphics[width=\linewidth]{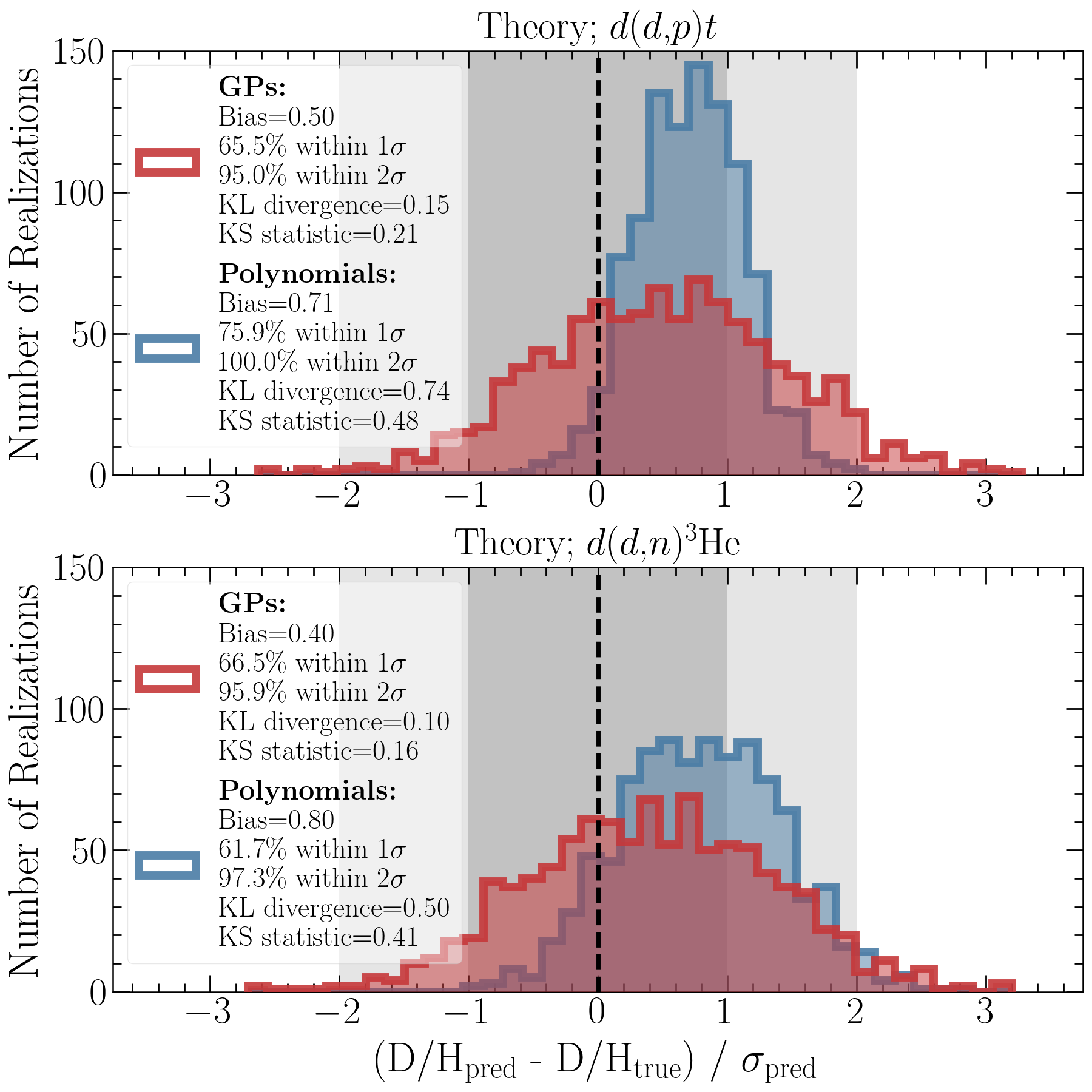}
	\end{minipage}
	\begin{minipage}{0.48\textwidth}
		\centering
		\includegraphics[width=\linewidth]{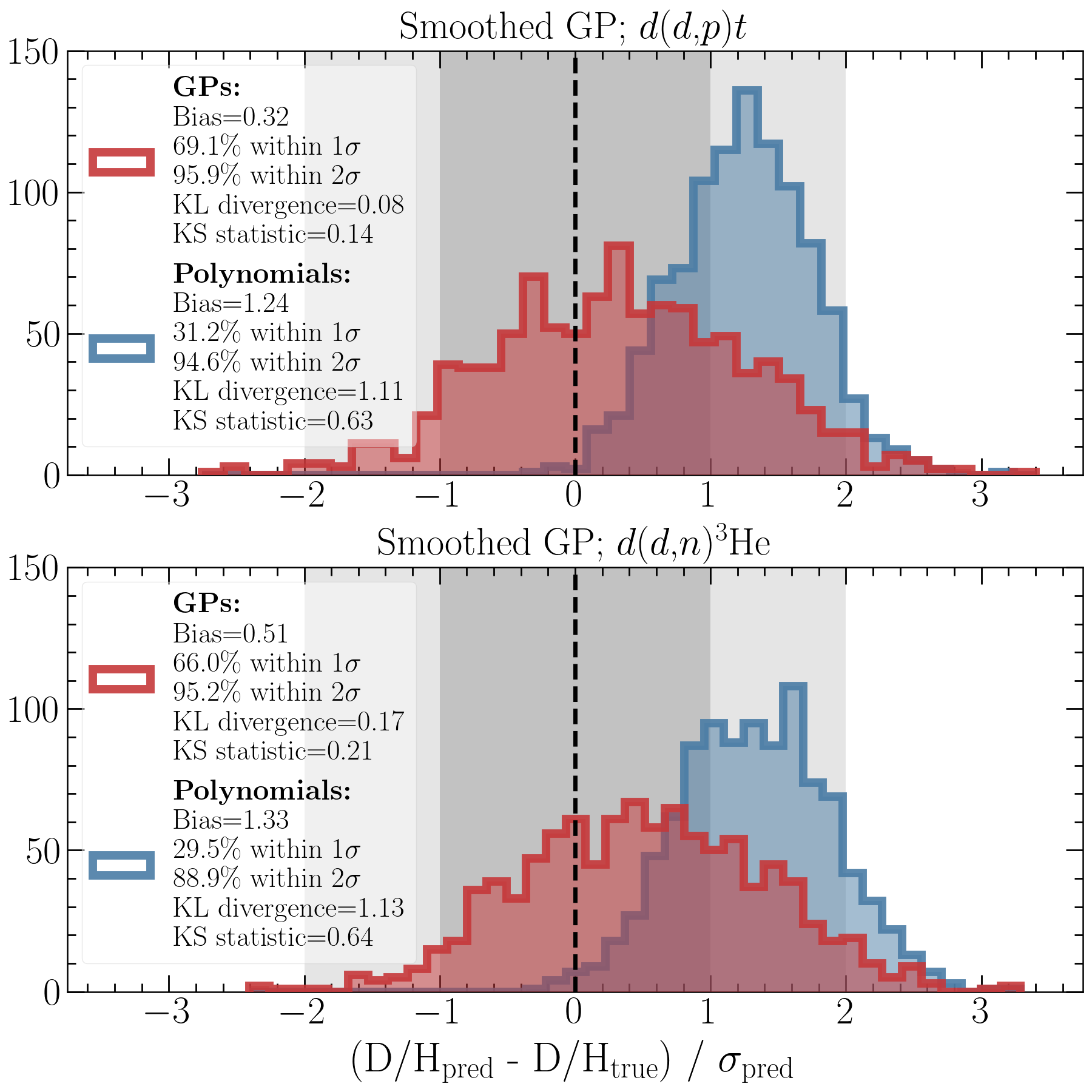}
	\end{minipage}
	\caption{Statistics for 1,000 Monte Carlo realizations of mock \ddp~and \ddn~data fit with both Gaussian process regression (red) and low-degree polynomials (blue). Polynomial coefficients are optimized with Eq.~\eqref{eqn:poly_chi2_single}. We use the Arai \textit{et al.}\ theory calculation~\cite{Arai2011} and the smoothed GP mean inferred from real data, smoothed and offset in energy, to generate the mock data. \textit{Top}: Reconstruction of the theory (left) and smoothed GP (right) generating functions through many data realizations, normalized to the generating functions. The uncertainty envelopes denote the standard deviation of best-fit S-factors. GPs recover the generating function well for the simpler theory and well enough for the more complicated smoothed GP case to make robust D/H predictions, outperforming polynomials throughout the BBN energy range. \textit{Bottom}: D/H predictions are shown as the difference between the predicted and true values, normalized by the predicted $1\sigma$ uncertainty. The bias, defined in Eq.~\eqref{eqn:bias}, and the percentage of realizations yielding a predicted D/H within $1$ and $2\sigma$ of the true value are displayed for each combination of generating function, fit procedure, and reaction. Additionally, we use the empirical results to approximate the full D/H distributions and quantify how similar these distributions are to $\mathcal{N}(0,1)$ with two values: the KL divergence~\cite{Kullback1951} and the KS statistic~\cite{Kolmogorov1933}.}
	\label{fig:mc_test_single}
\end{figure*}

To validate our procedure, we conduct Monte Carlo experiments with mock S-factor data at the same energies as the real data. We select a function $S_{\rm true}(E)$ to generate mock data and repeat the pipeline described above. An effective procedure should reproduce $S_{\rm true}(E)$ and make unbiased D/H predictions with robust uncertainties after many data realizations. 

We draw mock data corresponding to the five datasets of  Refs.~\cite{Krauss1987, Brown1990, Greife1995, Leonard2006}. The $k^{th}$ mock dataset is assigned a percent normalization uncertainty $\epsilon_k$ equal to the reported value of the corresponding real experiment. The $i^{th}$ data point of the $k^{th}$ dataset has some percent statistical noise $\sigma_{ik}$ also matching the reported value. At each energy $E$ measured by experiment $k$, data are drawn from 
\begin{equation}
	S_{ik}(E) = \alpha_k \left[S_{\rm true}(E)(1 + \beta_{ik})\right],
	\label{eqn:mock_gen}
\end{equation}
where $\alpha_k \sim \mathcal{N}(1, \epsilon_k)$ and $\beta_{ik} \sim \mathcal{N}(0, \sigma_{ik})$. Correlated systematic uncertainty scales the mock data after adding statistical noise. 

We assume two different generating functions for both \ddn~and \ddp:\footnote{For the MC test, we do not consider \dpg~individually, as the predicted $R(T)$ from all reaction networks---including the GP---agree well with one another (see Methods).} \textit{1}) the theory calculations of Ref.~\cite{Arai2011} and \textit{2}) the means for GP fits to real data, smoothed and offset slightly in energy such that the features of these functions are independent of mock data energies. While probably an unrealistic model for the true S-factor, the second choice tests performance for a more complicated generating function that produces mock data resembling the scatter of the real data. 

For both generating functions for \ddn~and \ddp,~1,000 mock realizations of the data are generated using Eq.~\eqref{eqn:mock_gen}. For each mock realization, we perform a GP regression, after which 1,000 samples are drawn from each GP, thermally-averaged, and passed into LINX. All other inputs are fixed. We obtain the mean $\mathrm{D/H}_{\rm pred}$ and standard deviation $\mathrm{\sigma}_{\rm pred}$ of the D/H distribution, as well as D/H from the thermal average of $S_{\rm true}(E)$, $\mathrm{D/H}_{\rm true}$.  In this and all subsequent analyses, we check that our results are converged for our chosen sample size.

To quantify the performance of our method, we define the bias as the expectation value of the difference between the predicted and true D/H values, normalized to the predicted uncertainty:
\begin{equation}
	\mathrm{Bias} = \mathbb{E}_\mathrm{mocks} \left[\frac{\mathrm{D/H}_{\rm pred} - \mathrm{D/H}_{\rm true}}{\sigma_{\rm pred}}\right].
	\label{eqn:bias}
\end{equation}
We approximate the expectation value with the average for 1,000 data realizations. 
\begin{figure}
	\centering
	\includegraphics[width=\columnwidth]{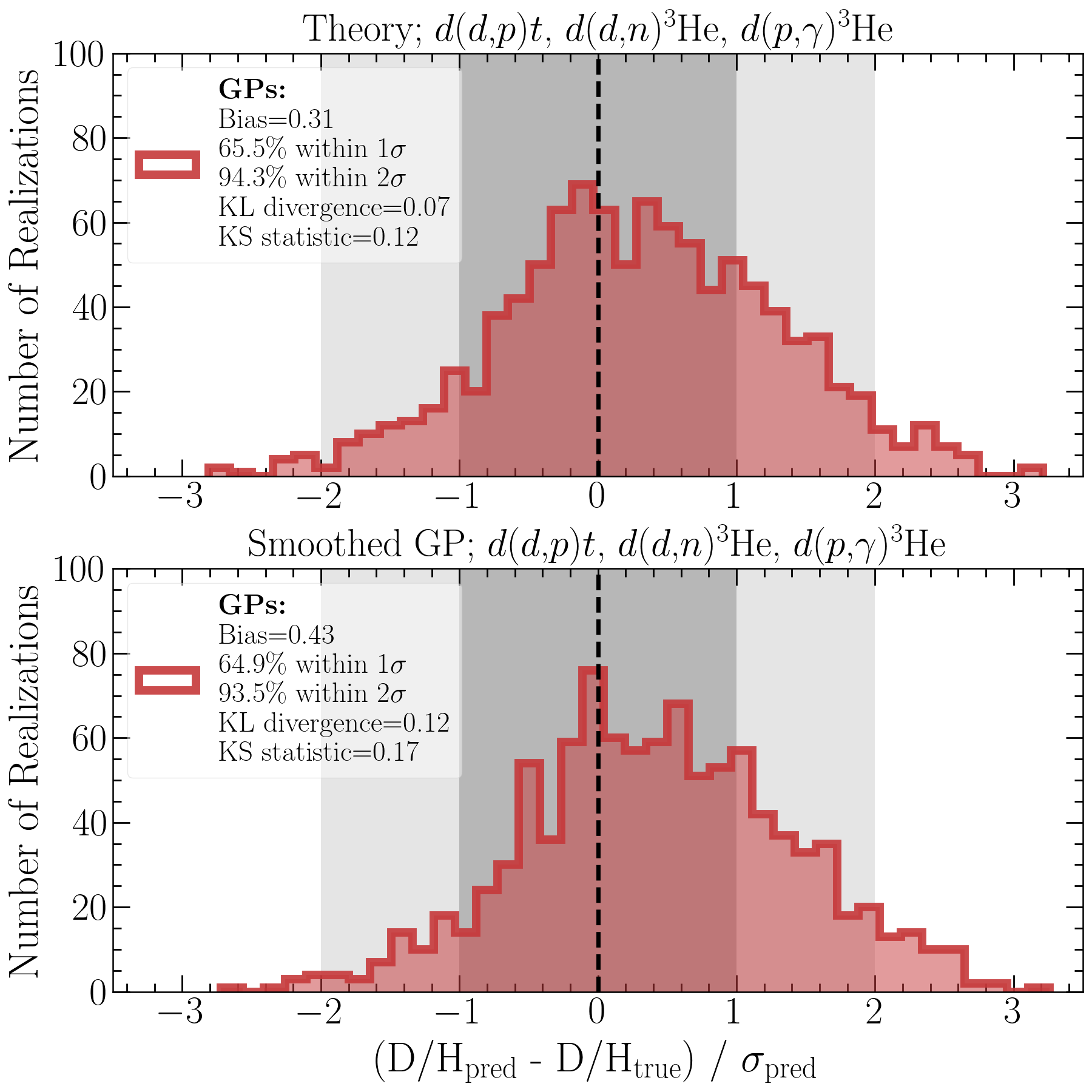}
	\caption{D/H predictions for 1,000 Monte Carlo realizations of mock \ddp,~\ddn,~and \dpg~data fit with Gaussian processes. All three mock reactions are sampled simultaneously. We consider generating functions (top) drawn from theory (Refs.~\cite{Arai2011, Marcucci2016}) and (bottom) smoothed versions of the GP mean for real data. D/H predictions are shown as the difference between the predicted and true values, normalized by the predicted $1\sigma$ uncertainty. The bias, defined in Eq.~\eqref{eqn:bias}, and the percentage of realizations yielding a predicted D/H within $1\sigma$ and $2\sigma$ of the true value are displayed for both generating functions. Additionally, we use the empirical results to approximate the full D/H distributions and quantify how similar these distributions are to $\mathcal{N}(0,1)$ with two values: the KL divergence~\cite{Kullback1951} and the KS statistic~\cite{Kolmogorov1933}. }
	\label{fig:mc_full}
\end{figure}

The top panels of Figure~\ref{fig:mc_test_single} show the statistics of the GP posterior means for both \ddp~generating functions. The performance of GPs is shown in red; blue results are discussed below. GPs are able to learn the smooth theory curve well at all energies up to $\sim\qty{0.3}{MeV}$. GPs exhibit less success for the smoothed GP curve, but perform well enough to make robust D/H predictions. We do not use any data beyond this energy in our analyses of real data or in mock tests, so the GP must extrapolate. This is at the tail end of the relevant energy range for these reactions (determined in this work to be $0.014$-$\qty{0.6}{MeV}$, see Methods for details), so GPs still make robust D/H predictions; we confirm the reliability of this extrapolation in Methods. However, a new experiment measuring the \ddn~and \ddp~S-factors up to \qty{0.6}{MeV}, with similar uncertainties to Ref.~\cite{Leonard2006}, would no longer require the GP to extrapolate and reduce our prediction uncertainties. 

Statistics of the D/H posteriors are shown in the bottom panels of Figure~\ref{fig:mc_test_single}. On average, the mean is in agreement with the true value, coming in at $\lesssim0.5\sigma$ greater than the true value for each scenario. The true D/H is contained within the $1\sigma$ error bar for 66\%--69\% of realizations, and within the $2\sigma$ error bar for $\sim95\%$ of realizations, supporting the robustness of the uncertainties reported in this work. 

Matching our analysis with real data more closely, we extend the Monte Carlo validation to include realizations of \dpg~and sample all three reactions simultaneously. Statistics of the D/H posteriors are shown in Figure~\ref{fig:mc_full}. D/H posteriors show similar bias and coverage as in Figure~\ref{fig:mc_test_single}, demonstrating that the robustness of the GPs extends to all three important reactions. 

\subsection{Analysis Results}

\begin{figure}
	\centering
	\includegraphics[width=\columnwidth]{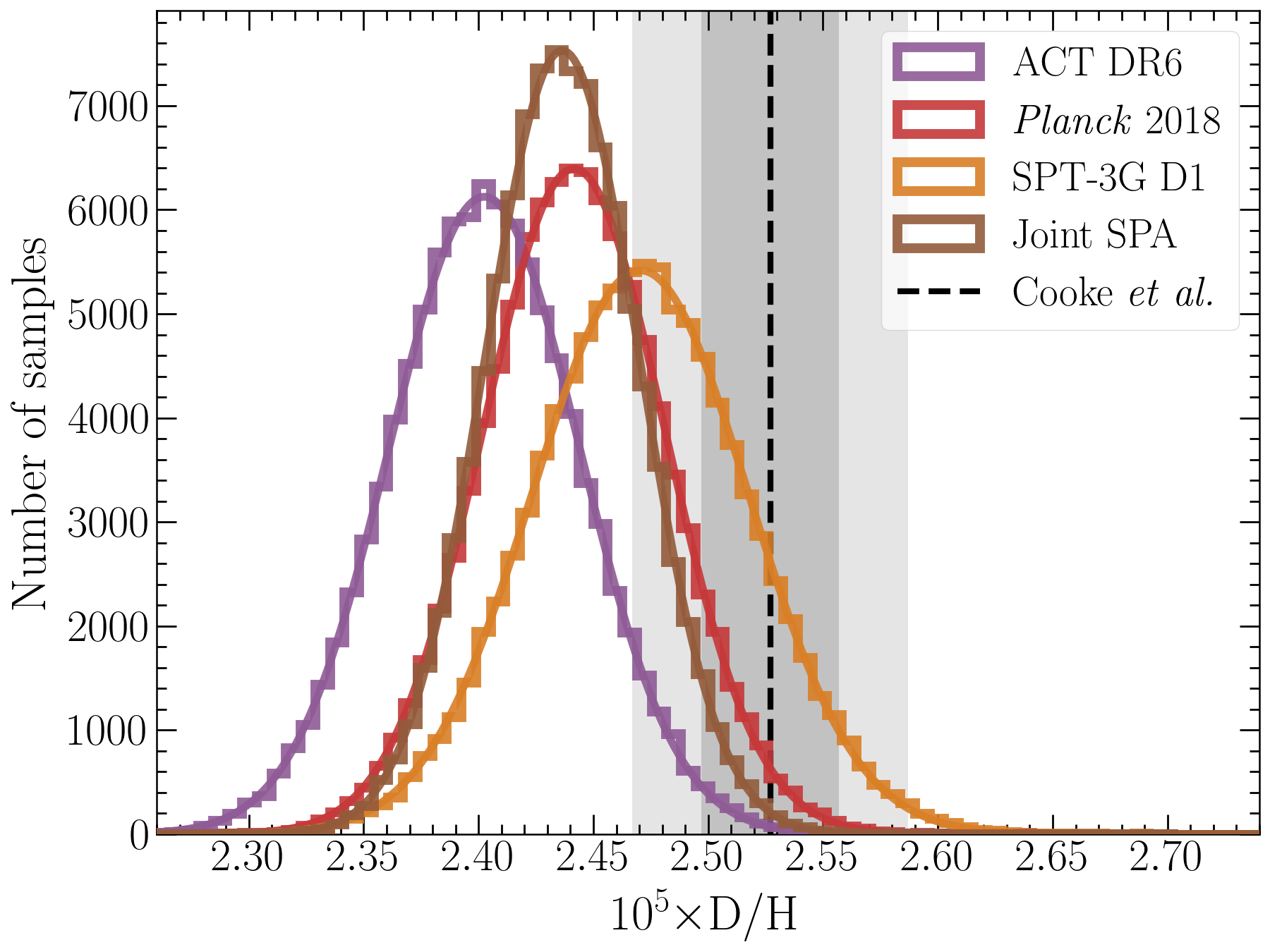}
	\caption{Deuterium abundance posteriors for four different CMB inferences of $\omega_b$: \textit{Planck} 2018~\cite{Planck2018}, ACT DR6~\cite{ACT2025a},  SPT-3G~\cite{SPT2025}, and a combined inference from all three (Joint SPA)~\cite{SPT2025}. Inferences including large-scale structure data are not considered. A Gaussian fit is overlayed on each. The measured D/H from Cooke \textit{et al.}~\cite{Cooke2018} is shown as shaded bands ($1\sigma$ and $2\sigma$ uncertainties). ACT prefers the highest $\omega_b$, giving the most discrepant D/H prediction. }
	\label{fig:vary_omegab}
\end{figure}

Having validated the GP method, we compute D/H with the real data. To provide a point of comparison with other reaction networks, we first fix $\omega_b$ to the \textit{Planck} central value. Then, we sample from each of the three GPs (and overall scalings for the other rates) 60,000 times. The mean and standard deviation of the subsequent D/H distribution give our D/H prediction. We obtain 
\begin{equation}
	10^5 \times\mathrm{D/H} = 2.442 \pm 0.029,
	\label{eqn:D_H_partial}
\end{equation}
$2.04\sigma$ below the Cooke \textit{et al.}\ measurement.
In full cosmological parameter estimation, our results and the Cooke \textit{et al.}\ measurement would prefer a lower value of $\omega_b$ than the \textit{Planck} central value.  However, we leave detailed cosmological parameter estimation to future work, as care is required to include the non-parametric functions obtained from the GP in a Bayesian inference framework involving all cosmological parameters.

We examine the consistency of the Cooke \textit{et al.}\ result with CMB data with our new method for predicting D/H. For \textit{Planck}~\cite{Planck2018}, ACT DR6~\cite{ACT2025a}, SPT-3G D1~\cite{SPT2025}, and a combination of all three (Joint SPA)~\cite{SPT2025}, we assume Gaussian distributions for the $\omega_b$ posteriors, displayed in Table~\ref{tab:vary_omegab}, and sample to marginalize over $\omega_b$. We increase to 100,000 samples and obtain D/H predictions $1.70\sigma$, $2.40\sigma$, $1.00\sigma$, and $1.98\sigma$ below the Cooke \textit{et al.}\ measurement respectively. D/H posteriors are shown in Table~\ref{tab:vary_omegab} and Figure~\ref{fig:vary_omegab}. Since ACT prefers the highest value for $\omega_b$, it predicts the greatest discrepancy. SPT predicts the closest agreement with the lowest $\omega_b$ and largest uncertainty.

\begin{table}[!t]
	\centering
	\begin{tabular}{cccc}
		\hline
		\thead{\textbf{CMB} \\\textbf{Experiment}} & $\mathbf{100 \omega_b}$ & $\mathbf{10^{5} \times}$\textbf{{D/H}} & \thead{\textbf{Agreement with} \\\textbf{Cooke \textit{et al.}}} \\
		\hline
		 \rule{0pt}{10pt}{\textit{Planck}~\cite{Planck2018}} & {$2.337\pm0.015$} & {$2.442\pm0.040$} & {$1.70\sigma$} \\
		{ACT~\cite{ACT2025b}} & {$2.259\pm0.017$} & {$2.403\pm0.042$} & {$2.40\sigma$} \\
		{SPT~\cite{SPT2025}} & {$2.221\pm0.020$} & {$2.471\pm0.047$} & {$1.00\sigma$} \\
		{Joint SPA~\cite{SPT2025}} & {$2.2398\pm0.0095$} & {$2.437\pm0.034$} & {$1.98\sigma$} \\[2pt]
		\hline
	\end{tabular}
	\caption{Different CMB inferences of $\omega_b$---\textit{Planck} 2018 TT,TE,EE+lowE+lensing~\cite{Planck2018}, ACT DR6~\cite{ACT2025a}, SPT-3G D1~\cite{SPT2025}, and the combination of \textit{Planck}, ACT, and SPT (SPA) in Ref.~\cite{SPT2025}---with corresponding D/H predictions using the GP pipeline. Uncertainties include those from nuclear reaction rates, neutron lifetime, and $\omega_b$.}
    \label{tab:vary_omegab}
\end{table}

\begin{table}[!t]
	\centering
	\begin{tabular}{ccc}
		\hline
		\textbf{Reaction Network} & $\mathbf{10^{5} \times}$\textbf{{D/H}} & \thead{\textbf{Agreement with} \\\textbf{Cooke \textit{et al.}}} \\
		\hline
          	\rule{0pt}{10pt}{\textbf{This work}} & {$\mathbf{2.442\pm0.040}$} & {$\mathbf{1.70\boldsymbol{\sigma}}$} \\
		{PRIMAT} & {$2.444\pm0.037$} & {$1.74\sigma$} \\
		{PArthENoPE} & {$2.512\pm0.042$} & {$0.29\sigma$} \\
		{PRyMordial/YOF} & {$2.520\pm0.104$} & {$0.06\sigma$} \\
		{Cooke \textit{et al.}\ (measurement)} & {$2.527\pm0.030$} & {---} \\[2pt]
		\hline
	\end{tabular}
	\caption{Comparison of D/H predictions for this work and for the base nuclear reaction networks of PRIMAT, PArthENoPE, and PRyMordial/YOF~\cite{Burns2024,Yeh:2020mgl} for \textit{Planck} $\omega_b$.~The PRyMordial/YOF network uses Ref.~\cite{Mossa2020} for \dpg~and Ref.~\cite{Xu2013} for \ddn~and \ddp.~The Cooke \textit{et al.}~\cite{Cooke2018} measurement is also shown. LINX is used for all predictions, and uncertainties include those from nuclear reaction rates, neutron lifetime, and $\omega_b$.}
    \label{tab:D_H}
\end{table}

\section{Discussion}
\label{sec:discussion}

We have demonstrated that our method to predict D/H can reproduce the appropriate D/H distribution in mock tests and obtained fiducial results including marginalization over CMB determinations of $\omega_b$. 

Ours is not the only literature prediction for the primordial deuterium abundance.  As discussed above, the PRIMAT and PArthENoPE reaction networks use methodologies distinct from those introduced in this work, and can also be used to predict D/H.  Additional choices of reaction network that sometimes appear in the literature include the PRyMordial/YOF reaction network of Ref.~\cite{Burns2024}; this network is similar to that of Ref.~\cite{Yeh:2020mgl} and uses S-factors taken from the NACREII database~\cite{Xu2013}.  

However, we emphasize that \textit{ours is the only network that has been explicitly validated to produce robust D/H predictions and uncertainties}.  Table~\ref{tab:D_H} shows the D/H predictions from these networks.  We find agreement with the PRIMAT prediction, and mild disagreement with the PArthENoPE prediction.  We agree with the PRyMordial/YOF network, though this is largely a byproduct of the large uncertainty associated with this network; our determination achieves higher precision and has been demonstrated to provide the correct coverage.

As our prediction and the PArthENoPE prediction have similar error bars, we aim to understand the cause of the disagreement in these two networks.  We repeat our Monte Carlo validation procedure for S-factor fits with low-degree polynomials to mimic the procedure used to obtain the PArthENoPE network. We assume the same generating functions as above and perform degree-2 (4) polynomial fits to mock \ddp~(\ddn) data, matching polynomial degree to those used in the PArthENoPE network. We also include mock data representing Refs.~\cite{McNeill1951, Schulte1972, RG1985, Tumino2014}, as these datasets are used to construct the PArthENoPE network. 

We consider fits minimizing a $\chi^2$ expression adapted from Ref.~\cite{DAgostini1994}:
\begin{multline}
		\chi^2 = \sum_{k\in\mathrm{sets}} \bigg[\sum_{i=1}^{N_k} \frac{(S_{\rm th}(E_{ik}, a_l) - \omega_kS_{ik})^2}{\omega_k^2 \sigma_{ik}^2} + \frac{(\omega_k-1)^2}{\epsilon_k^2}\bigg],
	\label{eqn:poly_chi2_single}
\end{multline}
where $N_k$ is the number of data points in dataset $k$, $S_{\rm th}$ is the S-factor fit at $E_{ik}$ with polynomial coefficients $a_l$ (\textit{i.e.}\ $S_{\rm th}=\sum_l a_lE^l$), and $S_{ik}$ is the measured value at $E_{ik}$. $\{\omega_k\}$ are fit parameters modeling systematic errors. The final term in Eq.~\eqref{eqn:poly_chi2_single} is a penalty term that effectively serves as a Gaussian prior on $\{\omega_k\}$ centered at $1$ with width of the reported systematic uncertainty:
\begin{equation}
	\chi^2 = \ln{p(S_{ik}|a_l, \omega_k)} + \ln{p(\omega_k)} + \mathrm{const.}
\end{equation}
Minimization gives the maximum \textit{a posteriori} (MAP) estimate for the fit parameters, assuming uniform priors on $\{a_l\}$. Eq.~\eqref{eqn:poly_chi2_single} differs from the form used in the PArthENoPE network~\cite{Serpico2004, Pisanti2026}, which includes $N_k$ penalty terms per \footnote{We thank Ofelia Pisanti for informing us of a typo in Eq. (8) in Ref.~\cite{Pisanti2021}.}. Then, minimization no longer gives the MAP estimate and may bias toward smaller values of $|1-\omega_k|$, leading to poorer performance than the MAP estimate in Eq.~\eqref{eqn:poly_chi2_single} (see Supplementary Information). Uncertainties in polynomial fits are propagated to reaction rate uncertainties in the same manner as Refs.~\cite{Serpico2004, Pisanti2021} (see Supplementary Information).

Figure~\ref{fig:mc_test_single} also displays the Monte Carlo results for the polynomial fits (blue). Polynomials do not match the success of GPs for generating function recovery and D/H predictions; D/H prediction bias increases from $0.32-0.50$ with GPs to $0.71-1.33$ for polynomials. In the top-left panel of Figure~\ref{fig:mc_test_single}, \ddp~S-factor predictions from degree-2 polynomials consistently predict high values for $E<\qty{0.1}{MeV}$ and low values for $E>\qty{0.1}{MeV}$, even for the relatively simple $S_{\rm true}$ based on the theory calculation of Ref.~\cite{Arai2011}. 

A fully unbiased fitting procedure with robust uncertainties would give a set of $(\mathrm{D/H}_{\rm pred} - \mathrm{D/H}_{\rm true}) / \sigma_{\rm pred}$ converging to $\mathcal{N}(0,1)$ for infinite samples. Approximating the full distributions with empirical results, we quantify how similar these distributions are to $\mathcal{N}(0,1)$ with the KL divergence~\cite{Kullback1951} and the KS statistic~\cite{Kolmogorov1933} (see Supplementary Information). Both show that the GP predictions are more similar to the expected distribution, with KL divergences a factor of $\sim5$ ($10$) and KS statistics a factor of $\sim2$ ($3$) smaller with GPs for the theory (smoothed GP) generating function. These tests favor GPs, suggesting that increased model complexity relative to low-degree polynomials is necessary to conduct these fits.

In the era of precision cosmology, models are increasingly confronted by experimental data with uncertainties at the percent-level, and errors of just a few percent can be the difference between a successful model and a rejected hypothesis.  A trustworthy determination of all cosmological parameters and measured quantities is critical for the success of the field.  The validated methodology that we present here provides a robust prediction for the primordial deuterium abundance, which is critical for BBN analyses or BBN combinations with baryon acoustic oscillation (BAO) data, galaxy full shape analyses, and the CMB.  When combined with measurements of primordial deuterium, our procedure can be used to infer the baryon density of the universe at percent-level precision; this analysis will be performed in future work.  These efforts are complementary to the CMB inference of the baryon density, and consistency between these two epochs is a critical pillar of the $\Lambda$CDM paradigm. This work enables such consistency checks---and probes of extensions to $\Lambda$CDM---by providing a reliable means to predict the primordial deuterium abundance.

\section*{Acknowledgments}
We thank Frank Golf, David W.\ Hogg, Mariangela Lisanti, Pankaj Mehta, Siddharth Mishra-Sharma, Jo Moscoso, Ofelia Pisanti, Joshua Ruderman, and Ben Safdi for helpful discussions. We thank Ofelia Pisanti for her kind assistance in helping us understand the PArthENoPE fitting procedure. We thank attendees of the CERN theory group cosmology seminar for provocative questions about this work.  We thank David W.\ Hogg for comments on a draft version of this manuscript.  T.L. and H.L. are supported by the U.S. Department of Energy under grant DE-SC0026297. In addition, T.L. and H.L. are supported by the Cecile K. Dalton Career Development Professorship, endowed by Boston University trustee Nathaniel Dalton and Amy Gottleib Dalton.  C.G. is supported by the Office of High Energy Physics of the U.S. Department of Energy under contract DE-AC02-05CH11231.  This work makes use of matplotlib~\cite{Hunter2007}, JAX~\cite{Bradbury2018}, NumPy~\cite{Harris2020}, SciPy~\cite{Virtanen2020}, and MINUIT~\cite{James1994}.

\bibliography{references}

\begin{thebibliography}{72}%
\makeatletter
\providecommand \@ifxundefined [1]{%
 \@ifx{#1\undefined}
}%
\providecommand \@ifnum [1]{%
 \ifnum #1\expandafter \@firstoftwo
 \else \expandafter \@secondoftwo
 \fi
}%
\providecommand \@ifx [1]{%
 \ifx #1\expandafter \@firstoftwo
 \else \expandafter \@secondoftwo
 \fi
}%
\providecommand \natexlab [1]{#1}%
\providecommand \enquote  [1]{``#1''}%
\providecommand \bibnamefont  [1]{#1}%
\providecommand \bibfnamefont [1]{#1}%
\providecommand \citenamefont [1]{#1}%
\providecommand \href@noop [0]{\@secondoftwo}%
\providecommand \href [0]{\begingroup \@sanitize@url \@href}%
\providecommand \@href[1]{\@@startlink{#1}\@@href}%
\providecommand \@@href[1]{\endgroup#1\@@endlink}%
\providecommand \@sanitize@url [0]{\catcode `\\12\catcode `\$12\catcode
  `\&12\catcode `\#12\catcode `\^12\catcode `\_12\catcode `\%12\relax}%
\providecommand \@@startlink[1]{}%
\providecommand \@@endlink[0]{}%
\providecommand \url  [0]{\begingroup\@sanitize@url \@url }%
\providecommand \@url [1]{\endgroup\@href {#1}{\urlprefix }}%
\providecommand \urlprefix  [0]{URL }%
\providecommand \Eprint [0]{\href }%
\providecommand \doibase [0]{https://doi.org/}%
\providecommand \selectlanguage [0]{\@gobble}%
\providecommand \bibinfo  [0]{\@secondoftwo}%
\providecommand \bibfield  [0]{\@secondoftwo}%
\providecommand \translation [1]{[#1]}%
\providecommand \BibitemOpen [0]{}%
\providecommand \bibitemStop [0]{}%
\providecommand \bibitemNoStop [0]{.\EOS\space}%
\providecommand \EOS [0]{\spacefactor3000\relax}%
\providecommand \BibitemShut  [1]{\csname bibitem#1\endcsname}%
\let\auto@bib@innerbib\@empty
\bibitem [{\citenamefont {{Aver \textit{et al.}}}(2026)}]{Aver2026}%
  \BibitemOpen
  \bibfield  {author} {\bibinfo {author} {\bibfnamefont {E.}~\bibnamefont
  {{Aver \textit{et al.}}}},\ }\bibfield  {title} {\bibinfo {title} {{``The LBT
  Y$_\mathrm{P}$ Project IV: A New Value of the Primordial Helium
  Abundance''}},\ }\href@noop {} {\  (\bibinfo {year} {2026})},\ \Eprint
  {https://arxiv.org/abs/2601.22238} {arXiv:2601.22238 [astro-ph.CO]}
  \BibitemShut {NoStop}%
\bibitem [{\citenamefont {{Yeh \textit{et al.}}}(2026)}]{Yeh2026}%
  \BibitemOpen
  \bibfield  {author} {\bibinfo {author} {\bibfnamefont {T.-H.}\ \bibnamefont
  {{Yeh \textit{et al.}}}},\ }\bibfield  {title} {\bibinfo {title} {{``The LBT
  Y$_\mathrm{P}$ Project V: Cosmological Implications of a New Determination of
  Primordial $^4$He''}},\ }\href@noop {} {\  (\bibinfo {year} {2026})},\
  \Eprint {https://arxiv.org/abs/2601.22239} {arXiv:2601.22239 [astro-ph.CO]}
  \BibitemShut {NoStop}%
\bibitem [{\citenamefont {{Aghanim \textit{et al.}}}(2020)}]{Planck2018}%
  \BibitemOpen
  \bibfield  {author} {\bibinfo {author} {\bibfnamefont {N.}~\bibnamefont
  {{Aghanim \textit{et al.}}}} (\bibinfo {collaboration} {Planck
  Collaboration}),\ }\bibfield  {title} {\bibinfo {title} {{``Planck 2018
  results. VI. Cosmological parameters''}},\ }\href
  {https://doi.org/10.1051/0004-6361/201833910} {\bibfield  {journal} {\bibinfo
   {journal} {{A\&A}}\ }\textbf {\bibinfo {volume} {641}},\ \bibinfo {pages}
  {A6} (\bibinfo {year} {2020})},\ \Eprint {https://arxiv.org/abs/1807.06209}
  {arXiv:1807.06209 [astro-ph.CO]} \BibitemShut {NoStop}%
\bibitem [{\citenamefont {{Calabrese \textit{et al.}}}()}]{ACT2025b}%
  \BibitemOpen
  \bibfield  {author} {\bibinfo {author} {\bibfnamefont {E.}~\bibnamefont
  {{Calabrese \textit{et al.}}}},\ }\bibfield  {title} {\bibinfo {title}
  {{``The Atacama Cosmology Telescope: DR6 constraints on extended cosmological
  models''}},\ }\href {https://doi.org/10.1088/1475-7516/2025/11/063}
  {\bibfield  {journal} {\bibinfo  {journal} {JCAP}\ }\textbf {\bibinfo
  {volume} {2025}}\bibfield  {number} {\bibinfo  {number} { (11)},\ \bibinfo
  {pages} {063}},\ }\Eprint {https://arxiv.org/abs/2503.14454}
  {arXiv:2503.14454 [astro-ph.CO]} \BibitemShut {NoStop}%
\bibitem [{\citenamefont {{Camphuis \textit{et al.}}}(2026)}]{SPT2025}%
  \BibitemOpen
  \bibfield  {author} {\bibinfo {author} {\bibfnamefont {E.}~\bibnamefont
  {{Camphuis \textit{et al.}}}} (\bibinfo {collaboration} {SPT-3G}),\
  }\bibfield  {title} {\bibinfo {title} {{``SPT-3G D1: CMB temperature and
  polarization power spectra and cosmology from 2019 and 2020 observations of
  the SPT-3G main field''}},\ }\href {https://doi.org/10.1103/7wt3-9v2yv}
  {\bibfield  {journal} {\bibinfo  {journal} {Phys. Rev. D}\ }\textbf {\bibinfo
  {volume} {113}},\ \bibinfo {pages} {083504} (\bibinfo {year} {2026})},\
  \Eprint {https://arxiv.org/abs/2506.20707} {arXiv:2506.20707 [astro-ph.CO]}
  \BibitemShut {NoStop}%
\bibitem [{\citenamefont {Cooke}\ \emph {et~al.}(2018)\citenamefont {Cooke},
  \citenamefont {Pettini},\ and\ \citenamefont {Steidel}}]{Cooke2018}%
  \BibitemOpen
  \bibfield  {author} {\bibinfo {author} {\bibfnamefont {R.~J.}\ \bibnamefont
  {Cooke}}, \bibinfo {author} {\bibfnamefont {M.}~\bibnamefont {Pettini}},\
  and\ \bibinfo {author} {\bibfnamefont {C.~C.}\ \bibnamefont {Steidel}},\
  }\bibfield  {title} {\bibinfo {title} {{``One percent determination of the
  primordial deuterium abundance''}},\ }\href
  {https://doi.org/10.3847/1538-4357/aaab53} {\bibfield  {journal} {\bibinfo
  {journal} {Astrophys. J.}\ }\textbf {\bibinfo {volume} {855}},\ \bibinfo
  {pages} {102} (\bibinfo {year} {2018})},\ \Eprint
  {https://arxiv.org/abs/1710.11129} {arXiv:1710.11129 [astro-ph.CO]}
  \BibitemShut {NoStop}%
\bibitem [{\citenamefont {Iliadis}(2015)}]{Iliadis2015}%
  \BibitemOpen
  \bibfield  {author} {\bibinfo {author} {\bibfnamefont {C.}~\bibnamefont
  {Iliadis}},\ }\href {https://doi.org/10.1002/9783527692668} {\emph {\bibinfo
  {title} {Nuclear Physics of Stars}}},\ \bibinfo {edition} {2nd}\ ed.\
  (\bibinfo  {publisher} {Wiley-VCH},\ \bibinfo {address} {Weinheim},\ \bibinfo
  {year} {2015})\BibitemShut {NoStop}%
\bibitem [{\citenamefont {Gariazzo}\ \emph {et~al.}(2022)\citenamefont
  {Gariazzo}, \citenamefont {{de Salas}}, \citenamefont {Pisanti},\ and\
  \citenamefont {Consiglio}}]{Gariazzo2022}%
  \BibitemOpen
  \bibfield  {author} {\bibinfo {author} {\bibfnamefont {S.}~\bibnamefont
  {Gariazzo}}, \bibinfo {author} {\bibfnamefont {P.}~\bibnamefont {{de
  Salas}}}, \bibinfo {author} {\bibfnamefont {O.}~\bibnamefont {Pisanti}},\
  and\ \bibinfo {author} {\bibfnamefont {R.}~\bibnamefont {Consiglio}},\
  }\bibfield  {title} {\bibinfo {title} {{``PArthENoPE revolutions''}},\ }\href
  {https://doi.org/10.1016/j.cpc.2021.108205} {\bibfield  {journal} {\bibinfo
  {journal} {Computer Physics Communications}\ }\textbf {\bibinfo {volume}
  {271}},\ \bibinfo {pages} {108205} (\bibinfo {year} {2022})},\ \Eprint
  {https://arxiv.org/abs/2103.05027} {arXiv:2103.05027 [astro-ph.IM]}
  \BibitemShut {NoStop}%
\bibitem [{\citenamefont {Pitrou}\ \emph {et~al.}(2018)\citenamefont {Pitrou},
  \citenamefont {Coc}, \citenamefont {Uzan},\ and\ \citenamefont
  {Vangioni}}]{Pitrou2018}%
  \BibitemOpen
  \bibfield  {author} {\bibinfo {author} {\bibfnamefont {C.}~\bibnamefont
  {Pitrou}}, \bibinfo {author} {\bibfnamefont {A.}~\bibnamefont {Coc}},
  \bibinfo {author} {\bibfnamefont {J.}~\bibnamefont {Uzan}},\ and\ \bibinfo
  {author} {\bibfnamefont {E.}~\bibnamefont {Vangioni}},\ }\bibfield  {title}
  {\bibinfo {title} {{``Precision big bang nucleosynthesis with improved
  helium-4 predictions''}},\ }\href
  {https://doi.org/10.1016/j.physrep.2018.04.005} {\bibfield  {journal}
  {\bibinfo  {journal} {Phys. Rep.}\ }\textbf {\bibinfo {volume} {754}},\
  \bibinfo {pages} {1} (\bibinfo {year} {2018})},\ \Eprint
  {https://arxiv.org/abs/1801.08023} {arXiv:1801.08023 [astro-ph.CO]}
  \BibitemShut {NoStop}%
\bibitem [{\citenamefont {Serpico}\ \emph {et~al.}()\citenamefont {Serpico},
  \citenamefont {Esposito}, \citenamefont {Iocco}, \citenamefont {Mangano},
  \citenamefont {Miele},\ and\ \citenamefont {Pisanti}}]{Serpico2004}%
  \BibitemOpen
  \bibfield  {author} {\bibinfo {author} {\bibfnamefont {P.~D.}\ \bibnamefont
  {Serpico}}, \bibinfo {author} {\bibfnamefont {S.}~\bibnamefont {Esposito}},
  \bibinfo {author} {\bibfnamefont {F.}~\bibnamefont {Iocco}}, \bibinfo
  {author} {\bibfnamefont {G.}~\bibnamefont {Mangano}}, \bibinfo {author}
  {\bibfnamefont {G.}~\bibnamefont {Miele}},\ and\ \bibinfo {author}
  {\bibfnamefont {O.}~\bibnamefont {Pisanti}},\ }\bibfield  {title} {\bibinfo
  {title} {{``Nuclear reaction network for primordial nucleosynthesis: a
  detailed analysis of rates, uncertainties and light nuclei yields''}},\
  }\href {https://doi.org/10.1088/1475-7516/2004/12/010} {\bibfield  {journal}
  {\bibinfo  {journal} {JCAP}\ }\textbf {\bibinfo {volume} {2004}}\bibfield
  {number} {\bibinfo  {number} { (12)},\ \bibinfo {pages} {010}},\ }\Eprint
  {https://arxiv.org/abs/0408076} {arXiv:0408076 [astro-ph]} \BibitemShut
  {NoStop}%
\bibitem [{\citenamefont {Pisanti}\ \emph {et~al.}(2021)\citenamefont
  {Pisanti}, \citenamefont {Mangano}, \citenamefont {Miele},\ and\
  \citenamefont {Mazzela}}]{Pisanti2021}%
  \BibitemOpen
  \bibfield  {author} {\bibinfo {author} {\bibfnamefont {O.}~\bibnamefont
  {Pisanti}}, \bibinfo {author} {\bibfnamefont {G.}~\bibnamefont {Mangano}},
  \bibinfo {author} {\bibfnamefont {G.}~\bibnamefont {Miele}},\ and\ \bibinfo
  {author} {\bibfnamefont {P.}~\bibnamefont {Mazzela}},\ }\bibfield  {title}
  {\bibinfo {title} {{``Primordial deuterium after LUNA: concordances and error
  budget''}},\ }\href {https://doi.org/10.1088/1475-7516/2021/04/020}
  {\bibfield  {journal} {\bibinfo  {journal} {JCAP}\ }\textbf {\bibinfo
  {volume} {2021}}\bibfield  {number} {\bibinfo  {number} { (04)},\ \bibinfo
  {pages} {020}},\ }\Eprint {https://arxiv.org/abs/2011.11537}
  {arXiv:2011.11537 [astro-ph.CO]} \BibitemShut {NoStop}%
\bibitem [{\citenamefont {{I\~nesta G\'omez}}\ \emph
  {et~al.}(2017)\citenamefont {{I\~nesta G\'omez}}, \citenamefont {Iliadis},\
  and\ \citenamefont {Coc}}]{Inesta2017}%
  \BibitemOpen
  \bibfield  {author} {\bibinfo {author} {\bibfnamefont {{\'A}.}~\bibnamefont
  {{I\~nesta G\'omez}}}, \bibinfo {author} {\bibfnamefont {C.}~\bibnamefont
  {Iliadis}},\ and\ \bibinfo {author} {\bibfnamefont {A.}~\bibnamefont {Coc}},\
  }\bibfield  {title} {\bibinfo {title} {{``Bayesian Estimation of
  Thermonuclear Reaction Rates for Deuterium+Deuterium Reactions''}},\ }\href
  {https://doi.org/10.3847/1538-4357/aa9025} {\bibfield  {journal} {\bibinfo
  {journal} {Astrophys. J.}\ }\textbf {\bibinfo {volume} {849}},\ \bibinfo
  {pages} {134} (\bibinfo {year} {2017})},\ \Eprint
  {https://arxiv.org/abs/1710.01647} {arXiv:1710.01647 [astro-ph.IM]}
  \BibitemShut {NoStop}%
\bibitem [{\citenamefont {Moscoso}\ \emph {et~al.}(2021)\citenamefont
  {Moscoso}, \citenamefont {{de Souza}}, \citenamefont {Coc},\ and\
  \citenamefont {Iliadis}}]{Moscoso2021}%
  \BibitemOpen
  \bibfield  {author} {\bibinfo {author} {\bibfnamefont {J.}~\bibnamefont
  {Moscoso}}, \bibinfo {author} {\bibfnamefont {R.~S.}\ \bibnamefont {{de
  Souza}}}, \bibinfo {author} {\bibfnamefont {A.}~\bibnamefont {Coc}},\ and\
  \bibinfo {author} {\bibfnamefont {C.}~\bibnamefont {Iliadis}},\ }\bibfield
  {title} {\bibinfo {title} {{``Bayesian Estimation of the
  D($p$,$\gamma$)$^3$He Thermonuclear Reaction Rate''}},\ }\href
  {https://doi.org/10.3847/1538-4357/ac1db0} {\bibfield  {journal} {\bibinfo
  {journal} {Astrophys. J.}\ }\textbf {\bibinfo {volume} {923}},\ \bibinfo
  {pages} {49} (\bibinfo {year} {2021})},\ \Eprint
  {https://arxiv.org/abs/2109.00049} {arXiv:2109.00049 [astro-ph.CO]}
  \BibitemShut {NoStop}%
\bibitem [{\citenamefont {Giovanetti}\ \emph {et~al.}(2025)\citenamefont
  {Giovanetti}, \citenamefont {Lisanti}, \citenamefont {Liu}, \citenamefont
  {Mishra-Sharma},\ and\ \citenamefont {Ruderman}}]{Giovanetti2025}%
  \BibitemOpen
  \bibfield  {author} {\bibinfo {author} {\bibfnamefont {C.}~\bibnamefont
  {Giovanetti}}, \bibinfo {author} {\bibfnamefont {M.}~\bibnamefont {Lisanti}},
  \bibinfo {author} {\bibfnamefont {H.}~\bibnamefont {Liu}}, \bibinfo {author}
  {\bibfnamefont {S.}~\bibnamefont {Mishra-Sharma}},\ and\ \bibinfo {author}
  {\bibfnamefont {J.~T.}\ \bibnamefont {Ruderman}},\ }\bibfield  {title}
  {\bibinfo {title} {{``Fast, differentiable, and extensible big bang
  nucleosynthesis package''}},\ }\href {https://doi.org/10.1103/f3tj-r882}
  {\bibfield  {journal} {\bibinfo  {journal} {Phys. Rev. D}\ }\textbf {\bibinfo
  {volume} {112}},\ \bibinfo {pages} {063531} (\bibinfo {year} {2025})},\
  \Eprint {https://arxiv.org/abs/2408.14538} {arXiv:2408.14538 [astro-ph.CO]}
  \BibitemShut {NoStop}%
\bibitem [{\citenamefont {Rasmussen}\ and\ \citenamefont
  {Williams}(2006)}]{Rasmussen2006}%
  \BibitemOpen
  \bibfield  {author} {\bibinfo {author} {\bibfnamefont {C.~E.}\ \bibnamefont
  {Rasmussen}}\ and\ \bibinfo {author} {\bibfnamefont {C.~K.~I.}\ \bibnamefont
  {Williams}},\ }\href {https://doi.org/10.7551/mitpress/3206.001.0001} {\emph
  {\bibinfo {title} {Gaussian Processes for Machine Learning}}}\ (\bibinfo
  {publisher} {MIT Press},\ \bibinfo {address} {Cambridge, MA},\ \bibinfo
  {year} {2006})\BibitemShut {NoStop}%
\bibitem [{\citenamefont {Cyburt}(2004)}]{Cyburt2004}%
  \BibitemOpen
  \bibfield  {author} {\bibinfo {author} {\bibfnamefont {R.~H.}\ \bibnamefont
  {Cyburt}},\ }\bibfield  {title} {\bibinfo {title} {{``Primordial
  nucleosynthesis for the new cosmology: Determining uncertainties and
  examining concordance''}},\ }\href
  {https://doi.org/10.1103/PhysRevD.70.023505} {\bibfield  {journal} {\bibinfo
  {journal} {Phys. Rev. D}\ }\textbf {\bibinfo {volume} {70}},\ \bibinfo
  {pages} {023505} (\bibinfo {year} {2004})}\BibitemShut {NoStop}%
\bibitem [{\citenamefont {Coc}\ and\ \citenamefont {Vangioni}(2010)}]{Coc2010}%
  \BibitemOpen
  \bibfield  {author} {\bibinfo {author} {\bibfnamefont {A.}~\bibnamefont
  {Coc}}\ and\ \bibinfo {author} {\bibfnamefont {E.}~\bibnamefont {Vangioni}},\
  }\bibfield  {title} {\bibinfo {title} {{``Big-Bang nucleosynthesis with
  updated nuclear data''}},\ }\href
  {https://doi.org/10.1088/1742-6596/202/1/012001} {\bibfield  {journal}
  {\bibinfo  {journal} {J. Phys.: Conf. Ser.}\ }\textbf {\bibinfo {volume}
  {202}},\ \bibinfo {pages} {012001} (\bibinfo {year} {2010})}\BibitemShut
  {NoStop}%
\bibitem [{\citenamefont {Krauss}\ \emph {et~al.}(1987)\citenamefont {Krauss},
  \citenamefont {Becker}, \citenamefont {Trautvetter},\ and\ \citenamefont
  {Rolfs}}]{Krauss1987}%
  \BibitemOpen
  \bibfield  {author} {\bibinfo {author} {\bibfnamefont {A.}~\bibnamefont
  {Krauss}}, \bibinfo {author} {\bibfnamefont {H.~W.}\ \bibnamefont {Becker}},
  \bibinfo {author} {\bibfnamefont {H.~P.}\ \bibnamefont {Trautvetter}},\ and\
  \bibinfo {author} {\bibfnamefont {C.}~\bibnamefont {Rolfs}},\ }\bibfield
  {title} {\bibinfo {title} {{``Low-energy fusion cross sections of $D$+$D$ and
  $D$+$^3$He reactions''}},\ }\href
  {https://doi.org/10.1016/0375-9474(87)90302-2} {\bibfield  {journal}
  {\bibinfo  {journal} {Nucl. Phys. A}\ }\textbf {\bibinfo {volume} {465}},\
  \bibinfo {pages} {150} (\bibinfo {year} {1987})}\BibitemShut {NoStop}%
\bibitem [{\citenamefont {Brown}\ and\ \citenamefont
  {Jarmie}(1990)}]{Brown1990}%
  \BibitemOpen
  \bibfield  {author} {\bibinfo {author} {\bibfnamefont {R.~E.}\ \bibnamefont
  {Brown}}\ and\ \bibinfo {author} {\bibfnamefont {N.}~\bibnamefont {Jarmie}},\
  }\bibfield  {title} {\bibinfo {title} {{``Differential cross sections at low
  energies for $^2$H($d$,$p$)$^3$H and $^2$H($d$,$n$)$^3$He''}},\ }\href
  {https://doi.org/10.1103/PhysRevC.41.1391} {\bibfield  {journal} {\bibinfo
  {journal} {Phys. Rev. C}\ }\textbf {\bibinfo {volume} {41}},\ \bibinfo
  {pages} {1391} (\bibinfo {year} {1990})}\BibitemShut {NoStop}%
\bibitem [{\citenamefont {Greife}\ \emph {et~al.}(1995)\citenamefont {Greife},
  \citenamefont {Gorris}, \citenamefont {Junker}, \citenamefont {Rolfs},\ and\
  \citenamefont {Zahnow}}]{Greife1995}%
  \BibitemOpen
  \bibfield  {author} {\bibinfo {author} {\bibfnamefont {U.}~\bibnamefont
  {Greife}}, \bibinfo {author} {\bibfnamefont {F.}~\bibnamefont {Gorris}},
  \bibinfo {author} {\bibfnamefont {M.}~\bibnamefont {Junker}}, \bibinfo
  {author} {\bibfnamefont {C.}~\bibnamefont {Rolfs}},\ and\ \bibinfo {author}
  {\bibfnamefont {D.}~\bibnamefont {Zahnow}},\ }\bibfield  {title} {\bibinfo
  {title} {{``Oppenheimer-Phillips effect and electron screening in $d$+$d$
  fusion reactions''}},\ }\href {https://doi.org/10.1007/BF01292792} {\bibfield
   {journal} {\bibinfo  {journal} {Z. Phys. A}\ }\textbf {\bibinfo {volume}
  {351}},\ \bibinfo {pages} {107} (\bibinfo {year} {1995})}\BibitemShut
  {NoStop}%
\bibitem [{\citenamefont {Leonard}\ \emph {et~al.}(2006)\citenamefont
  {Leonard}, \citenamefont {Karwowski}, \citenamefont {Brune}, \citenamefont
  {Fisher},\ and\ \citenamefont {Ludwig}}]{Leonard2006}%
  \BibitemOpen
  \bibfield  {author} {\bibinfo {author} {\bibfnamefont {D.~S.}\ \bibnamefont
  {Leonard}}, \bibinfo {author} {\bibfnamefont {H.~J.}\ \bibnamefont
  {Karwowski}}, \bibinfo {author} {\bibfnamefont {C.~R.}\ \bibnamefont
  {Brune}}, \bibinfo {author} {\bibfnamefont {B.~M.}\ \bibnamefont {Fisher}},\
  and\ \bibinfo {author} {\bibfnamefont {E.~J.}\ \bibnamefont {Ludwig}},\
  }\bibfield  {title} {\bibinfo {title} {{``Precision measurements of
  $^2$H($d$,$p$)$^3$He and $^2$H($d$,$n$)$^3$He total cross sections at Big
  Bang nucleosynthesis energies''}},\ }\href
  {https://doi.org/10.1103/PhysRevC.73.045801} {\bibfield  {journal} {\bibinfo
  {journal} {Phys. Rev. C}\ }\textbf {\bibinfo {volume} {73}},\ \bibinfo
  {pages} {045801} (\bibinfo {year} {2006})}\BibitemShut {NoStop}%
\bibitem [{\citenamefont {{Chen \textit{et al.}}}(2026)}]{Chen2026}%
  \BibitemOpen
  \bibfield  {author} {\bibinfo {author} {\bibfnamefont {Y.}~\bibnamefont
  {{Chen \textit{et al.}}}},\ }\bibfield  {title} {\bibinfo {title}
  {{``High-Precision Measurement of $D$($\gamma$,$n$)$p$ Photodisintegration
  Reaction and Implications for Big-Bang Nucleosynthesis''}},\ }\href
  {https://doi.org/10.1103/tbbt-s819} {\bibfield  {journal} {\bibinfo
  {journal} {Phys. Rev. Lett.}\ }\textbf {\bibinfo {volume} {136}},\ \bibinfo
  {pages} {052701} (\bibinfo {year} {2026})},\ \Eprint
  {https://arxiv.org/abs/2509.11743} {arXiv:2509.11743 [nucl-ex]} \BibitemShut
  {NoStop}%
\bibitem [{\citenamefont {Czarnecki}\ \emph {et~al.}(2018)\citenamefont
  {Czarnecki}, \citenamefont {Marciano},\ and\ \citenamefont
  {Sirlin}}]{Czarnecki2018}%
  \BibitemOpen
  \bibfield  {author} {\bibinfo {author} {\bibfnamefont {A.}~\bibnamefont
  {Czarnecki}}, \bibinfo {author} {\bibfnamefont {W.~J.}\ \bibnamefont
  {Marciano}},\ and\ \bibinfo {author} {\bibfnamefont {A.}~\bibnamefont
  {Sirlin}},\ }\bibfield  {title} {\bibinfo {title} {{``Neutron Lifetime and
  Axial Coupling Connection''}},\ }\href
  {https://doi.org/10.1103/PhysRevLett.120.202002} {\bibfield  {journal}
  {\bibinfo  {journal} {Phys. Rev. Lett.}\ }\textbf {\bibinfo {volume} {120}},\
  \bibinfo {pages} {202002} (\bibinfo {year} {2018})},\ \Eprint
  {https://arxiv.org/abs/1802.01804} {arXiv:1802.01804 [hep-ph]} \BibitemShut
  {NoStop}%
\bibitem [{\citenamefont {Arai}\ \emph {et~al.}(2011)\citenamefont {Arai},
  \citenamefont {Aoyama}, \citenamefont {Suzuki}, \citenamefont
  {Descouvemont},\ and\ \citenamefont {Baye}}]{Arai2011}%
  \BibitemOpen
  \bibfield  {author} {\bibinfo {author} {\bibfnamefont {K.}~\bibnamefont
  {Arai}}, \bibinfo {author} {\bibfnamefont {S.}~\bibnamefont {Aoyama}},
  \bibinfo {author} {\bibfnamefont {Y.}~\bibnamefont {Suzuki}}, \bibinfo
  {author} {\bibfnamefont {P.}~\bibnamefont {Descouvemont}},\ and\ \bibinfo
  {author} {\bibfnamefont {D.}~\bibnamefont {Baye}},\ }\bibfield  {title}
  {\bibinfo {title} {{``Tensor Force Manifestations in \textit{Ab Initio} Study
  of the $^2$H($d$,$\gamma$)$^4$He, $^2$H($d$,$p$)$^3$H, and
  $^2$H($d$,$n$)$^3$He Reactions''}},\ }\href
  {https://doi.org/10.1103/PhysRevLett.107.132502} {\bibfield  {journal}
  {\bibinfo  {journal} {Phys. Rev. Lett}\ }\textbf {\bibinfo {volume} {107}},\
  \bibinfo {pages} {132502} (\bibinfo {year} {2011})}\BibitemShut {NoStop}%
\bibitem [{\citenamefont {Kullback}\ and\ \citenamefont
  {Leibler}(1951)}]{Kullback1951}%
  \BibitemOpen
  \bibfield  {author} {\bibinfo {author} {\bibfnamefont {S.}~\bibnamefont
  {Kullback}}\ and\ \bibinfo {author} {\bibfnamefont {R.~A.}\ \bibnamefont
  {Leibler}},\ }\bibfield  {title} {\bibinfo {title} {{``On Information and
  Sufficiency''}},\ }\href {https://doi.org/10.1214/aoms/1177729694} {\bibfield
   {journal} {\bibinfo  {journal} {Ann. Math. Stat.}\ }\textbf {\bibinfo
  {volume} {22}},\ \bibinfo {pages} {79} (\bibinfo {year} {1951})}\BibitemShut
  {NoStop}%
\bibitem [{\citenamefont {Kolmogorov}(1933)}]{Kolmogorov1933}%
  \BibitemOpen
  \bibfield  {author} {\bibinfo {author} {\bibfnamefont {A.~N.}\ \bibnamefont
  {Kolmogorov}},\ }\bibfield  {title} {\bibinfo {title} {{``Sulla
  determinazione empirica di una legge di distribuzione''}},\ }\href@noop {}
  {\bibfield  {journal} {\bibinfo  {journal} {Giornale dell'Istituto Italiano
  degli Attuari}\ }\textbf {\bibinfo {volume} {4}},\ \bibinfo {pages} {83}
  (\bibinfo {year} {1933})}\BibitemShut {NoStop}%
\bibitem [{\citenamefont {Marcucci}\ \emph {et~al.}(2016)\citenamefont
  {Marcucci}, \citenamefont {Mangano}, \citenamefont {Kievsky},\ and\
  \citenamefont {Viviani}}]{Marcucci2016}%
  \BibitemOpen
  \bibfield  {author} {\bibinfo {author} {\bibfnamefont {L.~E.}\ \bibnamefont
  {Marcucci}}, \bibinfo {author} {\bibfnamefont {G.}~\bibnamefont {Mangano}},
  \bibinfo {author} {\bibfnamefont {A.}~\bibnamefont {Kievsky}},\ and\ \bibinfo
  {author} {\bibfnamefont {M.}~\bibnamefont {Viviani}},\ }\bibfield  {title}
  {\bibinfo {title} {{``Implication of the Proton-Deuteron Radiative Capture
  for Big Bang Nucleosynthesis''}},\ }\href
  {https://doi.org/10.1103/PhysRevLett.116.102501} {\bibfield  {journal}
  {\bibinfo  {journal} {Phys. Rev. Lett}\ }\textbf {\bibinfo {volume} {116}},\
  \bibinfo {pages} {102501} (\bibinfo {year} {2016})},\ \Eprint
  {https://arxiv.org/abs/1510.07877} {arXiv:1510.07877 [nucl-th]} \BibitemShut
  {NoStop}%
\bibitem [{\citenamefont {{Louis \textit{et al.}}}()}]{ACT2025a}%
  \BibitemOpen
  \bibfield  {author} {\bibinfo {author} {\bibfnamefont {T.}~\bibnamefont
  {{Louis \textit{et al.}}}},\ }\bibfield  {title} {\bibinfo {title} {{``The
  Atacama Cosmology Telescope: DR6 power spectra, likelihoods, and $\Lambda$CDM
  parameters''}},\ }\href {https://doi.org/10.1088/1475-7516/2025/11/062}
  {\bibfield  {journal} {\bibinfo  {journal} {JCAP}\ }\textbf {\bibinfo
  {volume} {2025}}\bibfield  {number} {\bibinfo  {number} { (11)},\ \bibinfo
  {pages} {062}},\ }\Eprint {https://arxiv.org/abs/2503.14452}
  {arXiv:2503.14452 [astro-ph.CO]} \BibitemShut {NoStop}%
\bibitem [{\citenamefont {Burns}\ \emph {et~al.}(2024)\citenamefont {Burns},
  \citenamefont {Tait},\ and\ \citenamefont {Valli}}]{Burns2024}%
  \BibitemOpen
  \bibfield  {author} {\bibinfo {author} {\bibfnamefont {A.-K.}\ \bibnamefont
  {Burns}}, \bibinfo {author} {\bibfnamefont {T.~M.}\ \bibnamefont {Tait}},\
  and\ \bibinfo {author} {\bibfnamefont {M.}~\bibnamefont {Valli}},\ }\bibfield
   {title} {\bibinfo {title} {{``PRyMordial: The First Three Minutes, Within
  and Beyond the Standard Model''}},\ }\href
  {https://doi.org/10.1140/epjc/s10052-024-12442-0} {\bibfield  {journal}
  {\bibinfo  {journal} {Eur. Phys. J. C}\ }\textbf {\bibinfo {volume} {84}},\
  \bibinfo {pages} {86} (\bibinfo {year} {2024})},\ \Eprint
  {https://arxiv.org/abs/2307.07061} {arXiv:2307.07061 [hep-ph]} \BibitemShut
  {NoStop}%
\bibitem [{\citenamefont {Yeh}\ \emph {et~al.}(2021)\citenamefont {Yeh},
  \citenamefont {Olive},\ and\ \citenamefont {Fields}}]{Yeh:2020mgl}%
  \BibitemOpen
  \bibfield  {author} {\bibinfo {author} {\bibfnamefont {T.-H.}\ \bibnamefont
  {Yeh}}, \bibinfo {author} {\bibfnamefont {K.~A.}\ \bibnamefont {Olive}},\
  and\ \bibinfo {author} {\bibfnamefont {B.~D.}\ \bibnamefont {Fields}},\
  }\bibfield  {title} {\bibinfo {title} {{``The impact of new
  $d$($p$,$\gamma$)$^3$He rates on Big Bang Nucleosynthesis''}},\ }\href
  {https://doi.org/10.1088/1475-7516/2021/03/046} {\bibfield  {journal}
  {\bibinfo  {journal} {JCAP}\ }\textbf {\bibinfo {volume} {03}},\ \bibinfo
  {pages} {046}},\ \Eprint {https://arxiv.org/abs/2011.13874} {arXiv:2011.13874
  [astro-ph.CO]} \BibitemShut {NoStop}%
\bibitem [{\citenamefont {{Mossa \textit{et al.}}}(2020)}]{Mossa2020}%
  \BibitemOpen
  \bibfield  {author} {\bibinfo {author} {\bibfnamefont {V.}~\bibnamefont
  {{Mossa \textit{et al.}}}},\ }\bibfield  {title} {\bibinfo {title} {{``The
  baryon density of the Universe from an improved rate of deuterium
  burning''}},\ }\href {https://doi.org/10.1038/s41586-020-2878-4} {\bibfield
  {journal} {\bibinfo  {journal} {Nature}\ }\textbf {\bibinfo {volume} {587}},\
  \bibinfo {pages} {210} (\bibinfo {year} {2020})}\BibitemShut {NoStop}%
\bibitem [{\citenamefont {Xu}\ \emph {et~al.}(2013)\citenamefont {Xu},
  \citenamefont {Takahashi}, \citenamefont {Goriely}, \citenamefont {Arnould},
  \citenamefont {Ohta},\ and\ \citenamefont {Utsunomiya}}]{Xu2013}%
  \BibitemOpen
  \bibfield  {author} {\bibinfo {author} {\bibfnamefont {Y.}~\bibnamefont
  {Xu}}, \bibinfo {author} {\bibfnamefont {K.}~\bibnamefont {Takahashi}},
  \bibinfo {author} {\bibfnamefont {S.}~\bibnamefont {Goriely}}, \bibinfo
  {author} {\bibfnamefont {M.}~\bibnamefont {Arnould}}, \bibinfo {author}
  {\bibfnamefont {M.}~\bibnamefont {Ohta}},\ and\ \bibinfo {author}
  {\bibfnamefont {H.}~\bibnamefont {Utsunomiya}},\ }\bibfield  {title}
  {\bibinfo {title} {{``NACRE II: an update of the NACRE compilation of
  charged-particle-induced thermonuclear reaction rates for nuclei with mass
  number $A<16$''}},\ }\href {https://doi.org/10.1016/j.nuclphysa.2013.09.007}
  {\bibfield  {journal} {\bibinfo  {journal} {Nucl. Phys. A}\ }\textbf
  {\bibinfo {volume} {918}},\ \bibinfo {pages} {61} (\bibinfo {year} {2013})},\
  \Eprint {https://arxiv.org/abs/1310.7099} {arXiv:1310.7099 [nucl-th]}
  \BibitemShut {NoStop}%
\bibitem [{\citenamefont {McNeill}\ and\ \citenamefont
  {Keyser}(1951)}]{McNeill1951}%
  \BibitemOpen
  \bibfield  {author} {\bibinfo {author} {\bibfnamefont {K.~G.}\ \bibnamefont
  {McNeill}}\ and\ \bibinfo {author} {\bibfnamefont {G.~M.}\ \bibnamefont
  {Keyser}},\ }\bibfield  {title} {\bibinfo {title} {{``The Relative
  Probabilities and Absolute Cross Sections of the D --- D Reactions''}},\
  }\href {https://doi.org/10.1103/PhysRev.81.602} {\bibfield  {journal}
  {\bibinfo  {journal} {Phys. Rev.}\ }\textbf {\bibinfo {volume} {81}},\
  \bibinfo {pages} {602} (\bibinfo {year} {1951})}\BibitemShut {NoStop}%
\bibitem [{\citenamefont {Schulte}\ \emph {et~al.}(1972)\citenamefont
  {Schulte}, \citenamefont {Cosack}, \citenamefont {Obst},\ and\ \citenamefont
  {Weil}}]{Schulte1972}%
  \BibitemOpen
  \bibfield  {author} {\bibinfo {author} {\bibfnamefont {R.~L.}\ \bibnamefont
  {Schulte}}, \bibinfo {author} {\bibfnamefont {M.}~\bibnamefont {Cosack}},
  \bibinfo {author} {\bibfnamefont {A.~W.}\ \bibnamefont {Obst}},\ and\
  \bibinfo {author} {\bibfnamefont {J.~L.}\ \bibnamefont {Weil}},\ }\bibfield
  {title} {\bibinfo {title} {{``$^2$H+ reactions from 1.96 to 6.20 MeV''}},\
  }\href {https://doi.org/10.1016/0375-9474(72)90093-0} {\bibfield  {journal}
  {\bibinfo  {journal} {Nucl. Phys. A}\ }\textbf {\bibinfo {volume} {192}},\
  \bibinfo {pages} {609} (\bibinfo {year} {1972})}\BibitemShut {NoStop}%
\bibitem [{\citenamefont {{First Research
  Group}}(1985{\natexlab{a}})}]{RG1985}%
  \BibitemOpen
  \bibfield  {author} {\bibinfo {author} {\bibnamefont {{First Research
  Group}}},\ }\href@noop {} {\bibfield  {journal} {\bibinfo  {journal} {Chin.
  J. Nucl. High Energy Phys.}\ }\textbf {\bibinfo {volume} {9}},\ \bibinfo
  {pages} {723} (\bibinfo {year} {1985}{\natexlab{a}})}\BibitemShut {NoStop}%
\bibitem [{\citenamefont {{Tumino \textit{et al.}}}(2014)}]{Tumino2014}%
  \BibitemOpen
  \bibfield  {author} {\bibinfo {author} {\bibfnamefont {A.}~\bibnamefont
  {{Tumino \textit{et al.}}}},\ }\bibfield  {title} {\bibinfo {title} {{``New
  determination of the $^2$H($d$,$p$)$^3$H and $^2$H($d$,$n$)$^3$He reaction
  rates at astrophysical energies''}},\ }\href
  {https://doi.org/10.1088/0004-637X/785/2/96} {\bibfield  {journal} {\bibinfo
  {journal} {Astrophys. J.}\ }\textbf {\bibinfo {volume} {785}},\ \bibinfo
  {pages} {96} (\bibinfo {year} {2014})}\BibitemShut {NoStop}%
\bibitem [{\citenamefont {D'Agostini}(1994)}]{DAgostini1994}%
  \BibitemOpen
  \bibfield  {author} {\bibinfo {author} {\bibfnamefont {G.}~\bibnamefont
  {D'Agostini}},\ }\bibfield  {title} {\bibinfo {title} {{``On the use of the
  covariance matrix to fit correlated data''}},\ }\href
  {https://doi.org/10.1016/0168-9002(94)90719-6} {\bibfield  {journal}
  {\bibinfo  {journal} {NIM-A}\ }\textbf {\bibinfo {volume} {346}},\ \bibinfo
  {pages} {306} (\bibinfo {year} {1994})}\BibitemShut {NoStop}%
\bibitem [{\citenamefont {Pisanti}()}]{Pisanti2026}%
  \BibitemOpen
  \bibfield  {author} {\bibinfo {author} {\bibfnamefont {O.}~\bibnamefont
  {Pisanti}},\ }\href@noop {} {}\bibinfo {howpublished} {private
  communication}\BibitemShut {NoStop}%
\bibitem [{\citenamefont {Hunter}(2007)}]{Hunter2007}%
  \BibitemOpen
  \bibfield  {author} {\bibinfo {author} {\bibfnamefont {J.~D.}\ \bibnamefont
  {Hunter}},\ }\bibfield  {title} {\bibinfo {title} {{``Matplotlib: A 2D
  graphics environment''}},\ }\href {https://doi.org/10.1109/MCSE.2007.55}
  {\bibfield  {journal} {\bibinfo  {journal} {Computing in Science \&
  Engineering}\ }\textbf {\bibinfo {volume} {9}},\ \bibinfo {pages} {90}
  (\bibinfo {year} {2007})}\BibitemShut {NoStop}%
\bibitem [{\citenamefont {Bradbury}\ \emph {et~al.}(2018)\citenamefont
  {Bradbury}, \citenamefont {Frostig}, \citenamefont {Hawkins}, \citenamefont
  {Johnson}, \citenamefont {Leary}, \citenamefont {Maclaurina}, \citenamefont
  {Necula}, \citenamefont {Paszke}, \citenamefont {VanderPlas}, \citenamefont
  {Wanderman-Milne},\ and\ \citenamefont {Zhang}}]{Bradbury2018}%
  \BibitemOpen
  \bibfield  {author} {\bibinfo {author} {\bibfnamefont {J.}~\bibnamefont
  {Bradbury}}, \bibinfo {author} {\bibfnamefont {R.}~\bibnamefont {Frostig}},
  \bibinfo {author} {\bibfnamefont {P.}~\bibnamefont {Hawkins}}, \bibinfo
  {author} {\bibfnamefont {M.~J.}\ \bibnamefont {Johnson}}, \bibinfo {author}
  {\bibfnamefont {C.}~\bibnamefont {Leary}}, \bibinfo {author} {\bibfnamefont
  {D.}~\bibnamefont {Maclaurina}}, \bibinfo {author} {\bibfnamefont
  {G.}~\bibnamefont {Necula}}, \bibinfo {author} {\bibfnamefont
  {A.}~\bibnamefont {Paszke}}, \bibinfo {author} {\bibfnamefont
  {J.}~\bibnamefont {VanderPlas}}, \bibinfo {author} {\bibfnamefont
  {S.}~\bibnamefont {Wanderman-Milne}},\ and\ \bibinfo {author} {\bibfnamefont
  {Q.}~\bibnamefont {Zhang}},\ }\bibfield  {title} {\bibinfo {title} {{``JAX:
  composable transformations of Python+NumPY programs''}},\ }\href
  {https://github.com/jax-ml/jax} {\  (\bibinfo {year} {2018})}\BibitemShut
  {NoStop}%
\bibitem [{\citenamefont {{Harris \textit{et al.}}}(2020)}]{Harris2020}%
  \BibitemOpen
  \bibfield  {author} {\bibinfo {author} {\bibfnamefont {C.~R.}\ \bibnamefont
  {{Harris \textit{et al.}}}},\ }\bibfield  {title} {\bibinfo {title} {{``Array
  programming with NumPy''}},\ }\href
  {https://doi.org/10.1038/s41586-020-2649-2} {\bibfield  {journal} {\bibinfo
  {journal} {Nature}\ }\textbf {\bibinfo {volume} {585}},\ \bibinfo {pages}
  {357} (\bibinfo {year} {2020})},\ \Eprint {https://arxiv.org/abs/2006.10256}
  {arXiv:2006.10256 [cs.MS]} \BibitemShut {NoStop}%
\bibitem [{\citenamefont {{Virtanen \textit{et al.}}}(2020)}]{Virtanen2020}%
  \BibitemOpen
  \bibfield  {author} {\bibinfo {author} {\bibfnamefont {P.}~\bibnamefont
  {{Virtanen \textit{et al.}}}},\ }\bibfield  {title} {\bibinfo {title}
  {{``SciPy 1.0: Fundamental Algorithms for Scientific Computing in
  Python''}},\ }\href {https://doi.org/10.1038/s41592-019-0686-2} {\bibfield
  {journal} {\bibinfo  {journal} {Nature Methods}\ }\textbf {\bibinfo {volume}
  {17}},\ \bibinfo {pages} {261} (\bibinfo {year} {2020})},\ \Eprint
  {https://arxiv.org/abs/1907.10121} {arXiv:1907.10121 [cs.MS]} \BibitemShut
  {NoStop}%
\bibitem [{\citenamefont {James}(1994)}]{James1994}%
  \BibitemOpen
  \bibfield  {author} {\bibinfo {author} {\bibfnamefont {F.}~\bibnamefont
  {James}},\ }\href@noop {} {\emph {\bibinfo {title} {{``MINUIT Function
  Minimization and Error Analysis: Reference Manual Version 94.1''}}}},\
  \bibinfo {type} {Tech. Rep.}\ \bibinfo {number} {CERN Program Library D506}\
  (\bibinfo  {institution} {CERN},\ \bibinfo {address} {Geneva},\ \bibinfo
  {year} {1994})\BibitemShut {NoStop}%
\bibitem [{\citenamefont {Ma}\ \emph {et~al.}(1997)\citenamefont {Ma},
  \citenamefont {Karwowski}, \citenamefont {Brune}, \citenamefont {Ayer},
  \citenamefont {Black}, \citenamefont {Blackmon},\ and\ \citenamefont
  {Ludwig}}]{Ma1997}%
  \BibitemOpen
  \bibfield  {author} {\bibinfo {author} {\bibfnamefont {L.}~\bibnamefont
  {Ma}}, \bibinfo {author} {\bibfnamefont {H.~J.}\ \bibnamefont {Karwowski}},
  \bibinfo {author} {\bibfnamefont {C.~R.}\ \bibnamefont {Brune}}, \bibinfo
  {author} {\bibfnamefont {Z.}~\bibnamefont {Ayer}}, \bibinfo {author}
  {\bibfnamefont {T.~C.}\ \bibnamefont {Black}}, \bibinfo {author}
  {\bibfnamefont {J.~C.}\ \bibnamefont {Blackmon}},\ and\ \bibinfo {author}
  {\bibfnamefont {E.~J.}\ \bibnamefont {Ludwig}},\ }\bibfield  {title}
  {\bibinfo {title} {{``Measurements of $^1$H($d$,$\gamma$)$^3$He and
  $^2$H($p$,$\gamma$)$^3$He at very low energies''}},\ }\href
  {https://doi.org/10.1103/PhysRevC.55.588} {\bibfield  {journal} {\bibinfo
  {journal} {Phys. Rev. C}\ }\textbf {\bibinfo {volume} {55}},\ \bibinfo
  {pages} {588} (\bibinfo {year} {1997})}\BibitemShut {NoStop}%
\bibitem [{\citenamefont {{Schmid \textit{et al.}}}(1997)}]{Schmid1997}%
  \BibitemOpen
  \bibfield  {author} {\bibinfo {author} {\bibfnamefont {G.~J.}\ \bibnamefont
  {{Schmid \textit{et al.}}}},\ }\bibfield  {title} {\bibinfo {title} {{``The
  $^2$H($\vec{p}$,$\gamma$)$^3$He and $^1$H($\vec{d}$,$\gamma$)$^3$He reactions
  below 80 keV''}},\ }\href {https://doi.org/10.1103/PhysRevC.56.2565}
  {\bibfield  {journal} {\bibinfo  {journal} {Phys. Rev. C}\ }\textbf {\bibinfo
  {volume} {56}},\ \bibinfo {pages} {2565} (\bibinfo {year}
  {1997})}\BibitemShut {NoStop}%
\bibitem [{\citenamefont {{Casella \textit{et al.}}}(2002)}]{Casella2002}%
  \BibitemOpen
  \bibfield  {author} {\bibinfo {author} {\bibfnamefont {C.}~\bibnamefont
  {{Casella \textit{et al.}}}},\ }\bibfield  {title} {\bibinfo {title}
  {{``First measurement of the $d$($p$,$\gamma$)$^3$He cross section down to
  the solar Gamow peak''}},\ }\href
  {https://doi.org/10.1016/S0375-9474(02)00749-2} {\bibfield  {journal}
  {\bibinfo  {journal} {Nucl. Phys. A}\ }\textbf {\bibinfo {volume} {706}},\
  \bibinfo {pages} {203} (\bibinfo {year} {2002})}\BibitemShut {NoStop}%
\bibitem [{\citenamefont {Ti{\v{s}}ma}\ \emph {et~al.}()\citenamefont
  {Ti{\v{s}}ma}, \citenamefont {Lipoglav{\v{s}}ek}, \citenamefont
  {Mihovilovi{\v{c}}}, \citenamefont {Markelj}, \citenamefont {Vencelj},\ and\
  \citenamefont {Vesi{\'{c}}}}]{Tisma2019}%
  \BibitemOpen
  \bibfield  {author} {\bibinfo {author} {\bibfnamefont {I.}~\bibnamefont
  {Ti{\v{s}}ma}}, \bibinfo {author} {\bibfnamefont {M.}~\bibnamefont
  {Lipoglav{\v{s}}ek}}, \bibinfo {author} {\bibfnamefont {M.}~\bibnamefont
  {Mihovilovi{\v{c}}}}, \bibinfo {author} {\bibfnamefont {S.}~\bibnamefont
  {Markelj}}, \bibinfo {author} {\bibfnamefont {M.}~\bibnamefont {Vencelj}},\
  and\ \bibinfo {author} {\bibfnamefont {J.}~\bibnamefont {Vesi{\'{c}}}},\
  }\bibfield  {title} {\bibinfo {title} {{``Experimental cross section and
  angular distribution of the $^2$H($p$,$\gamma$)$^3$He reaction at Big-Bang
  nucleosynthesis energies''}},\ }\href
  {https://doi.org/10.1140/epja/i2019-12816-1} {\bibfield  {journal} {\bibinfo
  {journal} {Eur. Phys. J. A}\ }\textbf {\bibinfo {volume} {55}},\ \bibinfo
  {pages} {137}}\BibitemShut {NoStop}%
\bibitem [{\citenamefont {{Turkat \textit{et al.}}}(2021)}]{Turkat2021}%
  \BibitemOpen
  \bibfield  {author} {\bibinfo {author} {\bibfnamefont {S.}~\bibnamefont
  {{Turkat \textit{et al.}}}},\ }\bibfield  {title} {\bibinfo {title}
  {{``Measurement of the $^2$H($p$,$\gamma$)$^3$He \textit{S} factor at
  265-1094 keV''}},\ }\href {https://doi.org/10.1103/PhysRevC.103.045805}
  {\bibfield  {journal} {\bibinfo  {journal} {Phys. Rev. C}\ }\textbf {\bibinfo
  {volume} {103}},\ \bibinfo {pages} {045805} (\bibinfo {year} {2021})},\
  \Eprint {https://arxiv.org/abs/2104.06914} {arXiv:2104.06914 [nucl-ex]}
  \BibitemShut {NoStop}%
\bibitem [{\citenamefont {Dessert}\ \emph {et~al.}(2024)\citenamefont
  {Dessert}, \citenamefont {Foster}, \citenamefont {Park},\ and\ \citenamefont
  {Safdi}}]{Dessert2024}%
  \BibitemOpen
  \bibfield  {author} {\bibinfo {author} {\bibfnamefont {C.}~\bibnamefont
  {Dessert}}, \bibinfo {author} {\bibfnamefont {J.~W.}\ \bibnamefont {Foster}},
  \bibinfo {author} {\bibfnamefont {Y.}~\bibnamefont {Park}},\ and\ \bibinfo
  {author} {\bibfnamefont {B.~R.}\ \bibnamefont {Safdi}},\ }\bibfield  {title}
  {\bibinfo {title} {{``Was There a 3.5 keV Line?''}},\ }\href
  {https://doi.org/10.3847/1538-4357/ad2612} {\bibfield  {journal} {\bibinfo
  {journal} {Astrophys. J.}\ }\textbf {\bibinfo {volume} {964}},\ \bibinfo
  {pages} {185} (\bibinfo {year} {2024})},\ \Eprint
  {https://arxiv.org/abs/2309.03254} {arXiv:2309.03254 [astro-ph.CO]}
  \BibitemShut {NoStop}%
\bibitem [{\citenamefont {Bulbul}\ \emph {et~al.}(2014)\citenamefont {Bulbul},
  \citenamefont {Markevitch}, \citenamefont {Foster}, \citenamefont {Smith},
  \citenamefont {Loewenstein},\ and\ \citenamefont {Randal}}]{Bulbul2014}%
  \BibitemOpen
  \bibfield  {author} {\bibinfo {author} {\bibfnamefont {E.}~\bibnamefont
  {Bulbul}}, \bibinfo {author} {\bibfnamefont {M.}~\bibnamefont {Markevitch}},
  \bibinfo {author} {\bibfnamefont {A.}~\bibnamefont {Foster}}, \bibinfo
  {author} {\bibfnamefont {R.~K.}\ \bibnamefont {Smith}}, \bibinfo {author}
  {\bibfnamefont {M.}~\bibnamefont {Loewenstein}},\ and\ \bibinfo {author}
  {\bibfnamefont {S.~W.}\ \bibnamefont {Randal}},\ }\bibfield  {title}
  {\bibinfo {title} {{``Detection of An Unidentified Emission Line in the
  Stacked X-ray spectrum of Galaxy Clusters''}},\ }\href
  {https://doi.org/10.1088/0004-637X/789/1/13} {\bibfield  {journal} {\bibinfo
  {journal} {Astrophys. J.}\ }\textbf {\bibinfo {volume} {789}},\ \bibinfo
  {pages} {13} (\bibinfo {year} {2014})},\ \Eprint
  {https://arxiv.org/abs/1402.2301} {arXiv:1402.2301 [astro-ph.CO]}
  \BibitemShut {NoStop}%
\bibitem [{\citenamefont {Boyarsky}\ \emph {et~al.}(2014)\citenamefont
  {Boyarsky}, \citenamefont {Ruchayskiy}, \citenamefont {Iakubovskyi},\ and\
  \citenamefont {Franse}}]{Boyarsky2014}%
  \BibitemOpen
  \bibfield  {author} {\bibinfo {author} {\bibfnamefont {A.}~\bibnamefont
  {Boyarsky}}, \bibinfo {author} {\bibfnamefont {O.}~\bibnamefont
  {Ruchayskiy}}, \bibinfo {author} {\bibfnamefont {D.}~\bibnamefont
  {Iakubovskyi}},\ and\ \bibinfo {author} {\bibfnamefont {J.}~\bibnamefont
  {Franse}},\ }\bibfield  {title} {\bibinfo {title} {{``Unidentified Line in
  X-Ray Spectra of the Andromeda Galaxy and Perseus Galaxy Cluster''}},\ }\href
  {https://doi.org/10.1103/PhysRevLett.113.251301} {\bibfield  {journal}
  {\bibinfo  {journal} {Phys. Rev. Lett.}\ }\textbf {\bibinfo {volume} {113}},\
  \bibinfo {pages} {251301} (\bibinfo {year} {2014})},\ \Eprint
  {https://arxiv.org/abs/1402.4119} {arXiv:1402.4119 [astro-ph.CO]}
  \BibitemShut {NoStop}%
\bibitem [{\citenamefont {Cappelluti}\ \emph {et~al.}(2018)\citenamefont
  {Cappelluti}, \citenamefont {Bulbul}, \citenamefont {Foster}, \citenamefont
  {Natarajan}, \citenamefont {Urry}, \citenamefont {Bautz}, \citenamefont
  {Civano}, \citenamefont {Miller},\ and\ \citenamefont
  {Smith}}]{Cappelluti2018}%
  \BibitemOpen
  \bibfield  {author} {\bibinfo {author} {\bibfnamefont {N.}~\bibnamefont
  {Cappelluti}}, \bibinfo {author} {\bibfnamefont {E.}~\bibnamefont {Bulbul}},
  \bibinfo {author} {\bibfnamefont {A.}~\bibnamefont {Foster}}, \bibinfo
  {author} {\bibfnamefont {P.}~\bibnamefont {Natarajan}}, \bibinfo {author}
  {\bibfnamefont {M.~C.}\ \bibnamefont {Urry}}, \bibinfo {author}
  {\bibfnamefont {M.~W.}\ \bibnamefont {Bautz}}, \bibinfo {author}
  {\bibfnamefont {F.}~\bibnamefont {Civano}}, \bibinfo {author} {\bibfnamefont
  {E.}~\bibnamefont {Miller}},\ and\ \bibinfo {author} {\bibfnamefont {R.~K.}\
  \bibnamefont {Smith}},\ }\bibfield  {title} {\bibinfo {title} {{``Searching
  for the 3.5 keV Line in the Deep Fields with Chandra: the 10 Ms
  observations''}},\ }\href {https://doi.org/10.3847/1538-4357/aaaa68}
  {\bibfield  {journal} {\bibinfo  {journal} {Astrophys. J.}\ }\textbf
  {\bibinfo {volume} {854}},\ \bibinfo {pages} {179} (\bibinfo {year}
  {2018})},\ \Eprint {https://arxiv.org/abs/1701.07932} {arXiv:1701.07932
  [astro-ph.CO]} \BibitemShut {NoStop}%
\bibitem [{\citenamefont {Gamow}(1928)}]{Gamow1928}%
  \BibitemOpen
  \bibfield  {author} {\bibinfo {author} {\bibfnamefont {G.}~\bibnamefont
  {Gamow}},\ }\bibfield  {title} {\bibinfo {title} {{``Zur Quantentheorie des
  Atomkernes''}},\ }\href {https://doi.org/10.1007/BF01343196} {\bibfield
  {journal} {\bibinfo  {journal} {Z. Physik}\ }\textbf {\bibinfo {volume}
  {51}},\ \bibinfo {pages} {204} (\bibinfo {year} {1928})}\BibitemShut
  {NoStop}%
\bibitem [{\citenamefont {{First Research
  Group}}(1985{\natexlab{b}})}]{RG_exfor}%
  \BibitemOpen
  \bibfield  {author} {\bibinfo {author} {\bibnamefont {{First Research
  Group}}},\ }\href
  {https://www-nds.iaea.org/exfor/servlet/X4sGetSubent?reqx=19047&subID=280025002}
  {\bibinfo {title} {Data file {EXFOR} {S0025.001}}},\ \bibinfo {howpublished}
  {IAEA Nuclear Data Section} (\bibinfo {year}
  {1985}{\natexlab{b}})\BibitemShut {NoStop}%
\bibitem [{\citenamefont {Ma}\ \emph {et~al.}(1999)\citenamefont {Ma},
  \citenamefont {Karwowski}, \citenamefont {Brune}, \citenamefont {Ayer},
  \citenamefont {Black}, \citenamefont {Blackmon},\ and\ \citenamefont
  {Ludwig}}]{Ma_exfor}%
  \BibitemOpen
  \bibfield  {author} {\bibinfo {author} {\bibfnamefont {L.}~\bibnamefont
  {Ma}}, \bibinfo {author} {\bibfnamefont {H.~J.}\ \bibnamefont {Karwowski}},
  \bibinfo {author} {\bibfnamefont {C.~R.}\ \bibnamefont {Brune}}, \bibinfo
  {author} {\bibfnamefont {Z.}~\bibnamefont {Ayer}}, \bibinfo {author}
  {\bibfnamefont {T.~C.}\ \bibnamefont {Black}}, \bibinfo {author}
  {\bibfnamefont {J.~C.}\ \bibnamefont {Blackmon}},\ and\ \bibinfo {author}
  {\bibfnamefont {E.~J.}\ \bibnamefont {Ludwig}},\ }\href
  {https://www-nds.iaea.org/exfor/servlet/X4sGetSubent?reqx=59364&subID=120539006}
  {\bibinfo {title} {Data file {EXFOR} {C0539.001}}},\ \bibinfo {howpublished}
  {IAEA Nuclear Data Section} (\bibinfo {year} {1999})\BibitemShut {NoStop}%
\bibitem [{\citenamefont {Warren}\ \emph {et~al.}(1963)\citenamefont {Warren},
  \citenamefont {Erdman}, \citenamefont {Robertson}, \citenamefont {Axen},\
  and\ \citenamefont {Macdonald}}]{Warren1963}%
  \BibitemOpen
  \bibfield  {author} {\bibinfo {author} {\bibfnamefont {J.~B.}\ \bibnamefont
  {Warren}}, \bibinfo {author} {\bibfnamefont {K.~L.}\ \bibnamefont {Erdman}},
  \bibinfo {author} {\bibfnamefont {L.~P.}\ \bibnamefont {Robertson}}, \bibinfo
  {author} {\bibfnamefont {D.~A.}\ \bibnamefont {Axen}},\ and\ \bibinfo
  {author} {\bibfnamefont {J.~R.}\ \bibnamefont {Macdonald}},\ }\bibfield
  {title} {\bibinfo {title} {{``Photodisintegration of He$^3$ near the
  Threshold''}},\ }\href {https://doi.org/10.1103/PhysRev.132.1691} {\bibfield
  {journal} {\bibinfo  {journal} {Phys. Rev.}\ }\textbf {\bibinfo {volume}
  {132}},\ \bibinfo {pages} {1691} (\bibinfo {year} {1963})}\BibitemShut
  {NoStop}%
\bibitem [{\citenamefont {Griffiths}\ \emph {et~al.}(1962)\citenamefont
  {Griffiths}, \citenamefont {Larson},\ and\ \citenamefont
  {Robertson}}]{Griffiths1962}%
  \BibitemOpen
  \bibfield  {author} {\bibinfo {author} {\bibfnamefont {G.~M.}\ \bibnamefont
  {Griffiths}}, \bibinfo {author} {\bibfnamefont {E.~A.}\ \bibnamefont
  {Larson}},\ and\ \bibinfo {author} {\bibfnamefont {L.~P.}\ \bibnamefont
  {Robertson}},\ }\bibfield  {title} {\bibinfo {title} {{``The capture of
  protons by deuterons''}},\ }\href {https://doi.org/10.1139/p62-045}
  {\bibfield  {journal} {\bibinfo  {journal} {Can. J. Phys.}\ }\textbf
  {\bibinfo {volume} {40}},\ \bibinfo {pages} {402} (\bibinfo {year}
  {1962})}\BibitemShut {NoStop}%
\bibitem [{\citenamefont {Griffiths}\ \emph {et~al.}(1963)\citenamefont
  {Griffiths}, \citenamefont {Lal},\ and\ \citenamefont
  {Scarfe}}]{Griffiths1963}%
  \BibitemOpen
  \bibfield  {author} {\bibinfo {author} {\bibfnamefont {G.~M.}\ \bibnamefont
  {Griffiths}}, \bibinfo {author} {\bibfnamefont {M.}~\bibnamefont {Lal}},\
  and\ \bibinfo {author} {\bibfnamefont {C.~D.}\ \bibnamefont {Scarfe}},\
  }\bibfield  {title} {\bibinfo {title} {{``The Reaction D($p$,$\gamma$)He$^3$
  Below 50 keV''}},\ }\href {https://doi.org/10.1139/p63-077} {\bibfield
  {journal} {\bibinfo  {journal} {Can. J. Phys.}\ }\textbf {\bibinfo {volume}
  {41}},\ \bibinfo {pages} {724} (\bibinfo {year} {1963})}\BibitemShut
  {NoStop}%
\bibitem [{\citenamefont {Geller}\ and\ \citenamefont
  {Muirhead}(1967)}]{Geller1967}%
  \BibitemOpen
  \bibfield  {author} {\bibinfo {author} {\bibfnamefont {K.~N.}\ \bibnamefont
  {Geller}}\ and\ \bibinfo {author} {\bibfnamefont {E.~G.}\ \bibnamefont
  {Muirhead}},\ }\bibfield  {title} {\bibinfo {title} {{``The
  $^2$H($p$,$\gamma$)$^3$He reaction at the breakup threshold''}},\ }\href
  {https://doi.org/10.1016/0375-9474(67)90721-X} {\bibfield  {journal}
  {\bibinfo  {journal} {Nucl. Phys. A}\ ,\ \bibinfo {pages} {397}} (\bibinfo
  {year} {1967})}\BibitemShut {NoStop}%
\bibitem [{\citenamefont {Bailey}\ \emph {et~al.}(1970)\citenamefont {Bailey},
  \citenamefont {Griffiths}, \citenamefont {Olivo},\ and\ \citenamefont
  {Helmer}}]{Bailey1970}%
  \BibitemOpen
  \bibfield  {author} {\bibinfo {author} {\bibfnamefont {G.~M.}\ \bibnamefont
  {Bailey}}, \bibinfo {author} {\bibfnamefont {G.~M.}\ \bibnamefont
  {Griffiths}}, \bibinfo {author} {\bibfnamefont {M.~A.}\ \bibnamefont
  {Olivo}},\ and\ \bibinfo {author} {\bibfnamefont {R.~L.}\ \bibnamefont
  {Helmer}},\ }\bibfield  {title} {\bibinfo {title} {{``Gamma-ray yields from
  the reaction D($p$,$\gamma$)$^3$He at low energies''}},\ }\href
  {https://doi.org/10.1139/p70-379} {\bibfield  {journal} {\bibinfo  {journal}
  {Can. J. Phys.}\ }\textbf {\bibinfo {volume} {48}},\ \bibinfo {pages} {3059}
  (\bibinfo {year} {1970})}\BibitemShut {NoStop}%
\bibitem [{\citenamefont {{Schmid \textit{et al.}}}(1999)}]{Schmid_exfor}%
  \BibitemOpen
  \bibfield  {author} {\bibinfo {author} {\bibfnamefont {G.~J.}\ \bibnamefont
  {{Schmid \textit{et al.}}}},\ }\href
  {https://www-nds.iaea.org/exfor/servlet/X4sGetSubent?reqx=59352&subID=120405009}
  {\bibinfo {title} {Data file {EXFOR} {C0405.001}}},\ \bibinfo {howpublished}
  {IAEA Nuclear Data Section} (\bibinfo {year} {1999})\BibitemShut {NoStop}%
\bibitem [{\citenamefont {{Acharya \textit{et al.}}}(2025)}]{SF32025}%
  \BibitemOpen
  \bibfield  {author} {\bibinfo {author} {\bibfnamefont {B.}~\bibnamefont
  {{Acharya \textit{et al.}}}},\ }\bibfield  {title} {\bibinfo {title}
  {{``Solar fusion III: New data and theory for hydrogen-burning stars''}},\
  }\href {https://doi.org/10.1103/8lm7-gs18} {\bibfield  {journal} {\bibinfo
  {journal} {Rev. Mod. Phys.}\ }\textbf {\bibinfo {volume} {97}},\ \bibinfo
  {pages} {035002} (\bibinfo {year} {2025})},\ \Eprint
  {https://arxiv.org/abs/2405.06470} {arXiv:2405.06470 [astro-ph.SR]}
  \BibitemShut {NoStop}%
\bibitem [{\citenamefont {Coc}\ \emph {et~al.}(2015)\citenamefont {Coc},
  \citenamefont {Petitjean}, \citenamefont {Uzan}, \citenamefont {Vangioni},
  \citenamefont {Descouvemont}, \citenamefont {Iliadis},\ and\ \citenamefont
  {Longland}}]{Coc2015}%
  \BibitemOpen
  \bibfield  {author} {\bibinfo {author} {\bibfnamefont {A.}~\bibnamefont
  {Coc}}, \bibinfo {author} {\bibfnamefont {P.}~\bibnamefont {Petitjean}},
  \bibinfo {author} {\bibfnamefont {J.}~\bibnamefont {Uzan}}, \bibinfo {author}
  {\bibfnamefont {E.}~\bibnamefont {Vangioni}}, \bibinfo {author}
  {\bibfnamefont {P.}~\bibnamefont {Descouvemont}}, \bibinfo {author}
  {\bibfnamefont {C.}~\bibnamefont {Iliadis}},\ and\ \bibinfo {author}
  {\bibfnamefont {R.}~\bibnamefont {Longland}},\ }\bibfield  {title} {\bibinfo
  {title} {{``New reaction rates for improved primordial D/H calculation and
  the cosmic evolution of deuterium''}},\ }\href
  {https://doi.org/10.1103/PhysRevD.92.123526} {\bibfield  {journal} {\bibinfo
  {journal} {Phys. Rev. D}\ }\textbf {\bibinfo {volume} {92}},\ \bibinfo
  {pages} {123526} (\bibinfo {year} {2015})},\ \Eprint
  {https://arxiv.org/abs/1511.03843} {arXiv:1511.03843 [astro-ph.CO]}
  \BibitemShut {NoStop}%
\bibitem [{\citenamefont {Shafieloo}\ \emph {et~al.}(2012)\citenamefont
  {Shafieloo}, \citenamefont {Kim},\ and\ \citenamefont
  {Linder}}]{Shafieloo2012}%
  \BibitemOpen
  \bibfield  {author} {\bibinfo {author} {\bibfnamefont {A.}~\bibnamefont
  {Shafieloo}}, \bibinfo {author} {\bibfnamefont {A.}~\bibnamefont {Kim}},\
  and\ \bibinfo {author} {\bibfnamefont {E.~V.}\ \bibnamefont {Linder}},\
  }\bibfield  {title} {\bibinfo {title} {{``Gaussian process cosmography''}},\
  }\href {https://doi.org/10.1103/PhysRevD.85.123530} {\bibfield  {journal}
  {\bibinfo  {journal} {Phys. Rev. D}\ ,\ \bibinfo {pages} {123530}} (\bibinfo
  {year} {2012})},\ \Eprint {https://arxiv.org/abs/1204.2272} {arXiv:1204.2272
  [astro-ph.CO]} \BibitemShut {NoStop}%
\bibitem [{\citenamefont {Haridasu}\ \emph {et~al.}()\citenamefont {Haridasu},
  \citenamefont {{Lukovi\'c}}, \citenamefont {Moresco},\ and\ \citenamefont
  {Vittorio}}]{Haridasu2018}%
  \BibitemOpen
  \bibfield  {author} {\bibinfo {author} {\bibfnamefont {B.~S.}\ \bibnamefont
  {Haridasu}}, \bibinfo {author} {\bibfnamefont {V.~V.}\ \bibnamefont
  {{Lukovi\'c}}}, \bibinfo {author} {\bibfnamefont {M.}~\bibnamefont
  {Moresco}},\ and\ \bibinfo {author} {\bibfnamefont {N.}~\bibnamefont
  {Vittorio}},\ }\bibfield  {title} {\bibinfo {title} {{``An improved
  model-independent assessment of the late-time cosmic expansion''}},\ }\href
  {https://doi.org/An improved model-independent assessment of the late-time
  cosmic expansion} {\bibfield  {journal} {\bibinfo  {journal} {JCAP}\ }\textbf
  {\bibinfo {volume} {2018}}\bibfield  {number} {\bibinfo  {number} { (10)},\
  \bibinfo {pages} {015}},\ }\Eprint {https://arxiv.org/abs/1805.03595}
  {arXiv:1805.03595 [astro-ph.CO]} \BibitemShut {NoStop}%
\bibitem [{\citenamefont {Ruchika}\ \emph {et~al.}(2025)\citenamefont
  {Ruchika}, \citenamefont {Mukherjee},\ and\ \citenamefont
  {Favale}}]{Ruchika2025}%
  \BibitemOpen
  \bibfield  {author} {\bibinfo {author} {\bibnamefont {Ruchika}}, \bibinfo
  {author} {\bibfnamefont {P.}~\bibnamefont {Mukherjee}},\ and\ \bibinfo
  {author} {\bibfnamefont {A.}~\bibnamefont {Favale}},\ }\bibfield  {title}
  {\bibinfo {title} {{``Revisiting Gaussian process reconstruction for
  cosmological inference: the generalised GP (Gen GP) framework''}},\
  }\href@noop {} {\  (\bibinfo {year} {2025})},\ \Eprint
  {https://arxiv.org/abs/2510.03742} {arXiv:2510.03742 [astro-ph.CO]}
  \BibitemShut {NoStop}%
\bibitem [{\citenamefont {{de Souza}}\ \emph {et~al.}(2025)\citenamefont {{de
  Souza}}, \citenamefont {{Sousa-Neto}}, \citenamefont {{Gonz\'alez}},\ and\
  \citenamefont {Alcaniz}}]{deSouza2025}%
  \BibitemOpen
  \bibfield  {author} {\bibinfo {author} {\bibfnamefont {R.}~\bibnamefont {{de
  Souza}}}, \bibinfo {author} {\bibfnamefont {A.}~\bibnamefont {{Sousa-Neto}}},
  \bibinfo {author} {\bibfnamefont {J.~E.}\ \bibnamefont {{Gonz\'alez}}},\ and\
  \bibinfo {author} {\bibfnamefont {J.}~\bibnamefont {Alcaniz}},\ }\bibfield
  {title} {\bibinfo {title} {{``A model-independent assessment of the late-time
  dark energy density evolution''}},\ }\href@noop {} {\  (\bibinfo {year}
  {2025})},\ \Eprint {https://arxiv.org/abs/2511.13666} {arXiv:2511.13666
  [astro-ph.CO]} \BibitemShut {NoStop}%
\bibitem [{\citenamefont {Gramacy}\ and\ \citenamefont
  {Lee}(2012)}]{Gramacy2012}%
  \BibitemOpen
  \bibfield  {author} {\bibinfo {author} {\bibfnamefont {R.~B.}\ \bibnamefont
  {Gramacy}}\ and\ \bibinfo {author} {\bibfnamefont {H.~K.~H.}\ \bibnamefont
  {Lee}},\ }\bibfield  {title} {\bibinfo {title} {{``Cases for the nugget in
  modeling computer experiments''}},\ }\href
  {https://doi.org/10.1007/s11222-010-9224-x} {\bibfield  {journal} {\bibinfo
  {journal} {Stat. Comput.}\ }\textbf {\bibinfo {volume} {22}},\ \bibinfo
  {pages} {713} (\bibinfo {year} {2012})},\ \Eprint
  {https://arxiv.org/abs/1007.4580} {arXiv:1007.4580 [stat.CO]} \BibitemShut
  {NoStop}%
\bibitem [{\citenamefont {Knapik}\ \emph {et~al.}(2011)\citenamefont {Knapik},
  \citenamefont {{van der Vaart}},\ and\ \citenamefont {{van
  Zanten}}}]{Knapik2011}%
  \BibitemOpen
  \bibfield  {author} {\bibinfo {author} {\bibfnamefont {B.~T.}\ \bibnamefont
  {Knapik}}, \bibinfo {author} {\bibfnamefont {A.}~\bibnamefont {{van der
  Vaart}}},\ and\ \bibinfo {author} {\bibfnamefont {J.~H.}\ \bibnamefont {{van
  Zanten}}},\ }\bibfield  {title} {\bibinfo {title} {{``Bayesian inverse
  problems with Gaussian priors''}},\ }\href
  {https://doi.org/10.48550/arXiv.1103.2692} {\bibfield  {journal} {\bibinfo
  {journal} {Ann. Stat.}\ }\textbf {\bibinfo {volume} {39}},\ \bibinfo {pages}
  {2626} (\bibinfo {year} {2011})},\ \Eprint {https://arxiv.org/abs/1103.2692}
  {arXiv:1103.2692 [math.ST]} \BibitemShut {NoStop}%
\bibitem [{\citenamefont {Yang}\ \emph {et~al.}(2017)\citenamefont {Yang},
  \citenamefont {Bhattacharya},\ and\ \citenamefont {Pati}}]{Yang2017}%
  \BibitemOpen
  \bibfield  {author} {\bibinfo {author} {\bibfnamefont {Y.}~\bibnamefont
  {Yang}}, \bibinfo {author} {\bibfnamefont {A.}~\bibnamefont {Bhattacharya}},\
  and\ \bibinfo {author} {\bibfnamefont {D.}~\bibnamefont {Pati}},\ }\bibfield
  {title} {\bibinfo {title} {{``Frequentist coverage and sup-norm convergence
  rate in Gaussian process regression''}},\ }\href@noop {} {\  (\bibinfo {year}
  {2017})},\ \Eprint {https://arxiv.org/abs/1708.04753} {arXiv:1708.04753
  [math.ST]} \BibitemShut {NoStop}%
\bibitem [{\citenamefont {Stone}(1974)}]{Stone1974}%
  \BibitemOpen
  \bibfield  {author} {\bibinfo {author} {\bibfnamefont {M.}~\bibnamefont
  {Stone}},\ }\bibfield  {title} {\bibinfo {title} {{``Cross-validatory Choice
  and Assessment of Statistical Predictions''}},\ }\href
  {https://doi.org/10.1111/j.2517-6161.1974.tb00994.x} {\bibfield  {journal}
  {\bibinfo  {journal} {Journal of the Royal Statistical Society: Series B
  (Methodological)}\ }\textbf {\bibinfo {volume} {36}},\ \bibinfo {pages} {111}
  (\bibinfo {year} {1974})}\BibitemShut {NoStop}%
\bibitem [{\citenamefont {Kingma}\ and\ \citenamefont {Ba}(2014)}]{Kingma2017}%
  \BibitemOpen
  \bibfield  {author} {\bibinfo {author} {\bibfnamefont {D.~P.}\ \bibnamefont
  {Kingma}}\ and\ \bibinfo {author} {\bibfnamefont {J.}~\bibnamefont {Ba}},\
  }\bibfield  {title} {\bibinfo {title} {{``Adam: A Method for Stochastic
  Optimization''}},\ }\href@noop {} {\  (\bibinfo {year} {2014})},\ \Eprint
  {https://arxiv.org/abs/1412.6980} {arXiv:1412.6980 [cs.LG]} \BibitemShut
  {NoStop}%
\end{thebibliography}%

\section{Methods}
\label{sec:methods}

Theoretical predictions for the primordial deuterium abundance relative to hydrogen D/H require fits to nuclear reaction S-factor data, the most important reactions being \ddp,~\ddn,~and \dpg.~We use Gaussian processes to construct probability distributions for these three S-factors consistent with experimental data. To do this, we model the S-factors for a vector of energies where we want to predict at as a multivariate Gaussian, then update this prior (Eq.~\eqref{eqn:gp_partitioned}) with the experimentally observed S-factor values. The posterior is also a multivariate Gaussian (Eq.~\eqref{eqn:gp_conditional}), from which we draw S-factor samples. 

As with any fitting procedure, there are choices we must make when constructing these posteriors, such as which experimental data to include and how to incorporate both statistical noise and systematic normalization uncertainties reported by these experiments. Additionally, we must make principled decisions regarding the kernel, the function that defines correlation between the S-factor at different energies. 

In this Section, we discuss these details. We begin with the fits to \ddn~and \dpg~data used in our fiducial analysis (\ddp~is discussed in the main text) in Section~\ref{sec:fits}. We then discuss 1) the relevant energy range for these reactions (Section~\ref{sec:energy_range}); 2) how it informs the datasets we choose to include (and exclude) in our analysis (Section~\ref{sec:datasets}), and 3) how correlated data are included in the GPs (Section~\ref{sec:correlated_gps}). We then motivate our kernel choice, highlighting the properties we require from the kernel (Section~\ref{sec:kernel_choice}). We conclude by considering alternative choices to those made in the fiducial analysis and calculate their impact on D/H predictions(Section~\ref{sec:further_checks}). The following discussion draws from Ref.~\cite{Rasmussen2006}, and we refer the reader to this text for a comprehensive overview of GPs.  

\subsection{Fits to S-Factor Data}
\label{sec:fits}

\begin{figure}
    	\centering
    	\begin{minipage}{.48\textwidth}
        		\centering
        		\includegraphics[width=\linewidth]{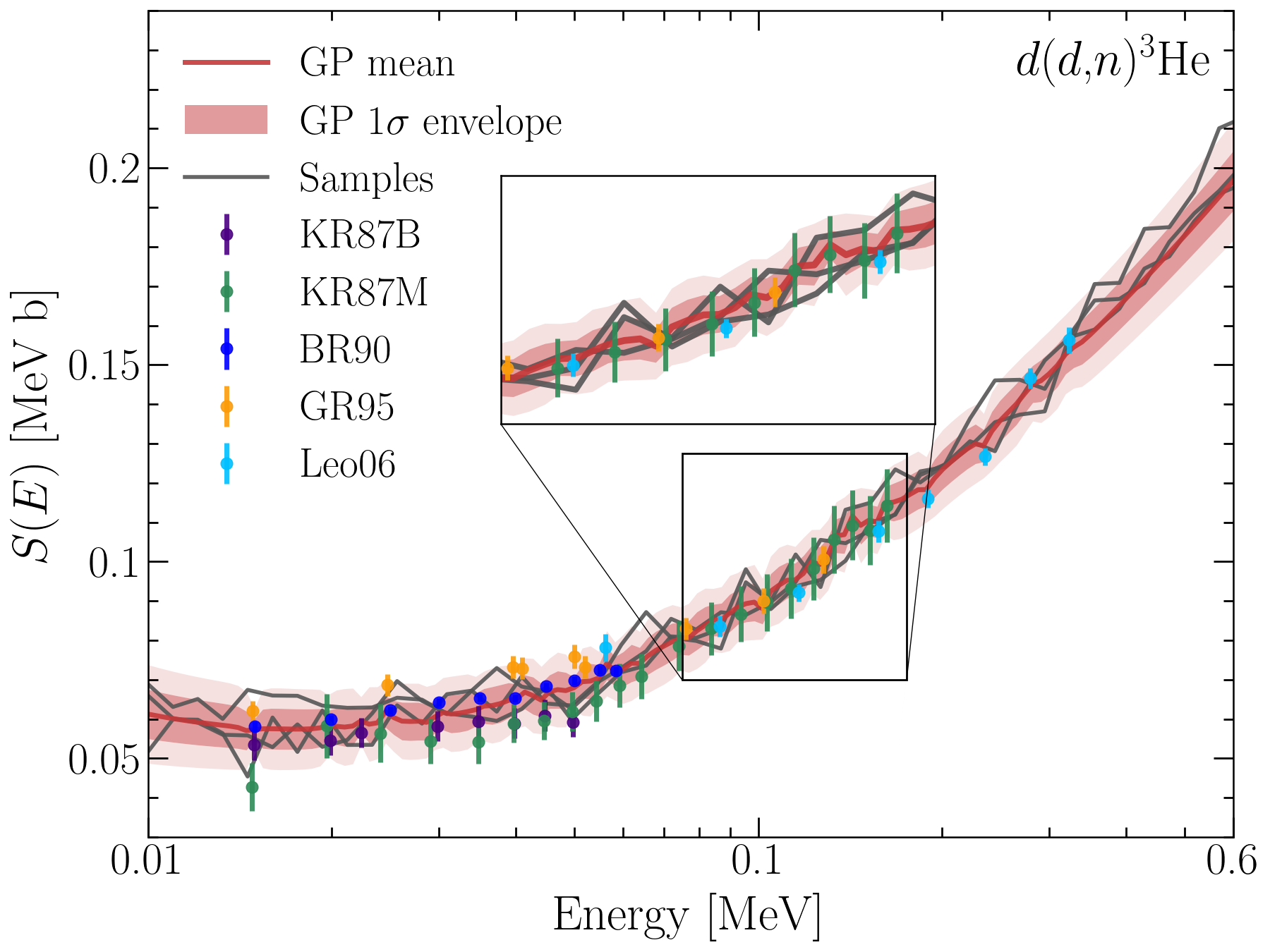}
    	\end{minipage}
    	\hspace{0mm}
    	\begin{minipage}{.48\textwidth}
        		\centering
        		\includegraphics[width=\linewidth]{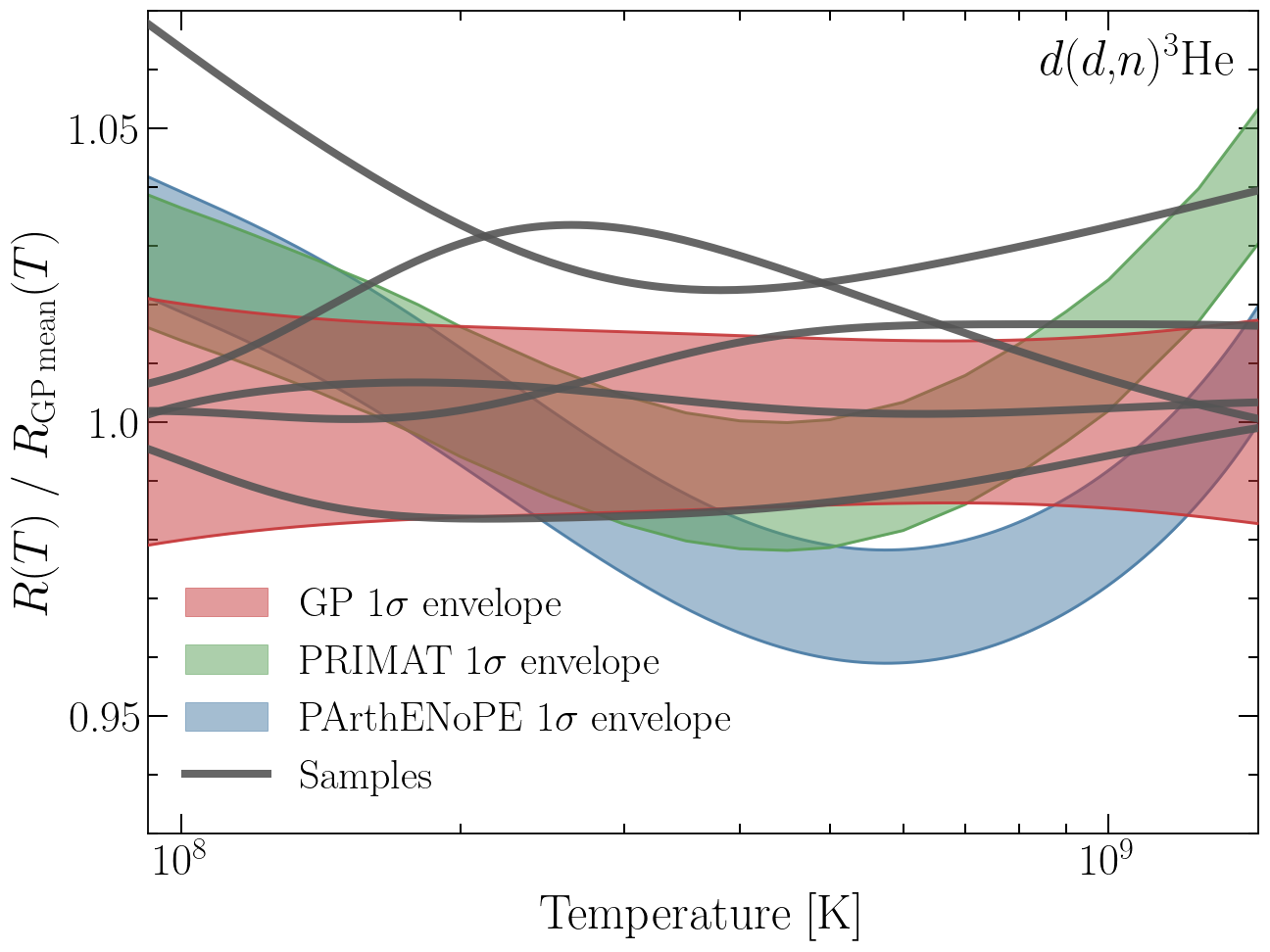}
    	\end{minipage}
    	\hspace{0mm}
    	\begin{minipage}{.48\textwidth}
        		\centering
        		\includegraphics[width=\linewidth]{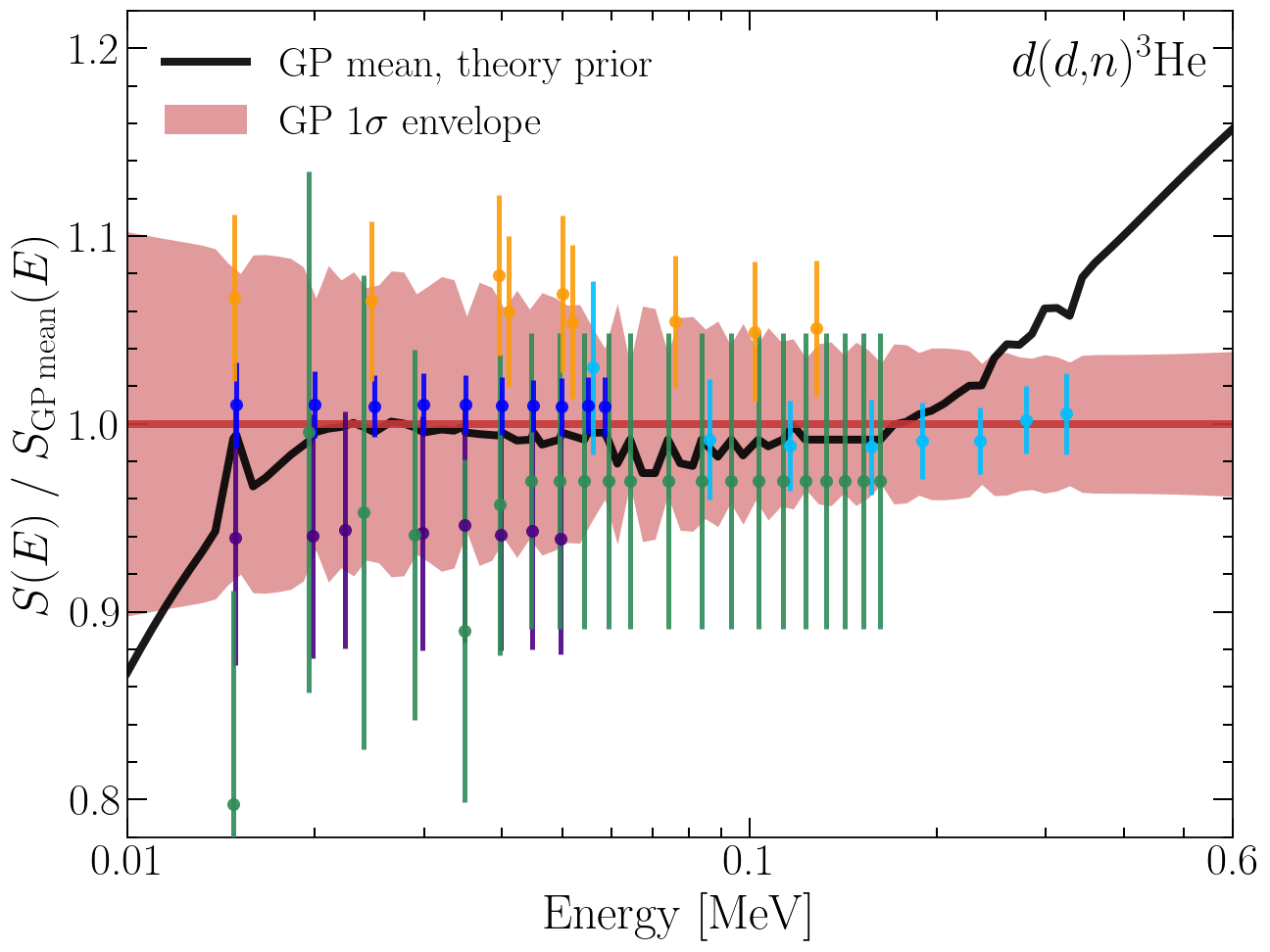}
    	\end{minipage}
	\caption{Gaussian process regression on \ddn~S-factor data. \textit{Top}: GP S-factor mean, $1\sigma$, and $2\sigma$ envelopes plotted along with data used for regression~\cite{Krauss1987, Brown1990, Greife1995, Leonard2006} (red). Three sample draws are overlayed (grey). \textit{Middle}: The GP $1\sigma$ reaction rate envelope and five samples thermally averaged with Eq.~\eqref{eqn:thermal_average} in the relevant BBN temperature range (red). The $1\sigma$ rate envelopes used in PRIMAT~\cite{Pitrou2018} (green) and PArthENoPE~\cite{Gariazzo2022} (blue) are shown for comparison. All curves are normalized to the GP mean rate. \textit{Bottom}: The GP mean S-factor with a prior mean of the calculation of Ref.~\cite{Arai2011} (black) and experimental data (colored points, same as top panel), normalized to the fiducial GP mean, with the fiducial $1\sigma$ envelope shown in red. The GP mean lies the same distance away from each highly correlated data points of BR90 and KR87M. }
	\label{fig:ddhe3n_samples}
\end{figure}

For \ddn~and \ddp,~we assume the S-factor data follow Gaussian distributions defined by the reported central values and uncertainties. We perform GP regression on data from Refs.~\cite{Krauss1987, Brown1990, Greife1995, Leonard2006} using a zero prior mean for our fiducial result. 

The top panel of Figure~\ref{fig:ddhe3n_samples} shows the GP fit to \ddn~S-factor data. Many features of the GP posterior for \ddn~are qualitatively similar to the GP posterior for \ddp~(Figure~\ref{fig:ddtp_samples}). As for \ddp, the posterior is consistent with scatter between experiments and not dominated by the small uncertainties of Refs.~\cite{Brown1990} (dark blue) and~\cite{Leonard2006} (light blue). 

Both the mean and function samples appear jagged from the small-$\nu$ Mat\'ern kernel component we add to the kernel (see Section~\ref{sec:kernel_choice} below). These small-scale fluctuations are smoothed from the thermal average, giving the smooth rate samples shown in the middle panel of Figure~\ref{fig:ddhe3n_samples}. While the agreement with the PRIMAT network is not as good as for the \ddp~rates, our \ddn~rate is nevertheless more similar to the rate in the PRIMAT network than the PArthENoPE network. Our rate uncertainties are larger than both.

\begin{figure}
    	\centering
    	\begin{minipage}{.48\textwidth}
        		\centering
        		\includegraphics[width=\linewidth]{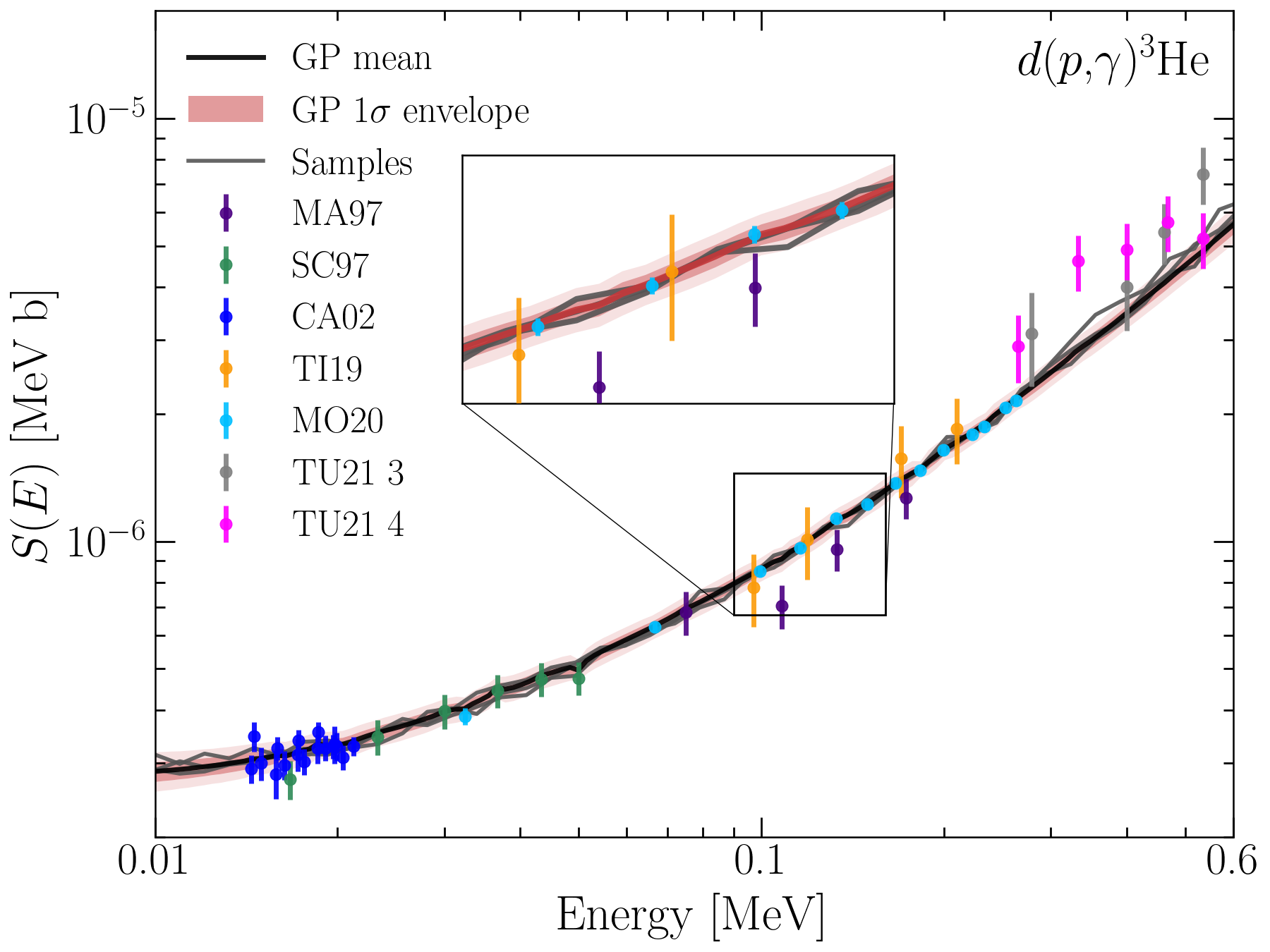}
    	\end{minipage}
    	\hspace{0mm}
    	\begin{minipage}{.48\textwidth}
        		\centering
        		\includegraphics[width=\linewidth]{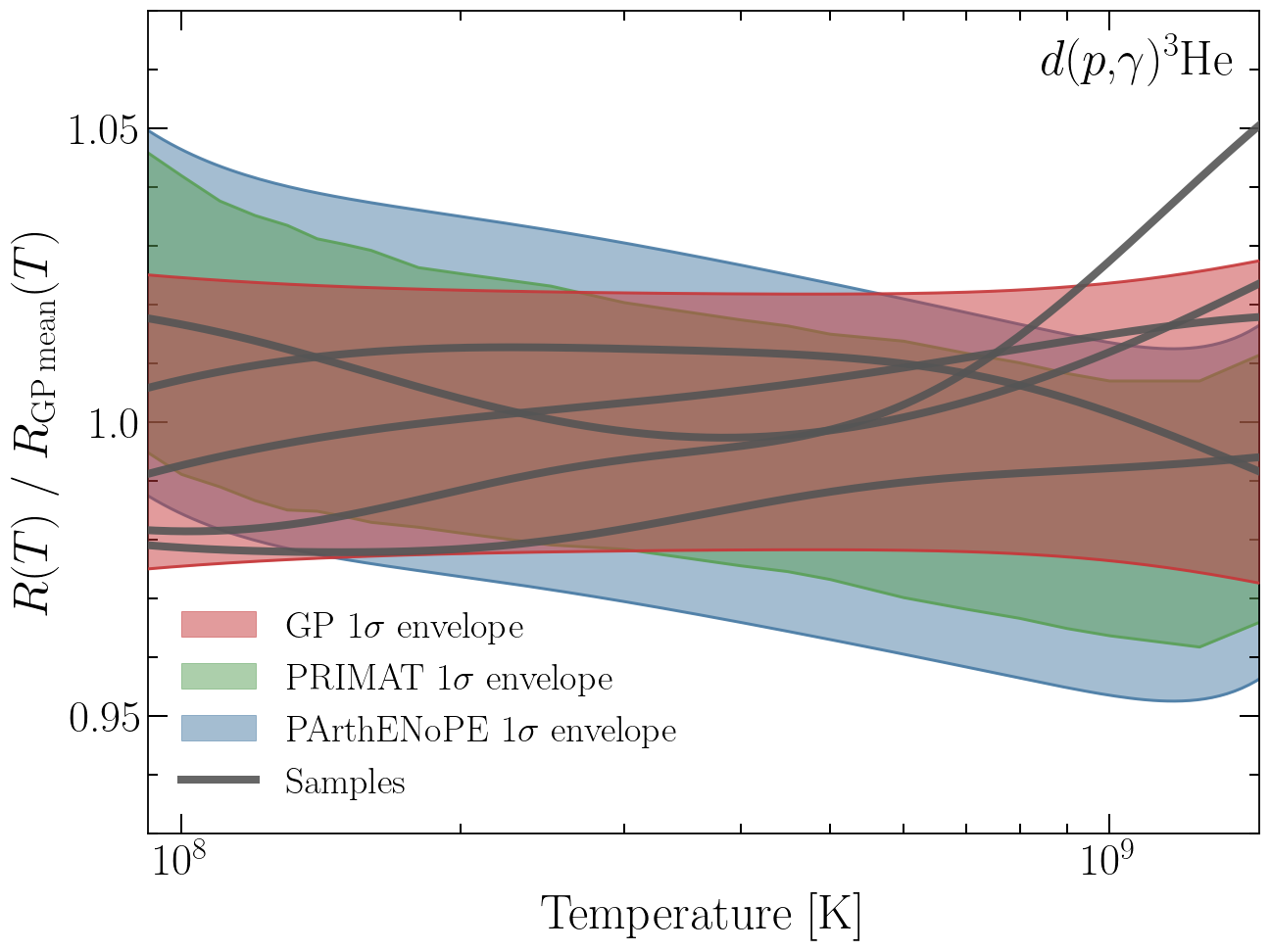}
    	\end{minipage}
	\caption{Gaussian process regression on \dpg~S-factor data. \textit{Top}: GP S-factor mean, $1\sigma$, and $2\sigma$ envelopes (red) plotted along with data used for regression~\cite{Ma1997, Schmid1997, Casella2002, Tisma2019, Mossa2020, Turkat2021}. Three sample draws are overlayed (grey). \textit{Bottom}: The GP $1\sigma$ reaction rate envelope (red) and five samples thermally averaged with Eq.~\eqref{eqn:thermal_average} in the relevant BBN temperature range (grey). The $1\sigma$ rate envelopes used in PRIMAT~\cite{Pitrou2018} (green) and PArthENoPE~\cite{Gariazzo2022} (blue) are shown for comparison. All curves are normalized to the GP mean rate.}
	\label{fig:dphe3g_samples}
\end{figure}

The \dpg~S-factor more naturally lies on a logarithmic scale, and any fit to these data must climb nearly two orders of magnitude in less than $2 \ln (E/\si{MeV})$. Therefore, we perform GP regression assuming the \dpg~S-factor data follow lognormal distributions, \textit{i.e.}\ $\ln S$ follows a normal distribution.

For \dpg,~we perform GP regression on data from Refs.~\cite{Ma1997, Schmid1997, Casella2002, Tisma2019, Mossa2020, Turkat2021}, again with a zero prior mean for our fiducial result. Figure~\ref{fig:dphe3g_samples} shows the GP fit to \dpg~S-factor data and corresponding reaction rates. There is less energy overlap and disagreement between different experiments for this reaction. This enables the high-precision data collected by the LUNA collaboration~\cite{Mossa2020} (light blue) to dominate the fit between $0.07-\qty{0.3}{MeV}$. Samples are again somewhat jagged from the small-$\nu$ Mat\'ern kernel component, although this is less obvious from the more narrow GP posterior. The data of Turkat \textit{et al.}\ ~\cite{Turkat2021} (pink and gray) have large total uncertainties and correlations. Since there is some overlap between these and the highest-energy LUNA point, the GP is confident that they are likely higher than the underlying S-factor, and the posterior mean remains below these datasets up to $\qty{0.6}{MeV}$. 

As shown in the bottom panel, the GP result is similar to the rates in both the PRIMAT and PArthENoPE networks throughout the BBN temperature range. The agreement between all three is driven by the precision of LUNA data. 

To demonstrate that experimental uncertainties---especially correlations between data points---are handled properly in the GP framework, we show the experimental data normalized to the GP mean for \ddn~in the bottom panel of Figure~\ref{fig:ddhe3n_samples}. In energy regions measured by multiple experiments, we expect that the GP mean lies closest to data with the smallest uncertainties. Four datasets measure the S-factor at low energies, between $0.014$ and $\qty{0.06}{MeV}$. In this energy range, the GP mean is closest to data with the smallest uncertainties, in this case the Brown \textit{et al.}\ data (dark blue). Additionally, perfectly correlated data, where systematic normalization uncertainties dominate over statistical uncertainty, should always be the same distance away from the GP mean. The data of Brown \textit{et al.}\ (dark blue) and high-energy data of Krauss \textit{et al.}\ (green) have negligible statistical uncertainties, and the GP mean always lies the same distance away from these data as desired.

\subsection{Relevant BBN Energy Range}
\label{sec:energy_range}

When performing a fit to data $y$ over an independent variable $x$ where the physically important values of $y$ are confined to a narrow range of $x$ values, an important analysis choice has to be made about whether to include data well outside of the physically important range. 
This choice can have a significant impact on the final conclusion of a study, and more data is not obviously better.  For example, Ref.~\cite{Dessert2024} found that evidence for the $\qty{3.5}{keV}$ line---detected at high significance by \textit{e.g.}\ Refs.~\cite{Bulbul2014, Boyarsky2014, Cappelluti2018} in a variety of datasets---disappears when the energy window for the fit to the line is narrowed. Another example of an analysis that demonstrates the importance of dataset choice is the recent analysis of metal-poor HII regions in Ref.~\cite{Aver2026}, which chose to intentionally narrow their dataset by excluding the HeI $\lambda$5016 and HeI $\lambda$7281 emission lines, in order to avoid contamination by suspected systematics. A similar choice about what data to include must be made for our analysis.

BBN is mostly complete within a narrow window of temperatures between $10^8$--\qty{e9}{\kelvin} (or, in natural units, $10^{-2}$--\qty{e-1}{\mega\eV}).  The integrand in Eq.~\eqref{eqn:thermal_average} for the rate $R(T)$ involves a Boltzmann suppression $e^{-E/k_BT}$, which diminishes the role of energies $E\gg k_BT$. Meanwhile, $S(E)$ is smaller for energies $E\ll k_BT$ than at energies $E\sim k_BT$, reducing the contribution of smaller energies over a narrow integration range.  Therefore, obtaining a good fit to $S(E)$ is only important in a finite range of energies to estimate the integral in Eq.~\eqref{eqn:thermal_average}. Because the GP regression assumes constant hyperparameters across the entire energy range of interest, it is beneficial to choose as narrow of an energy range as possible, since extending the energy range of our fits places greater strain on this assumption. For example, the \ddn~and \ddp~data of Ref.~\cite{Schulte1972} at energies $\gtrsim\qty{1}{MeV}$ likely lie well outside of the range of relevant energies given the relevant temperatures. Including these data forces the GP to attempt to accommodate a wider range of energies, and may introduce new systematics, especially given that the reported uncertainties are much smaller than a percent. Ref.~\cite{Leonard2006} raises similar concerns and opts to multiply their uncertainties by a factor of 5.

On the other hand, at low energies, the S-factor data (particularly for \ddp~and to a lesser degree \ddn) become increasingly impacted by electron screening effects~\cite{Greife1995, Tumino2014}. 
It is therefore extremely important to determine the energy range---both a low-energy and high-energy cut-off---over which \ddn,~\ddp,~and \dpg~are relevant for BBN. In doing so, we also identify concrete energy ranges for future S-factor measurements to target in order to best reduce predicted D/H uncertainties.

The history of identifying this energy range is older than BBN theory itself, starting from the Gamow window (\cite{Gamow1928}; see, \textit{e.g.}\ Ref.~\cite{Iliadis2015}). However, there is no consensus regarding what the energy range should be. Ref.~\cite{Leonard2006} claims that one must include data up to roughly $\qty{2.75}{MeV}$ for the integral in Eq.~\eqref{eqn:thermal_average} to converge beyond $T=\qty{2}{GK}$. On the other hand, Ref.~\cite{Pisanti2021} evaluates the change in the primordial deuterium abundance when varying the S-factor throughout these energies and claim that the most important energy range for D/H extends to $\qty{0.24}{MeV}$ (though they still integrate over a wider range of energies).  Crucially, no study takes into account the current experimental precision of D/H observations quantitatively.  The observational uncertainty provides a natural scale to evaluate the importance of D/H theoretical error incurred by neglecting $S(E)$ outside of a given energy window.

\begin{figure}
    	\centering
    	\begin{minipage}{.48\textwidth}
        		\centering
        		\includegraphics[width=\linewidth]{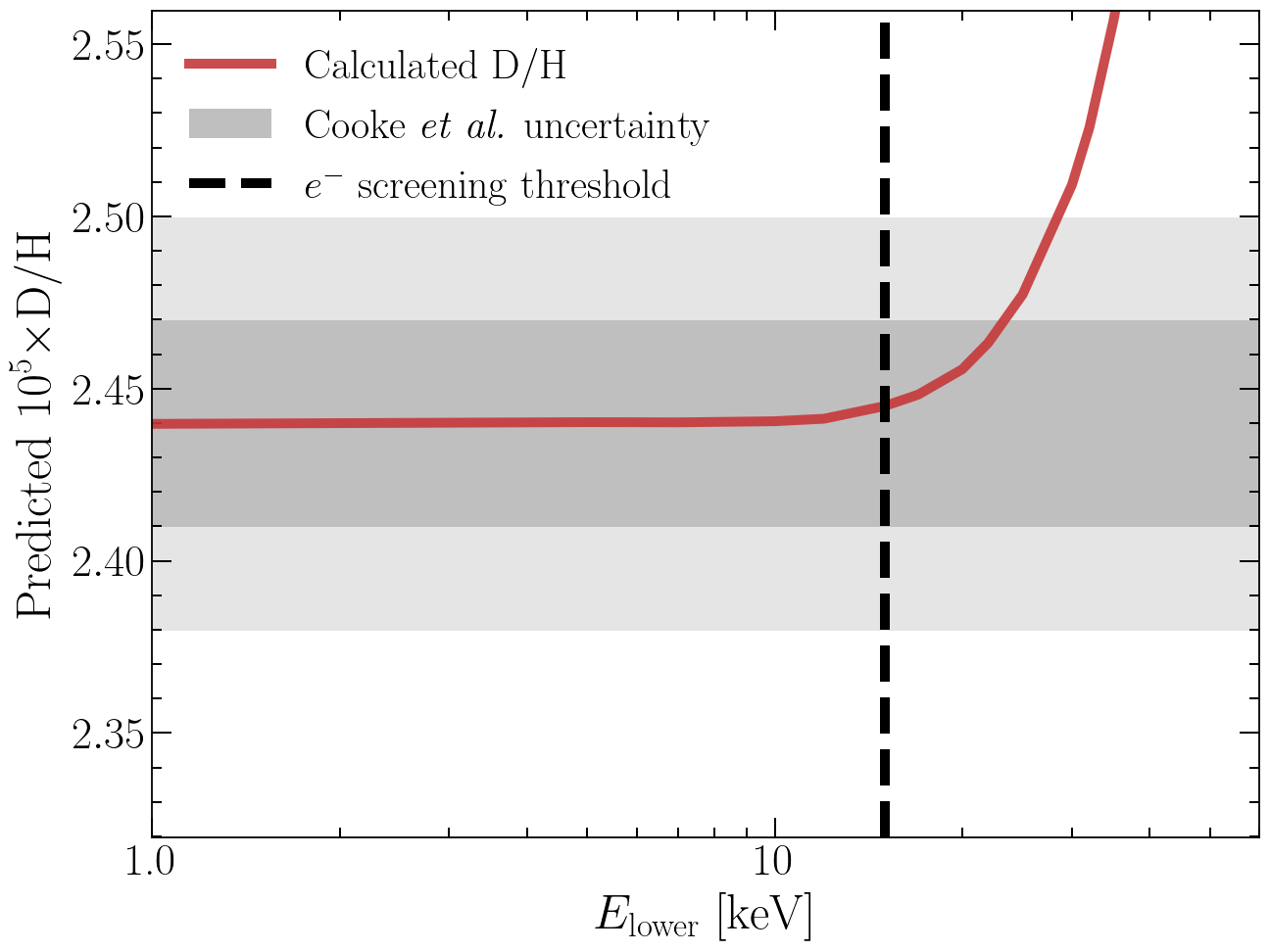}
    	\end{minipage}
    	\hspace{0mm}
    	\begin{minipage}{.48\textwidth}
        		\centering
        		\includegraphics[width=\linewidth]{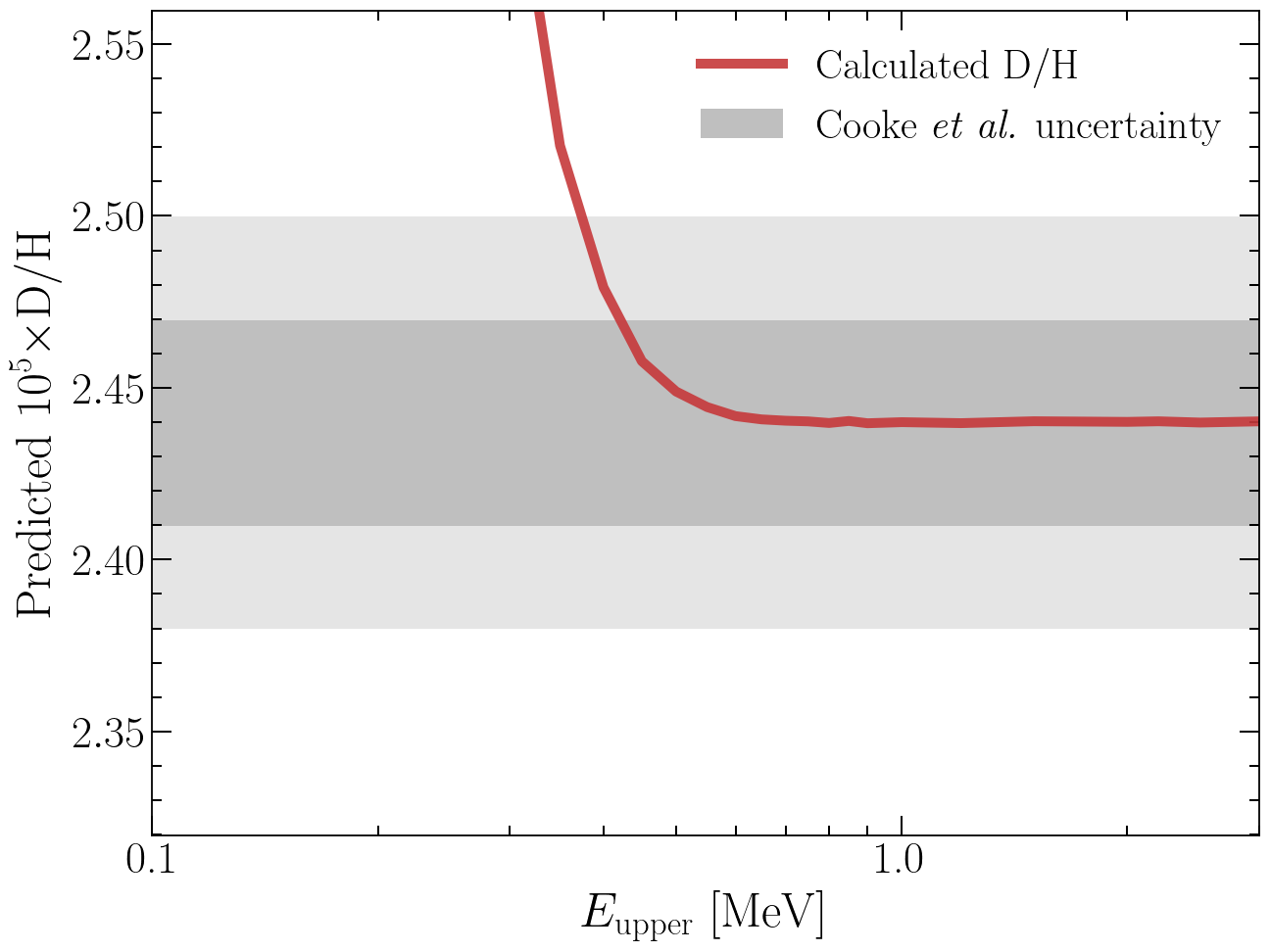}
    	\end{minipage}
	\caption{Predicted D/H as a function of lower and upper bounds (red) for the thermal average integral in Eq.~\eqref{eqn:thermal_average}. The horizontal axis denotes the value of the lower and upper energy bounds tested. The GP means for the \ddp,~\ddn,~and \dpg~S-factors are separately truncated from below (top) and from above (bottom). The observational $1\sigma$ and $2\sigma$ uncertainties from Ref.~\cite{Cooke2018} are shown in grey, centered at the deuterium prediction for the unbounded case. }
	\label{fig:energy_cutoff}
\end{figure}

To settle this question, we compute this energy range by calculating the change in D/H when shifting the bounds of integration for the thermal average and compare the magnitude of this change to the experimental precision of Ref.~\cite{Cooke2018}, \textit{i.e.}\ for each reaction, we define
\begin{multline}
		R(T; E_{\rm lower}, E_{\rm upper}) = \sqrt{\frac{8}{\pi \mu_{12}}} \frac{N_A}{(k_B T)^{3/2}} \\
		\times \int_{E_{\rm lower}}^{E_{\rm upper}} dE \, e^{-2 \pi \eta} S(E) e^{-E/k_B T},
\end{multline}
where the fiducial $R(T) = R(T; 0, \infty)$. To estimate these energy cutoffs $E_{\rm lower}$ and $E_{\rm upper}$, we compute D/H using the GP means for \ddn,~\ddp,~and \dpg,~changing the upper or lower bounds of the integral for each reaction together. This is equivalent to setting the S-factor to zero above $E_{\rm upper}$ or below $E_{\rm lower}$. Figure~\ref{fig:energy_cutoff} shows the predicted D/H as a function of $E_{\rm upper}$ ($E_{\rm lower}$) from $0.1 - \qty{3}{MeV}$ ($1- \qty{60}{keV}$). 

Let $\mathrm{D/H}_{\rm full}$ denote the predicted value with $R(T;0,\infty)$, approximated as $R(T;\qty{1.0}{keV}, \qty{3.0}{MeV})$. From this procedure, we choose $E_{\rm upper}$ ($E_{\rm lower}$) such that increasing (decreasing) the integral cutoff yields a D/H value differing from $\mathrm{D/H}_{\rm full}$ by no more than $0.1\sigma$ of Ref.~\cite{Cooke2018}. We find $E_{\rm lower}=\qty{14}{keV}$ and $E_{\rm upper}=\qty{0.6}{MeV}$. $E_{\rm lower}$ is approximately at the electron screening threshold, reported to be $\sim\qty{15}{keV}$ for \ddp~\cite{Greife1995, Tumino2014} (the reaction most heavily impacted), so we can safely exclude data below this energy and ignore electron screening effects. 

\subsection{Dataset Selection}
\label{sec:datasets}

\begin{table*}
	\begin{tabular}{cccccc}
		\hline
		\textbf{Dataset} & \textbf{Abbreviation} & \textbf{Reaction(s)} & \thead{\textbf{Statistical}\\ \textbf{Noise}\quad} & \thead{\textbf{Systematic} \\ \textbf{Uncertainty}} & \thead{\textbf{Fiducial} \\ \textbf{Analysis?}} \\
		\hline
          	\rule{0pt}{10pt}{Krauss 1987 Bochum~\cite{Krauss1987}} & {KR87B} & {\quad\ddp,~\ddn\quad} & {Table 2} & {0.064} & {Yes} \\
		{Krauss 1987 M\"unster~\cite{Krauss1987}} & {KR87M} & {\ddp,~\ddn} & {Table 2} & {0.0812} & {Yes} \\
		{Brown 1990~\cite{Brown1990}} & {BR90} & {\ddp,~\ddn} & {Table II} & {0.013} & {Yes} \\
		{Greife 1995~\cite{Greife1995}} & {GR95} & {\ddp,~\ddn} & {Table 2$^a$} & {0.033} & {Yes} \\
		{Leonard 2006~\cite{Leonard2006}} & {Leo06} & {\ddp,~\ddn} & {Table III} & {0.01$^b$} & {Yes} \\ 
		{Greife 1995 Close Geometry~\cite{Greife1995}} & {---} & {\ddp} & {Table 3$^a$} & {---} & {No} \\
		{McNeill 1951~\cite{McNeill1951}$^c$} & {MN51} & {\ddp,~\ddn} & {Table II} & {0.111$^d$} & {No} \\ 
		{Schulte 1972~\cite{Schulte1972}} & {SC72} & {\ddp,~\ddn} & {Table 3$^e$} & {Table 3} & {No} \\
		{First Research Group 1985~\cite{RG1985}} & {RG58} & {\ddp,~\ddn} & {\cite{RG_exfor}} & {\cite{RG_exfor}$^f$} & {No} \\
		{Tumino 2014~\cite{Tumino2014}} & {TU14} & {\ddp,~\ddn} & {\quad Main text$^a$\quad} & {0.0316} & {No} \\
		{Ma 1997~\cite{Ma1997}} & {MA97} & {\dpg} & {\cite{Ma_exfor}} & {0.09} & {Yes} \\ 
		{Schmid 1997~\cite{Schmid1997}} & {SC97} & {\dpg} & {Figure 13$^g$} & {0.09} & {Yes} \\
		{Casella 2002~\cite{Casella2002}} & {CA02} & {\dpg} & {Table 1} & {\quad0.036 -- 0.053$^h$\quad} & {Yes} \\
		{Ti{\v{s}}ma 2019~\cite{Tisma2019}} & {TI19} & {\dpg} & {\cite{Moscoso2021}$^i$} & {0.10} & {Yes} \\
		{Mossa 2020~\cite{Mossa2020}} & {MO20} & {\dpg} & {Table 1} & {Table 1} & {Yes} \\
		{Turkat 2021 Target No. 3~\cite{Turkat2021}} & {TU21 3} & {\dpg} & {Table II} & {0.08$^j$} & {Yes} \\
		{Turkat 2021 Target No. 4~\cite{Turkat2021}} & {TU21 4} & {\dpg} & {Table II} & {0.05} & {Yes} \\
		{Warren 1963~\cite{Warren1963}$^k$} & {---} & {\dpg} & {Table II$^a$} & {Table II} & {No} \\ 
		{Griffiths 1962~\cite{Griffiths1962}$^l$} & {---} & {\dpg} & {Table I$^a$} & {---} & {No} \\
		{Griffiths 1963~\cite{Griffiths1963}} & {---} & {\dpg} & {Figure 6} & {---} & {No} \\
		{Geller 1967~\cite{Geller1967}} & {---} & {\dpg} & {---} & {---} & {No} \\
		{Bailey 1970~\cite{Bailey1970}} & {---} & {\dpg} & {Figure 1} & {---} & {No} \\[2pt]
		\hline
	\end{tabular}
	\caption{Reported S-factor experimental uncertainties used in our analysis, entering as $\Sigma$ in Eq.~\eqref{eqn:gp_conditional} and $\sigma_{ik}$, $\epsilon$ in Eq.~\eqref{eqn:mock_gen} to generate mock data for our Monte Carlo analysis. $a$: Statistical noise is isolated by subtracting the reported systematics out from the total reported errors. $b$: The $1\%$ overall systematic uncertainty is summed in quadrature with the reported fit covariance in Table V. $c$: Refs.~\cite{McNeill1951, Schulte1972, RG1985, Tumino2014} are only used for Monte Carlo validation tests on polynomials. $d$: Originally authors summed error sources linearly, we sum them in quadrature. $e$: We scale the small reported errors by a factor of 5 and use the smallest as the overall systematic error percentage. $f$: We use the largest reported systematic uncertainties for both reactions as constant for mock data generation. $g$: Statistical noise isolated by subtracting the reported systematics from the error bars presented in Figure 13. S-factor data from Ref.~\cite{Schmid_exfor}. $h$: The lowest- and highest-energy data points are assigned $5.3\%$ and $3.6\%$ systematic uncertainties respectively. Systematics for the intermediate points are determined from a linear interpolation. We use a constant $4.5\%$ for all points during mock data generation. $i$: Thanks to Ref.~\cite{Moscoso2021}, who provide these data from correspondence with the authors. $j$: Targets 3 and 4 collectively share an overall $12\%$ systematic uncertainty in addition to those shown here, which is incorporated into our analysis. $k$: Ref.~\cite{Warren1963} is only included in a supplemental analysis. $l$: Refs.~\cite{Griffiths1962, Griffiths1963, Geller1967, Bailey1970} are not included in any analysis and are presented here for completeness. }
         \label{tab:exp_uncertainties}
\end{table*}

Choosing which datasets to include in any S-factor fit is important---for the reactions analyzed in this work, no one experiment spans the whole relevant energy range, so multiple datasets must be combined. However, some datasets may not be well-suited for our analysis. In this work, the following principles are generally used to select datasets for our fiducial analysis:
\begin{itemize}
	\item{Old (over half a century) datasets overlapping with newer experiments, but with larger uncertainties than newer experiments, are excluded.}
	\item{Data outside of the relevant energy range is excluded (see discussion in Section~\ref{sec:energy_range}).}
	\item{Datasets with unknown correlations with other experiments are excluded.}
	\item{If none of the above conditions apply to a dataset, it is included.}
	\item{If only total errors are reported, the percent error of the data point with the smallest error is conservatively assumed to be the systematic normalization error.}
\end{itemize}

All datasets we consider for our analysis are shown in Table~\ref{tab:exp_uncertainties}, in which we discuss our assumptions for the statistical and systematic uncertainties of each. For \ddp~and \ddn,~we use data between $\qty{14}{keV}$ and $\qty{0.6}{MeV}$ from the five independent datasets of Refs.~\cite{Krauss1987, Brown1990, Greife1995, Leonard2006}. The two datasets of Ref.~\cite{Krauss1987} are collected in different locations and are uncorrelated. Ref.~\cite{Leonard2006} report the covariance matrix for the Legendre polynomial fit that gives the total cross section, in which all data points are correlated to some degree. We incorporate these correlations in the GP, which no other analysis has done thus far. Including reported correlations between \ddn~and \ddp~data is left to future work. 

For \ddp,~we again restrict to data between $\qty{14}{keV}$ and $\qty{0.6}{MeV}$. For our fiducial result, we use the six independent datasets of Refs.~\cite{Ma1997, Schmid1997, Casella2002, Tisma2019, Mossa2020, Turkat2021} in line with those used for Solar Fusion III~\cite{SF32025}. Turkat \textit{et al.}\ present two datasets that share an overall normalization~\cite{Turkat2021}; this correlation is fully included in the GP. 

We exclude the following experiments from our analysis:
\begin{itemize}
	\item{The additional low-energy dataset of Ref.~\cite{Greife1995} (Greife 1995 Close Geometry) is collected to measure electron screening and is below $E_{\rm lower}$.}
	\item{The data of McNeill \textit{et al.}~\cite{McNeill1951} has large uncertainties and overlaps with more recent experiments.}
	\item{The data of Schulte \textit{et al.}~\cite{Schulte1972} is above $E_{\rm upper}$, and their sub-percent reported uncertainties may introduce new systematics in the GP.}
	\item{Despite our best efforts, we were not able to recover data of Ref.~\cite{RG1985} (First Research Group) from the plots displayed.  Additionally, the full text is not available in English and so it is difficult to verify their procedure.}
	\item{The Trojan Horse data of Tumino \textit{et al.}~\cite{Tumino2014} must be normalized to existing data. This inherently produces correlations between these data and the other experiments, which is important to include in these fits. This normalization procedure is not described in enough detail to include in the GP; with more details of the normalization, these data could be used in the GP. Additionally, Ref.~\cite{Coc2015} flag this dataset, claiming that the ratio between the \ddn~and \ddp~S-factors is well-constrained by theory. While other experiments match this prediction reasonably well, the data of Tumino \textit{et al.} do not. This result is independent of the normalization to existing data.}
	\item{The data of Griffiths \textit{et al.} (1963)~\cite{Griffiths1963} and Bailey \textit{et al.}~\cite{Bailey1970} overlap with more recent data and have much larger uncertainties.}
	\item{The data of Griffiths \textit{et al.} (1962)~\cite{Griffiths1962} is over half a century old and overlaps with more recent data.}
	\item{Warren \textit{et al.}~\cite{Warren1963} and Geller \textit{et al.}~\cite{Geller1967} measure energies above $E_{\rm upper}$. }
\end{itemize}

\subsection{Correlated Data in GPs}
\label{sec:correlated_gps}

In the partitioned GP prior we place on the S-factors, we add the covariance matrix of experimental data $\Sigma$ to the kernel in the top-left block (see Eq.~\eqref{eqn:gp_partitioned}). For \ddn,~and \ddp,~we assume that S-factor data follow a multivariate normal distribution. $\Sigma$ is block-diagonal, as only data within the same experiment are correlated. For some experiment $k$, consider two data points $i$ and $j$ with statistical noise $\sigma_{ik}$ and $\sigma_{jk}$ respectively. The majority of experiments report a single percent normalization uncertainty $\epsilon_k$, so these points have normalization uncertainty $\epsilon_k S_{ik}$ and $\epsilon_k S_{jk}$. Summing the statistical and normalization uncertainties in quadrature, the $ii^{th}$ component of $\Sigma$ is 
\begin{equation}
    \Sigma_{ii} = \sigma_{ik}^2+\epsilon_k^2S_{ik}^2.
\end{equation}
The off-diagonal entry $\Sigma_{ij}$ only includes the normalization uncertainties of these points:
\begin{equation}
    \Sigma_{ij} = \epsilon_k^2S_{ik}S_{jk}.
    \label{eqn:offdiag}
\end{equation}
Ref.~\cite{Leonard2006} provides a full covariance matrix for their data, which is used in place of Eq.~\eqref{eqn:offdiag}. Including Gaussian statistical uncertainty as a diagonal $\Sigma$ is a standard procedure \cite{Rasmussen2006}; our inclusion of correlations with off-diagonal elements is similar to other examples in the literature (for some examples in astrophysics and cosmology, see \cite{Shafieloo2012, Haridasu2018, Ruchika2025, deSouza2025}). We showed qualitatively that the GP captures these correlations desired in the discussion of GP regression on \ddn~data in Section~\ref{sec:fits}, and confirmed quantitatively the robustness of this procedure through our Monte Carlo tests.

As we discussed in Section~\ref{sec:fits}, the \dpg~S-factor values span nearly two orders of magnitude over an energy range of less than $2 \ln (E/\si{MeV})$. Therefore, we perform GP regression assuming the \dpg~S-factor data follow lognormal distributions, \textit{i.e.}\ $\ln S$ follows a normal distribution. 

Since we model $\ln{S}$ as a Gaussian process for \dpg,~we must be more careful when translating experimental data to $\ln S$. As with \ddn~and \ddp,~we take the reported central value $S_\mathrm{central}$ and squared percent uncertainty to be the median and $\mathrm{Var}[S]/\mathbb{E}[S]^2$ of the S-factor distributions. 
For normally-distributed S-factors, these directly correspond to the mean and variance. For $\ln S$, the reported central value is the peak of the distribution. Then, we interpret reported experimental data similarly for all three reactions, the only difference being whether $S$ or $\ln S$ follows a normal distribution. The mean and variance of the data distribution in log space $\mu_\mathrm{data}$ and $\sigma_\mathrm{data}$ are 
 \begin{equation}
 	\begin{split}
	 	\mu_\mathrm{data} = \ln (S_\mathrm{central} / \unit{MeV}), \\
		\sigma^2_\mathrm{data} = \ln{\left( 1 + \frac{\mathrm{Var}[S]}{\mathbb{E}[S]^2} \right)}.
	\end{split}
    \label{eqn:lognormal}
\end{equation}
For $\sigma_\mathrm{data} \leq 0.1$, the lognormal and Gaussian distributions are very similar. This is the case for nearly all data considered, supporting our choice of the lognormal distribution for \dpg. 

From Eq.~\eqref{eqn:lognormal}, it follows that the $ii^{th}$ component of $\Sigma$ in the $\ln S$ case is 
\begin{equation}
    \Sigma_{ii} = \ln{\left(1+\frac{\sigma_{ik}^2}{S_{ik}^2} + \epsilon_k^2 \right)},
\end{equation}
and the $ij^{th}$ component is 
\begin{equation}
    \Sigma_{ij} = \ln{\left(1 + \epsilon_k^2\right)}.
\end{equation}

\subsection{Kernel Choice} 
\label{sec:kernel_choice}

To model S-factors as Gaussian processes, we must define a kernel function to set correlations between S-factor values at different energies; the kernel enters into Eqs.~\eqref{eqn:gp_partitioned} and~\eqref{eqn:gp_conditional}. In setting these correlations, the kernel dictates the properties functions drawn from the GP will have. Thus, any kernel choice makes implicit assumptions regarding the set of functions used to fit S-factors, but these are more general and flexible than selecting a particular functional form.

The kernel is a function of two energies and includes hyperparameters $\boldsymbol{\theta}$ that must be chosen in a principled manner. These encode how strongly or weakly correlated the drawn S-factor values are as a function of some distance measure between different energy values. Throughout this work, we define the distance $d$ between two energies $E_i$ and $E_j$ as
\begin{equation}
	d(E_i, E_j) = \left|\ln(E_i/\unit{MeV}) - \ln(E_j/\unit{MeV})\right|=\left|\ln{\frac{E_i}{E_j}}\right|.
\end{equation}
We use log energies to fit data spanning nearly two orders of magnitude in $E$. Perhaps the most common kernel choice is the squared exponential:
\begin{equation}
	k_{\rm SE}(d; \sigma, \ell) = \sigma e^{-\frac{d^2}{2\ell^2}},
\end{equation}
where $\sigma$ is a hyperparameter controlling the amplitude of correlations and $\ell$ is a hyperparameter representing the characteristic length scale of correlations. The squared exponential is an appropriate kernel choice when the S-factor is expected to be smooth with correlations that only depend on energy differences. The Mat\'ern class of kernels relaxes the smoothness assumption while still including a characteristic length scale in a similar manner:
\begin{equation}
	k_{\rm{M},\nu}(d; \sigma, \ell) =  \sigma \, \frac{2^{1-\nu}}{\Gamma(\nu)} \left( \frac{\sqrt{2\nu} d}{\ell} \right)^\nu K_\nu\left( \frac{\sqrt{2\nu} d}{\ell} \right), 
\end{equation}
where $K_\nu$ is the modified Bessel function of the second kind~\cite{Rasmussen2006}. $\nu$ is a positive parameter that controls the smoothness of functions drawn from the GP.  Decreasing $\nu$ gives less smooth functions, while the $\nu\to\infty$ limit converges to the squared exponential kernel.

One might expect the squared exponential (or high-$\nu$ Mat\'ern) kernel to be sufficient to model these S-factors; since \ddp,~\ddn,~and \dpg~contain no known resonances in the BBN energy range~\cite{Xu2013}, the S-factors are expected to be globally smooth. However, datasets are internally correlated, and different experiments disagree by $\gtrsim1\sigma$ at some energies. Therefore, if only a smooth kernel is used, the data will require an extremely small length scale $\ell$, causing S-factor samples to fluctuate rapidly, exceeding $\qty{1}{\mega\eV\barn}$ and even going negative.

To allow GPs to fit small-scale fluctuations necessary for agreement between disparate datasets while preserving large-scale correlations, we adopt a kernel that is a combination of both a small-$\nu$ Mat\'ern kernel and the squared exponential as our fiducial choice:
\begin{equation}
	k(d; \boldsymbol{\theta}) = k_{\rm SE}(d; \sigma_0, \ell_0) + k_{\rm{M},\nu=1/4}(d; \sigma_1, \ell_1)
	\label{eqn:kernel}
\end{equation}
with hyperparameters $\boldsymbol{\theta} = \{\sigma_0, \ell_0, \sigma_1, \ell_1\}$. 
Eq.~\eqref{eqn:kernel} has two pieces: the squared exponential is a ``global component'' characterizing the shape of the S-factors throughout the entire BBN energy range, while the Mat\'ern-1/4 is a ``local component'' that allows small-scale fluctuations to account for scatter within the real data. 
We set $\nu=1/4$ to add small-scale correlations without introducing extra statistical noise at each point. In choosing Eq.~\eqref{eqn:kernel}, we assume that correlations depend only on differences between energies and that hyperparameters are constant throughout the entire BBN energy range. 

The fiducial kernel choice leads to S-factor sample draws that exhibit small-scale fluctuations as a function of energy (see the top panels of Figures~\ref{fig:ddtp_samples},~\ref{fig:ddhe3n_samples},~and \ref{fig:dphe3g_samples}). 
While individual samples likely do not resemble the true S-factor (which should be globally smooth), such samples are acceptable or even advantageous for two reasons.
First, undersmoothed GP kernels---such as a delta-function term (the $\nu \to 0$ limit of the Mat\'ern kernel) added to a squared exponential---have been shown to yield posteriors with better coverage, avoiding the underestimation of uncertainties that can arise from smoothed kernels~\cite{Gramacy2012, Knapik2011, Yang2017}. 
This choice therefore may play a role in achieving the reliable coverage that we see in our MC study of the GP method (see Section~\ref{sec:MC}).
Second, our GP draws for the S-factor are first thermally averaged to obtain the reaction rate $R(T)$, the quantity that actually enters BBN calculations. 
Our $R(T)$ draws are in fact smooth as a function of $T$ (see \textit{e.g.}\ the bottom panel of Figure~\ref{fig:ddtp_samples}). 

\subsection{Hyperparameter Optimization}

After selecting a kernel, hyperparameters $\boldsymbol{\theta}$ must be chosen in a principled manner before using Eq.~\eqref{eqn:gp_conditional} to draw samples. A common approach, adopted in this work, is to define and optimize a likelihood from the experimental data. For many applications, optimizing the marginal likelihood to observe the data given $\boldsymbol{\theta}$ is a well-motivated choice. However, some S-factor data we consider are in mild disagreement with others. Optimizing the marginal likelihood yields posteriors that can overfit to experiments with very small uncertainties, and we want to mitigate weighting individual datasets over others in this analysis. 

We achieve this with cross-validation~\cite{Stone1974}, in which a subset of the training data is left out from the rest of the set and the probability of observing those points given the remaining data is maximized (for further discussion of cross-validation with GPs, see Ref.~\cite{Rasmussen2006}). In Leave-One-Out (LOO) cross-validation, another common choice for GP hyperparameter selection, data points are left out one at a time. LOO-CV leads to extreme overfitting during optimization, as the kernels we consider have the flexibility (with very small correlation lengths) to perfectly fit every data point.

In this work, we select kernel hyperparameters by maximizing a Leave-Dataset-Out (LDO) pseudo-likelihood. For a set of hyperparameters $\boldsymbol{\theta}$, we leave each dataset $k$ out one at a time. We compute the probability to observe the left-out dataset given the GP conditioned on the remaining datasets, which come from Eq.~\eqref{eqn:gp_conditional}. We use the sum of the log predictive probabilities to evaluate performance and update hyperparameters:
\begin{equation}
	\mathcal{L}_{\rm LDO} = \sum_{k\in\text{sets}} \ln p(\boldsymbol{S}_k | \boldsymbol{S}_{-k}, \boldsymbol{\theta}),
	\label{eqn:full_ldo}
\end{equation}
where $\boldsymbol{S}_k$ and $\boldsymbol{S}_{-k}$ denote the S-factor measurements of dataset $k$ and all other datasets respectively. The log probability to observe dataset $k$ given the multivariate Gaussian conditioned on the other datasets is 
\begin{equation}
	\begin{split}
		- 2\ln p(\boldsymbol{S}_k | \boldsymbol{S}_{-k}, \boldsymbol{\theta}) = (\boldsymbol{S}_k - \boldsymbol{\mu}_{\boldsymbol{S}_k | \boldsymbol{S}_{-k}})^T \times \\
		\Sigma_{\boldsymbol{S}_k | \boldsymbol{S}_{-k}}^{-1} (\boldsymbol{S}_k - \boldsymbol{\mu}_{\boldsymbol{S}_k | \boldsymbol{S}_{-k}}) + \\
		\ln \det  \Sigma_{\boldsymbol{S}_k | \boldsymbol{S}_{-k}} + {\rm const}, 
	\end{split}
	\label{eqn:ldo}
\end{equation}
where $\mu_{\boldsymbol{S}_k | \boldsymbol{S}_{-k}}$ and $\Sigma_{\boldsymbol{S}_k | \boldsymbol{S}_{-k}}$ are the conditional mean and covariance from Eq.~\eqref{eqn:gp_conditional}. Eqs.~\eqref{eqn:full_ldo} and~\eqref{eqn:ldo} form a more general version of Leave-One-Out (LOO) cross-validation, expanded to include multiple left-out points. 

Each dataset receives one contribution to Eq.~\eqref{eqn:full_ldo}, weighting them evenly during optimization and mitigating the risk of overfitting to datasets with small uncertainties. Additionally, there are energy regions in which the GPs must interpolate or extrapolate. 
Since Eq.~\eqref{eqn:ldo} directly evaluates the performance of GPs conditioned on $\boldsymbol{S}_{-k}$ at energies not represented in $\boldsymbol{S}_{-k}$, it penalizes models that generalize poorly. 

As these kernels have four hyperparameters, we use gradient descent for optimization. To avoid convergence to local minima in hyperparameter space, we use \textit{Adam}~\cite{Kingma2017}, a gradient descent algorithm incorporating momentum when updating hyperparameters. We consider only fixed choices for the kernel hyperparameters set by these optimizations. Code for our analysis will be released with the full cosmological analysis to follow. 

\subsection{Further Checks}
\label{sec:further_checks}

We perform the following checks to test the sensitivity of our D/H prediction to the various choices made throughout our analysis:
\begin{itemize}
	\item{Different hyperparameter optimization.}
	\item{Pushing $E_{\rm upper}$ to $\qty{2}{MeV}$ for \dpg. }
	\item{Using the theory calculations of Refs.~\cite{Arai2011} and~\cite{Marcucci2016} as the GP prior means. }
	\item{Different kernel choices from Eq.~\eqref{eqn:kernel}.}
    \item{Checking the impact of extrapolation beyond \qty{0.3}{\mega\eV}.}
\end{itemize}
Throughout this section, the reader should keep in mind that our fiducial prediction uncertainty for D/H is 1.6\%. 

\textbf{Different hyperparameter optimization.}---LDO cross-validation differs from another common hyperparameter optimization method, in which the marginal likelihood of observing the known data given a particular choice of hyperparameters is maximized. Equivalently, one minimizes the negative log likelihood: 
\begin{equation}
	-2\ln{p\left( \boldsymbol{S} | \boldsymbol{\theta} \right)} = \boldsymbol{S}^T K_{11}^{-1} \boldsymbol{S} + \ln{|K_{11}|} + {\rm const.}
	\label{eqn:mle}
\end{equation}
Using Eq.~\eqref{eqn:mle} neither weights individual experiments evenly nor evaluates predictive power in regions without data. Repeating our analysis with Eq.~\eqref{eqn:mle} predicts a D/H only $\sim0.3\%$ greater than the fiducial result with slightly smaller uncertainties. Uncertainties decrease since datasets are not evenly weighted during optimization. That these differences are small relative to the prediction uncertainty (1.6\%) demonstrates the robustness of our result to the choice of optimization procedure. We still prefer LDO cross-validation for our fiducial result, as it is the more principled approach.

\begin{figure}
    	\centering
	\includegraphics[width=\linewidth]{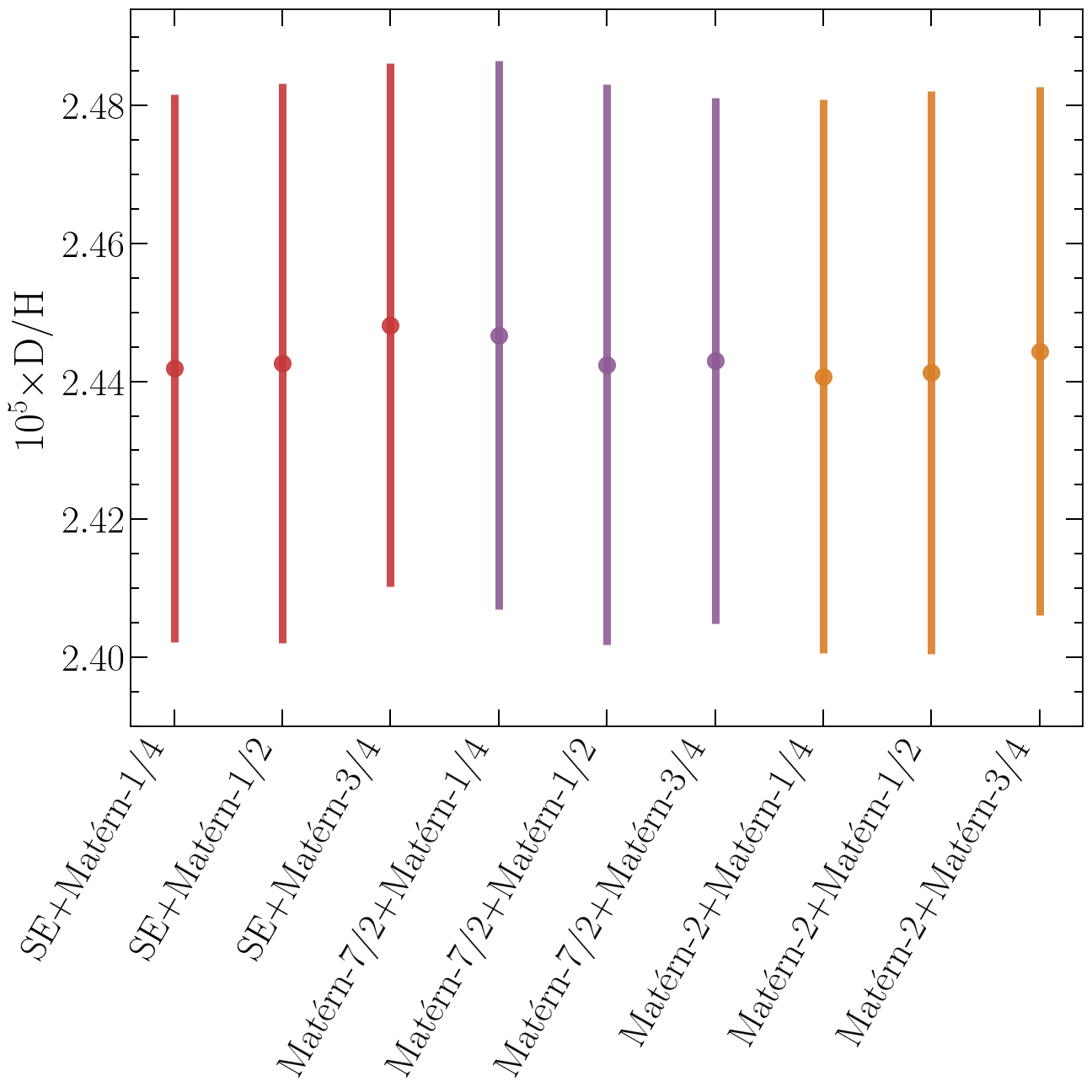}
	\caption{D/H predictions from GP regression on \ddp,~\ddn,~and \dpg~data for different kernel choices. Kernels are labeled \textit{global}+\textit{local}. All other inputs are the same as our fiducial analysis presented in Results, including marginalization over the \textit{Planck} baryon density.}
	\label{fig:kernel_comp}
\end{figure}

\textbf{Increasing $\boldsymbol{E_{\rm upper}}$ for $\boldsymbol{d}$($\boldsymbol{p}$,$\boldsymbol{\gamma}$)$^\mathbf{3}$He}.---Increasing $E_{\rm upper}$ for \ddn~and \ddp~would only add the data of Refs.~\cite{Schulte1972} and~\cite{Tumino2014} to our analysis, which we exclude for other reasons (see the discussion above). On the other hand, increasing $E_{\rm upper}$ to $\qty{2}{MeV}$ for \dpg~adds a new dataset (Ref.~\cite{Warren1963}) and additional data points from Ref.~\cite{Turkat2021}. Including these data in our \dpg~GP increases the predicted D/H by $0.25\%$ without changing the predicted uncertainty. Again, this shift is small relative to the prediction uncertainty, showing that our result is robust to different $E_{\rm upper}$. 

\textbf{Different prior means}.---The theory calculations of Refs.~\cite{Arai2011} and~\cite{Marcucci2016} give some prior information for these S-factors, so we repeat our analysis for GP regression on \ddn,~\ddp,~and \dpg~data with their results as the GP prior means ($\boldsymbol{\mu}$ in Eq.~\eqref{eqn:gp_conditional}). The bottom panel of Figure~\ref{fig:ddhe3n_samples} shows how this impacts the \ddn~GP mean. Where there are many data points, both the fiducial GP mean and the theory prior GP mean agree well within $1\sigma$. Large changes are only present in the extrapolation region---the theory prior pushes the extrapolation to higher values since the theory calculation predicts a slightly higher S-factor that measured by experiment. D/H lowers by $0.37\%$, and the prediction uncertainty slightly decreases as this extra information makes the GP more confident during extrapolation. 

\textbf{Different kernel choices}.---To investigate the sensitivity of our result to different choices, we additionally consider Mat\'ern-7/2 and Mat\'ern-2 for the global component and Mat\'ern-1/2 and Mat\'ern-3/4 for the local component. Figure~\ref{fig:kernel_comp} shows the D/H prediction for these different kernels: three global and three local components give nine variations in total. Sampling \ddp,~\ddn,~and \dpg~GPs and marginalizing over the other reactions, the \textit{Planck} baryon density, and the neutron lifetime give the uncertainties shown in Figure~\ref{fig:kernel_comp}. The D/H prediction has little sensitivity to the particular kernel choice relative to the size of these uncertainties, changing the central values by at most $0.12\%$. 

\begin{figure}
    	\centering
	\includegraphics[width=\linewidth]{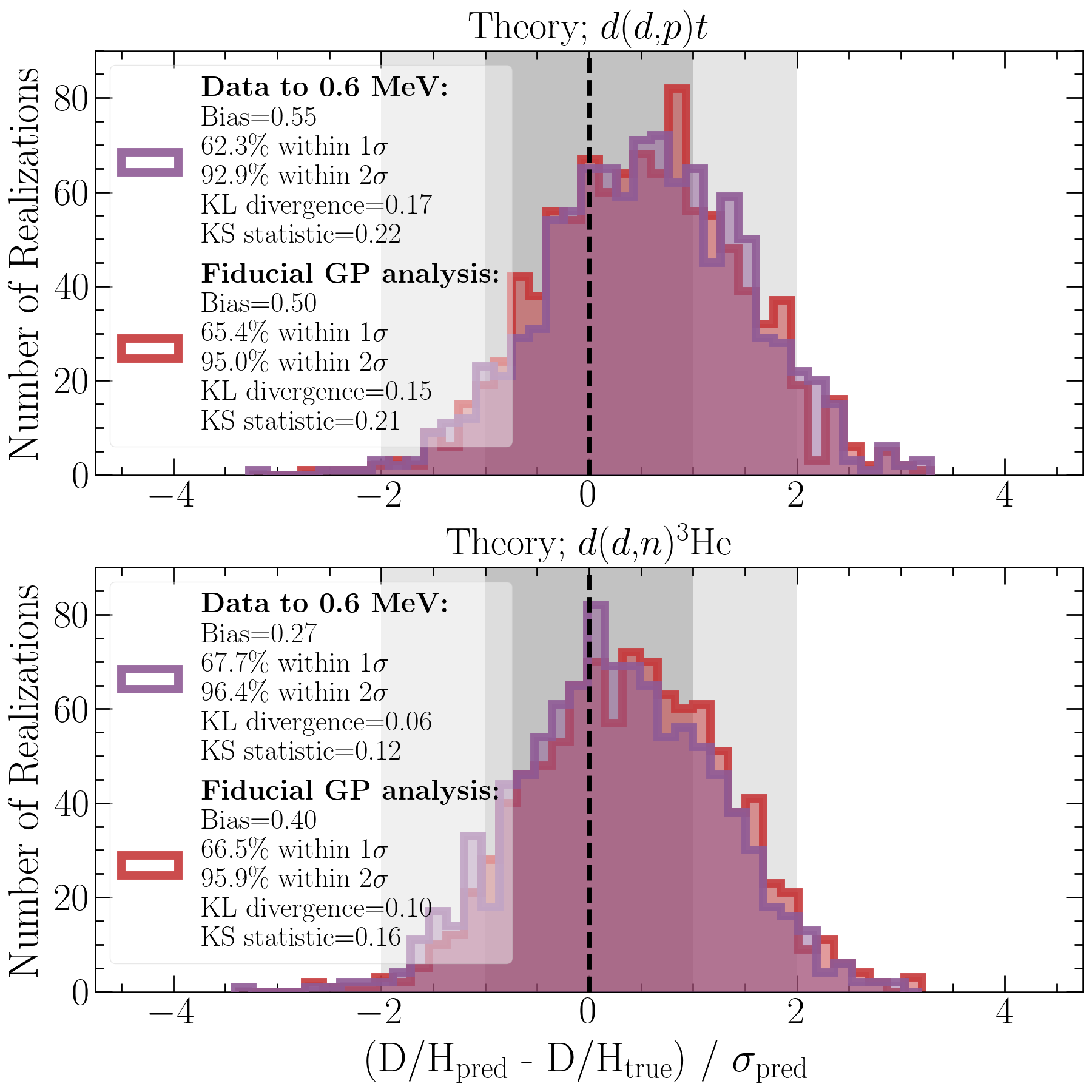}
	\caption{D/H predictions for 1,000 Monte Carlo realizations of mock \ddp~and \ddn~data fit with Gaussian processes. We use the Arai \textit{et al.}\ theory calculation~\cite{Arai2011} to generate the mock data. We add an additional mock dataset (purple), with data extending to $\qty{0.6}{MeV}$ that matches the precision of Ref.~\cite{Leonard2006}. We show the fiducial Monte Carlo analysis without this added dataset (red) for comparison. }
	\label{fig:mc_add}
\end{figure}

While all of these kernels work for this analysis, we must still make a choice for the fiducial analysis. We choose the squared exponential for the global component such that all non-differentiability comes from the small-$\nu$ Mat\'ern local component. For the local component, we choose Mat\'ern-1/4 from consideration of the properties of the Mat\'ern kernel. For fixed correlation length $\ell$, correlations between the S-factor at different energies decay more quickly as $\nu$ is lowered. In the $\nu\to0$ limit, $k(d)$ goes to $0$ unless $d=0$, and the Mat\'ern kernel becomes noise added equally to each point. 
We directly know experimental uncertainties, so we avoid adding extra statistical noise equally to each data point. Therefore, we cannot make $\nu$ arbitrarily small. 

$\nu$ cannot be too large either; setting $\nu\gg1$ does not produce the required small-scale fluctuations. Even for $\nu\geq0.5$, correlations decay slowly enough that fluctuations span energy ranges larger than fluctuations between data. The local component should only account for fluctuations in the data, and we want to mitigate artifacts arising from increasing the range of these small-scale fluctuations. The intermediate $\nu=1/4$ provides the desired behavior, but we emphasize that our analysis is robust to other choices.

\textbf{Checking extrapolation.}---To check the impact of extrapolation at energies greater than $\qty{0.3}{MeV}$ on our D/H predictions, we repeat the Monte Carlo analysis from Section~\ref{sec:MC} with GP regression. We add a new mock dataset at eight equally-spaced energies between $0.25$ and $\qty{0.6}{MeV}$ matching the reported uncertainties of Ref.~\cite{Leonard2006} and generate mock data using the theory calculation of Ref.~\cite{Arai2011} as in our fiducial MC analysis. Now, the GP does not need to extrapolate for any region in the relevant BBN energy range. In Figure~\ref{fig:mc_add}, we show the resulting D/H predictions for \ddp~and \ddn~for the theory data generating functions. D/H predictions do not significantly change for \ddp,~and there is moderate bias reduction from $0.40$ to $0.27$ for \ddn~with this extra mock dataset. This result suggests that extrapolation to high energies likely does not significantly bias our result. However, the S-factor may behave differently than the theory prediction of Ref.~\cite{Arai2011} above $\qty{0.3}{MeV}$; additional data measuring the \ddn~and \ddp~S-factors up to $\qty{0.6}{MeV}$ would certainly improve our analysis. This additional mock dataset does not eliminate bias in our procedure, indicating that the bias is not a result of our extrapolation; see Supplementary Information for further discussion.

\newpage

\onecolumngrid

\section*{Supplementary Information}
\setcounter{subsection}{0} 
\label{sec:supp}

\subsection{Comparing Distributions}

We compare the performance of GPs with polynomials to make accurate D/H predictions from fits to mock S-factor data in the main body. For each mock data realization, we calculate the predicted $\mathrm{D/H}_{\rm pred} \pm \sigma_{\rm pred}$ and compare these to the D/H for the underlying function that generated the data $\mathrm{D/H}_{\rm true}$. We do this for 1,000 mocks, and approximate $(\mathrm{D/H}_{\rm pred} - \mathrm{D/H}_{\rm true})/\sigma_{\rm pred}$ distributions as the resulting empirical sample set (see Figure~\ref{fig:mc_test_single}). 

A fully unbiased fitting procedure with robust uncertainties would produce a set of $(\mathrm{D/H}_{\rm pred} - \mathrm{D/H}_{\rm true})/\sigma_{\rm pred}$ converging to $\mathcal{N}(0,1)$ for infinite mock realizations. We introduce two quantities to assess the similarity of the empirical distributions to $\mathcal{N}(0,1)$: the Kullback--Leibler (KL) divergence~\cite{Kullback1951} and the Kolmogorov--Smirnov (KS) statistic~\cite{Kolmogorov1933}. We use the quantities as measures of proximity to $\mathcal{N}(0,1)$, enabling a quantitative comparison between GPs and polynomials across the scenarios shown in Figure~\ref{fig:mc_test_single}. Both are non-negative, vanishing only when the compared distributions are identical. However, they compare different features of the distributions. 

The KL divergence from a reference distribution $Q$ to a distribution $P$ is defined as 
\begin{equation}
	D_{\rm KL}(P \| Q) = \sum_i P_i \ln{\frac{P_i}{Q_i}}.
\end{equation}
Here, the sum runs over histogram bins in Figure~\ref{fig:mc_test_single}. $P$ is the empirical distribution and $Q$ is $\mathcal{N}(0,1)$. $D_{\rm KL}$ quantifies how distinguishable the empirical distribution is from $\mathcal{N}(0,1)$, with larger values indicating greater disagreement. We find little sensitivity of $D_{KL}$ values to bin width for $\geq10$ bins. 

The KS statistic is defined as 
\begin{equation}
	D_{\rm KS} = \sup_x \left| F(x) - G(x) \right|,
\end{equation}
where $F(x)$ is the empirical cumulative distribution function (CDF) and $G(x)$ is the $\mathcal{N}(0,1)$ CDF. The supremum is taken over all $x$, so $D_\text{KS}$ isolates the point at which the two CDFs differ the most. The CDFs $F(x)$ and $G(x)$ run from $0$ to $1$, so $D_{\rm KS}\in[0,1)$. Again, larger values indicate greater disagreement, and $D_\text{KS}$ is zero only when the two distributions are the same. $D_{\rm KS}$ is therefore most sensitive to localized deviations, such as an excess or lack of probability in a particular region. 

Since $D_\text{KL}$ is an integrated quantity weighted by the empirical distribution, it is more sensitive to broad, systematic differences. On the other hand, $D_\text{KS}$ highlights the largest local deviation regardless of how the rest of the distribution compares. Together, they provide a more comprehensive picture of distributional agreement.

In all scenarios considered in Figure~\ref{fig:mc_test_single}, both the KL divergence and KS statistic are consistently smaller for GPs than for polynomials, with KL divergences a factor of $\sim 5$ ($10$) and KS statistics a factor of $\sim 2$ ($3$) smaller for the theory (smoothed GP) generating function. Since $D_\text{KS}$ is bounded between $0$ and $1$, improvement by an $\mathcal{O}(1)$ factor is significant. Thus, we quantitatively confirm that GPs produce D/H predictions whose distributions are closer to the expected $\mathcal{N}(0,1)$ behavior of an unbiased procedure with well-calibrated uncertainties.

\subsection{Polynomial Regression on Correlated Data}

In the main body, we argue that, while polynomial S-factor fitting systematically over-predicts D/H, Eq.~\eqref{eqn:poly_chi2_single} should be used to fit correlated data from multiple experiments. For the rates in the PArthENoPE network, the following expression is used instead~\cite{Serpico2004,Pisanti2026}: 
\begin{equation}
	\chi^2 = \sum_{k\in\mathrm{sets}} \sum_{i=1}^{N_k} \left(\frac{(S_{\rm th}(E_{ik}, a_l) - \omega_kS_{ik})^2}{\omega_k^2 \sigma_{ik}^2} + \frac{(\omega_k-1)^2}{\epsilon_k^2}\right).
	\label{eqn:parth_chi2}
\end{equation}
With multiple $\omega_k$ penalty terms per dataset, this expression no longer corresponds with maximum \textit{a posteriori} parameter selection but is still valid as a penalized likelihood. Here, we argue that Eq.~\eqref{eqn:parth_chi2} yields estimators for $\{\hat{\omega}_k\}$ that do not match the expected distribution when $N_k$ is large. 

\begin{figure}
	\centering
	\includegraphics[width=0.7\linewidth]{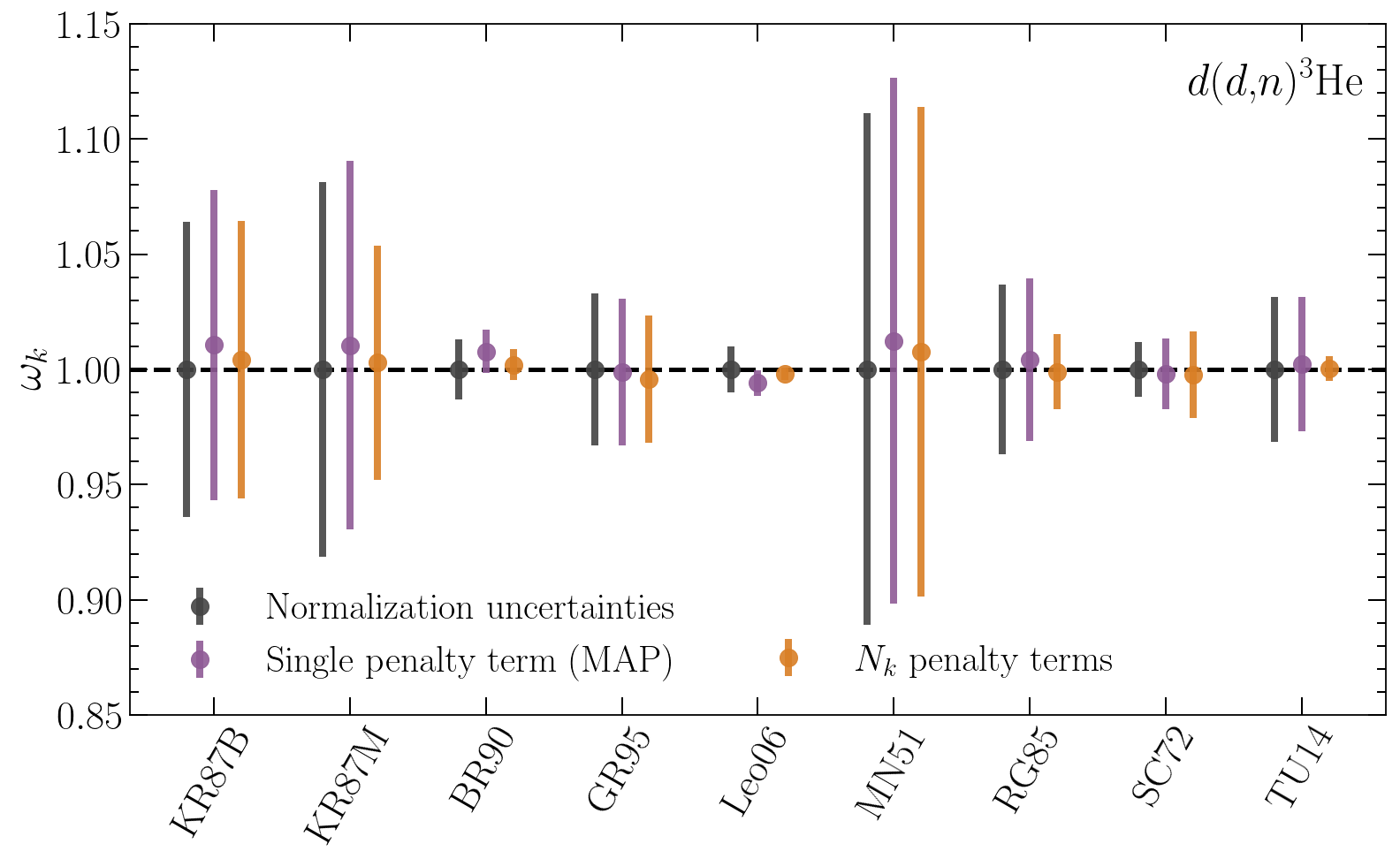}
	\caption{Values of $\{\omega_k\}$ for 1,000 realizations of mock \ddn~data generated by the theory curve of Ref.~\cite{Arai2011}. For the orange points, Eq.~\eqref{eqn:parth_chi2} is used to fit the mock data. For a dataset $k$ with $N_k$ points, this penalized likelihood contains $N_k$ penalty terms per dataset. For the purple points, Eq.~\eqref{eqn:poly_chi2_single} (yielding the maximum \textit{a posteriori} estimate) is used to fit the mock data. Grey points show the reported normalization uncertainties for each experiment, which are used to generate mock data. }
	\label{fig:parth_omegas}
\end{figure}

Consider a simpler case, where a single experiment performs $N$ correlated measurements $\{x_i\}$ of a single parameter $x_{\rm th}$, each with the same statistical variance $\sigma^2$ and percent systematic uncertainty $\epsilon$. Using the expression in Eq.~\eqref{eqn:parth_chi2}, the log likelihood is
\begin{equation}
	L = \sum_{i=1}^{N} \frac{(x_i-x_{th}/\omega)^2}{\sigma^2} + \frac{N(\omega-1)^2}{\epsilon^2}
\end{equation}
up to a constant, with $\omega$ being a nuisance parameter. We can calculate the best-fit estimators $\hat{x}_{\rm th}$ and $\hat{\omega}$ for $x_{\rm th}$ and $\omega$ respectively analytically. Beginning with $\hat{x}_{\rm th}$,
\begin{equation}
	\frac{\partial L}{\partial x_{th}} \bigg|_{\hat{x}_{th}, \hat{\omega}} = \sum_i \frac{2}{\sigma^2} (x_i - \hat{x}_{th}/\hat{\omega}) (-1/\hat{\omega}) = 0 \quad \Longrightarrow \quad \hat{x}_{th} = \hat{\omega} \mu,
	\label{eqn:x_th}
\end{equation}
where $\mu = \sum_i x_i / N$ is the experimental mean; $\hat{x}_{th}$ is successfully unbiased. Repeating this procedure for $\omega$:
\begin{equation}
	\frac{\partial L}{\partial \omega} \bigg|_{\hat{x}_{th}, \hat{\omega}} = \sum_i \frac{2}{\sigma^2} (x_i - \hat{x}_{th}/\hat{\omega}) \left(\frac{\hat{x}_{th}}{\hat{\omega}^2}\right) + \frac{2N}{\epsilon^2} (\hat{\omega} - 1) = 0. \nonumber
\end{equation}
Since $\sum_i (x_i - \hat{x}_{th}/\hat{\omega})=0$ from Eq.~\eqref{eqn:x_th}, we obtain $\hat{\omega} = 1$, which again is expected. Building a confidence interval around this estimate gives the error for $\omega$. Performing a Taylor expansion of $L$ at the minimum, where it is approximately quadratic, one obtains
\begin{equation}
	\sigma_\omega^{-2} \approx \mathbb{E} \left( \frac{1}{2} \frac{\partial^2 L}{\partial \omega^2} \bigg|_{\hat{x}_{th}, \hat{\omega}} \right) \nonumber
\end{equation}
Taking the derivative, it follows that
\begin{equation}
	\sigma_\omega^2 = \left( \frac{N}{\sigma^2} \langle\mu^2\rangle + \frac{N}{\epsilon^2} \right)^{-1} = \left( 1+ \frac{N}{\sigma^2} \langle\mu\rangle^2 + \frac{N}{\epsilon^2} \right)^{-1}.
\end{equation}
Consider the case where the penalty term for $\omega$ is relatively large. Then, one would expect that the uncertainty in the penalty term to be comparable to the normalization uncertainty: $\sigma^2_\omega \approx \epsilon^2$. However, we actually recover
\begin{equation}
	\sigma_\omega^2 \approx \epsilon^2 / N.
	\label{eqn:sigma_omega}
\end{equation}
In the large $N$ limit, the uncertainties on the scaling $\omega$ become suppressed by $1/N$. Therefore, in the multiple dataset case, we might expect datasets with many data points to underestimate systematic uncertainties for those datasets. For example, the \ddn~and \ddp~dataset of Tumino \textit{et al}. (TU14)~\cite{Tumino2014} contains 75 data points, and the dataset of the First Research Group (RG85)~\cite{RG1985} contains 50 data points, much larger than the 6 data points of Leonard \textit{et al.} (Leo06)~\cite{Leonard2006}. 

\begin{figure}
	\begin{minipage}{0.48\textwidth}
		\centering
		\includegraphics[width=\linewidth]{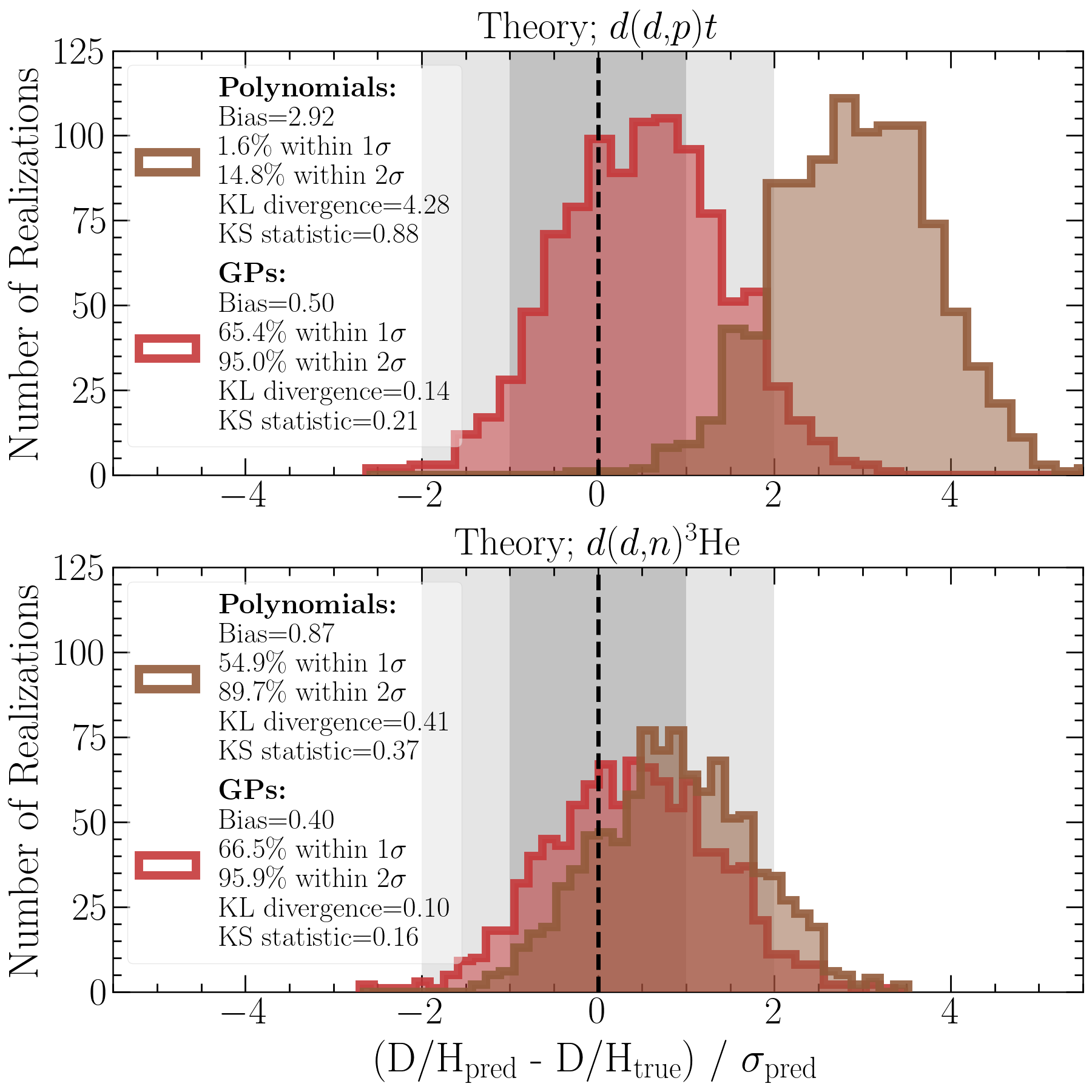}
	\end{minipage}
	\begin{minipage}{0.48\textwidth}
		\centering
		\includegraphics[width=\linewidth]{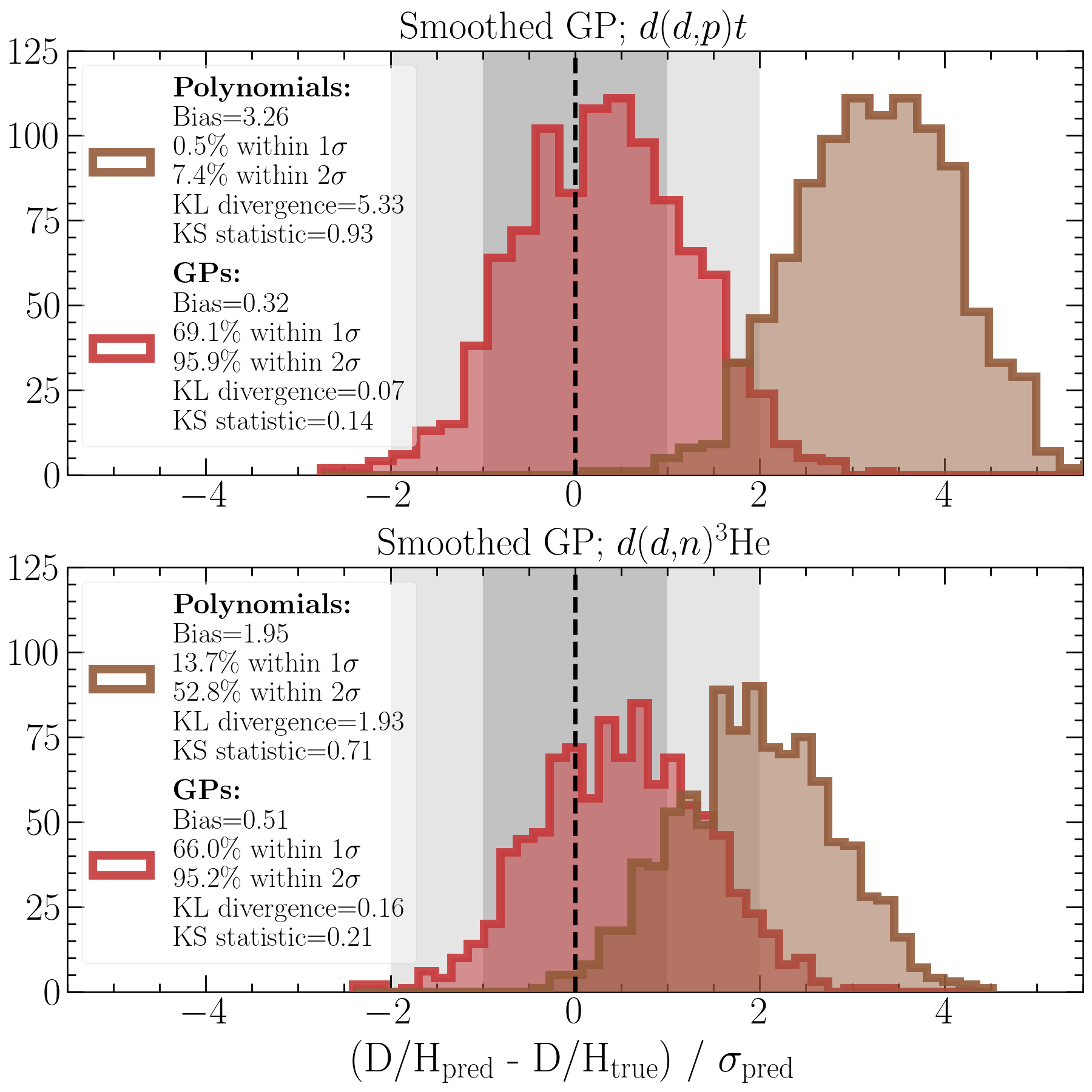}
	\end{minipage}
	\caption{D/H prediction statistics for 1,000 Monte Carlo realizations of mock \ddp~and \ddn~data fit with both Gaussian process regression (red) and low-degree polynomials (brown), similar to Figure~\ref{fig:mc_test_single} but for polynomial coefficients optimized using Eq.~\eqref{eqn:parth_chi2}. The Arai \textit{et al.}\ theory calculation~\cite{Arai2011} and the GP mean from real data, smoothed and offset in energy, are used to generate the mock data. D/H predictions are shown as the difference between the predicted and true values, normalized by the predicted $1\sigma$ uncertainty. }
	\label{fig:mc_test}
\end{figure}

We repeat the Monte Carlo analysis for polynomials with mock data described in the main body using Eq.~\eqref{eqn:parth_chi2}, thus replicating the PArthENoPE network more closely. In addition to the generating function reconstruction and D/H predictions, we track the values of $\{\omega_k\}$ for each mock dataset and each realization to check whether $\omega_k$ are pushed closer to 1 for datasets with many points. 

Figure~\ref{fig:parth_omegas} shows the distributions of $\{\omega_k\}$ obtained for degree-4 polynomial fits to mock \ddn~data generated by the theory curve of Ref.~\cite{Arai2011}, along with the reported experimental normalization uncertainties used to generate the mock data in gray. $\{\omega_k\}$ from fits using a single penalty term (Eq.~\eqref{eqn:poly_chi2_single}) are shown in purple, while orange points are from fits with $N_k$ penalty terms per dataset (Eq.~\eqref{eqn:parth_chi2}). The $\omega_k$ terms account for normalization uncertainties; their distributions from many realizations of the data should therefore be comparable to $\mathcal{N}(1,\epsilon_k$), where $\epsilon_k$ is the reported normalization uncertainty of experiment $k$. For mock datasets with $\lesssim10$ data points, there is not much difference between their $\omega_k$ distributions. This is the small-$N$ limit of Eq.~\eqref{eqn:sigma_omega}. The $\omega_k$ distributions for the mock datasets corresponding with RG85~\cite{RG1985} and TU14~\cite{Tumino2014} roughly match the reported normalization uncertainties ($\sim4\%$ and $\sim3\%$ respectively) when a single prior term is used per dataset (the MAP estimate from Eq.~\eqref{eqn:poly_chi2_single}). These two distributions become considerably narrower when multiple penalty terms are used (Eq.~\eqref{eqn:parth_chi2}) used, which was expected from the analytic result above. 

Figure~\ref{fig:mc_test} shows the reconstructed generating functions for \ddp~and the D/H predictions for polynomials fit to both the theory and smoothed GP generating functions for \ddn~and \ddp~(see Results) using Eq.~\eqref{eqn:parth_chi2}. This method consistently performs poorer at D/H recovery than GPs and also the penalized likelihood with a single penalty term per dataset (the MAP estimate). For example, \ddp~data generated by the simple theory curve over-predicting D/H by $\sim3\sigma$ on average with only $14.8\%$ of realizations within $2\sigma$ of the true value. Additionally, both the KL divergence and KS statistic indicate that these distributions differ even more from the desired $\mathcal{N}(0,1)$ corresponding with an unbiased, robust fitting procedure. 

\subsection{Monte Carlo Tests with Uncorrelated Data}

Through our Monte Carlo validation tests, we find that fitting mock \ddp,~\ddn,~and \dpg~S-factor data with GP regression predicts D/H biased $\sim0.3-0.5\sigma$ above the true value on average. This bias is important to note; however, it is small compared to the size of the predicted uncertainties, and the $1\sigma$ and $2\sigma$ predicted uncertainties include $\sim68\%$ and $\sim95\%$ of realizations respectively. In Section~\ref{sec:further_checks}, we showed that in our MC analysis, adding mock data to cover the entire relevant BBN energy range (\textit{i.e.}\ up to $\qty{0.6}{MeV}$) does not remove this bias, despite the GPs no longer needing to extrapolate. 

Our handling of correlated experimental S-factor data, described in Section~\ref{sec:correlated_gps}, could be the underlying cause for this persistent positive bias in D/H predictions. Ref.~\cite{DAgostini1994} found that fitting to correlated data by minimizing a $\chi^2$ with nonzero off-diagonal entries in the covariance matrix yields estimators that are biased towards lower values. We do not minimize a $\chi^2$ for GPs, but we do include a covariance matrix with correlations during regression (see Eq.~\eqref{eqn:gp_partitioned}). A similar effect could produce this slight bias in our analysis; the S-factor reconstructions for GPs are, on average, slightly lower than the generating function, which is clearly observed in the top-left panel of Figure~\ref{fig:mc_test_single}. This slightly lower \ddp~and \ddn~S-factors that we recover in our MC study leads to less deuterium burning during BBN and higher D/H, biasing our result toward closer agreement with the Cooke \textit{et al.}\ measurement assuming \textit{Planck} $\omega_b$. 

To further investigate the cause of this bias, we repeat the \ddn~and \ddp~Monte Carlo validation tests for GPs but remove any correlations between data points by setting $\epsilon=0$ in Eq.~\eqref{eqn:mock_gen}. Since only statistical noise is included for these tests, we remove mock data corresponding with the M\"unster dataset of Ref.~\cite{Krauss1987} for \ddn,~and both datasets of Ref.~\cite{Krauss1987} for \ddp,~as these data have no reported statistical uncertainties (cited as $<1\%$). Mock data still cover the same energy range with this choice. We use the theory calculation of Ref.~\cite{Arai2011} to generate mock data realizations. 

\begin{figure}
	\centering
	\begin{minipage}{0.48\textwidth}
		\centering
		\includegraphics[width=\linewidth]{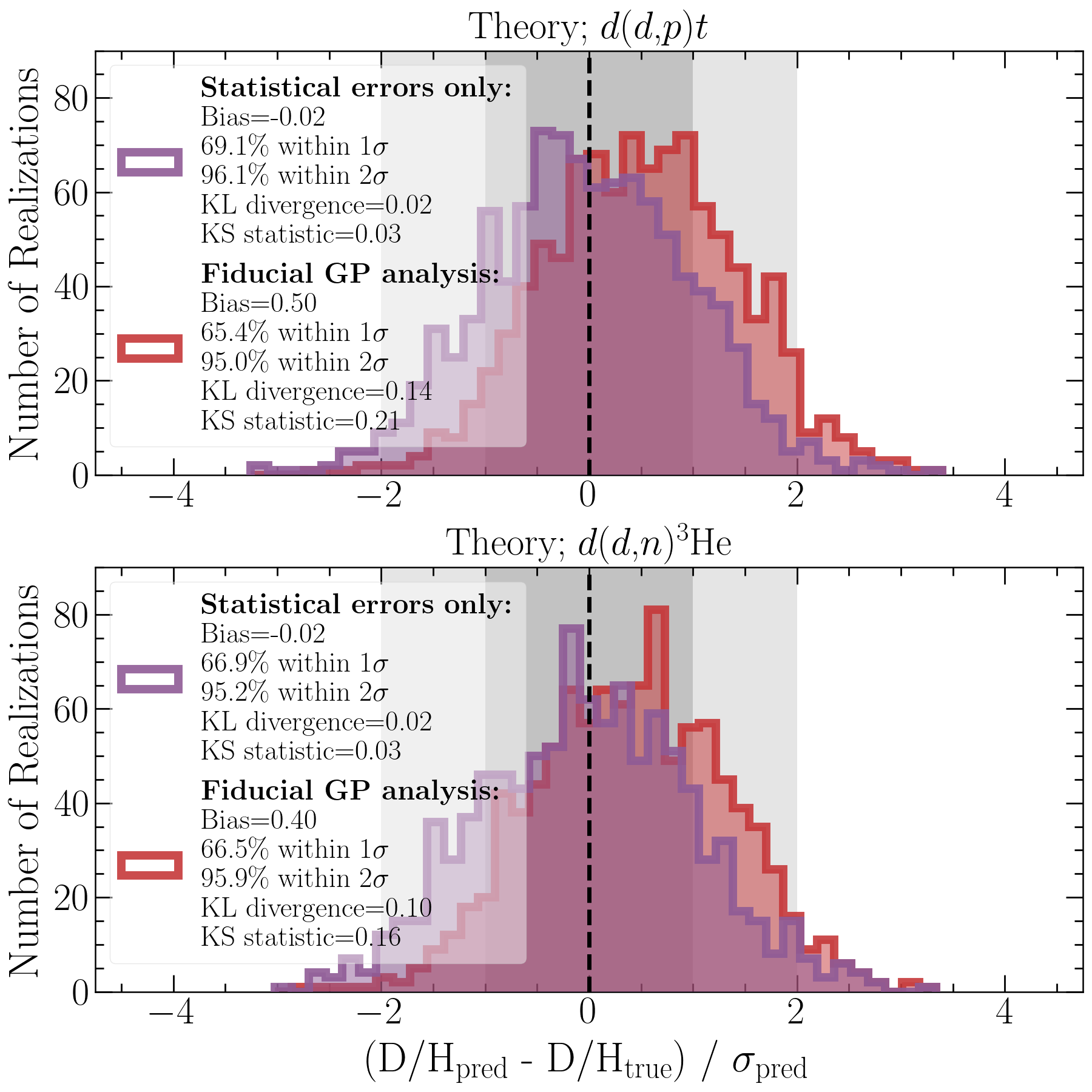}
		\caption{D/H predictions for 1,000 Monte Carlo realizations of mock \ddp~and \ddn~data fit with Gaussian processes. We use the Arai \textit{et al.}\ theory calculation~\cite{Arai2011} to generate the mock data. We remove systematic uncertainties from the data generation process and GP regression (purple) and compare with the fiducial Monte Carlo analysis (red). }
		\label{fig:mc_stat}
	\end{minipage}
	\hspace{0.02\textwidth}
	\begin{minipage}{0.48\textwidth}
		\centering
		\includegraphics[width=\linewidth]{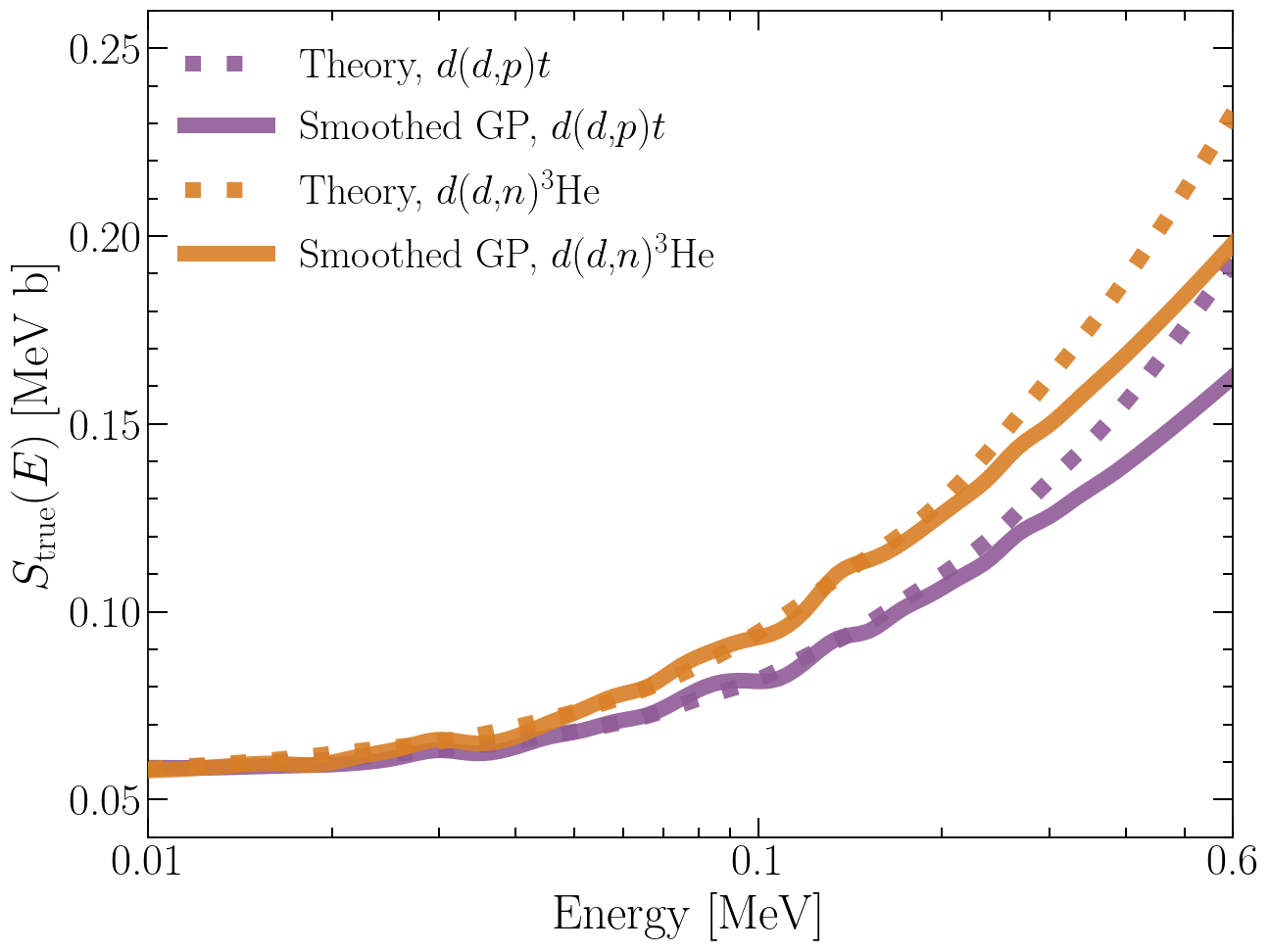}
		\caption{Functions used to generate mock \ddp~(purple) and \ddn~(orange) data for the Monte Carlo validation tests. Functions used to generate mock \ddp~data are shown in purple, and those used to generate mock \ddn~data are shown in orange. The theory calculations of Ref.~\cite{Arai2011} appear as dotted curves, and the smoothed GP means from fits to the real data are the solid curves. }
		\label{fig:gen_func}
	\end{minipage}
\end{figure}

D/H predictions (purple) from the Monte Carlo validation tests, only including statistical uncertainties in data generation and GP regression, are shown in Figure~\ref{fig:mc_stat}. GP regression on both \ddp~and \ddn~mock data now yields unbiased D/H predictions; the magnitude of the bias for both cases is now $\leq0.02$. Predicted uncertainties remain robust. 

Again, we approximate the $(\mathrm{D/H}_{\rm pred} - \mathrm{D/H}_{\rm true})/\sigma_{\rm pred}$ distributions with the empirical sets shown in Figure~\ref{fig:mc_stat} and calculate the KL divergence and KS statistic (see above). We use these quantities to assess the similarity between these distributions and $\mathcal{N}(0,1)$---the expected distribution for $(\mathrm{D/H}_{\rm pred} - \mathrm{D/H}_{\rm true})/\sigma_{\rm pred}$ with an unbiased, robust fitting procedure. Both are lower than when correlations are included, with the KL divergence between the empirical distribution dropping from $0.14$ ($0.10$) to $0.02$ ($0.02$) for \ddp~(\ddn).~Similarly, the KS statistic decreases from $0.21$ ($0.16$) to $0.03$ ($0.03$) for \ddp~(\ddn).

We therefore find that removing correlations from the mock data and GPs leads to unbiased D/H predictions with well-calibrated uncertainties. We conclude that the small bias in our fiducial Monte Carlo results may be a consequence of our handling of correlations. However, our implementation otherwise handles correlated data as desired, discussed in conjunction with Figure~\ref{fig:ddhe3n_samples} in Section~\ref{sec:fits}, with perfectly correlated data always lying the same distance away from the GP mean. 
We remind the reader that even with the present method, however, that our MC analysis shows that the bias is small, with the reported error bars providing excellent coverage. Further refinement of our method may reduce our bias, but we leave this to future work. 

\subsection{Additional Monte Carlo Details}

Figure~\ref{fig:gen_func} displays the functions used to generate mock data in our Monte Carlo analyses throughout this work. The dotted curves are the theory calculations of Ref.~\cite{Arai2011}. The solid curves are the smoothed GP means from regression on real data; these are offset in energy to remove correlations between small-scale features and the energies of the data. The generating functions for \ddp~are shown in purple, with those for \ddn~in orange.

To propagate uncertainties from polynomial fits through to reaction rate uncertainties, we adopt the same procedure as Refs.~\cite{Serpico2004, Pisanti2021}:
\begin{equation}
	\Delta R^2(T) = \sqrt{\chi^2_\nu} \int_0^\infty dE'\, K(E', T) \int_0^\infty dE\, K(E,T) \sum_{i,j} \frac{\partial S_{\rm th}(E',a)}{\partial a_i} \bigg|_{\hat{a}} \frac{\partial S_{\rm th}(E,a)}{\partial a_j}\bigg|_{\hat{a}} \mathrm{cov}(a_i,a_j),
\end{equation}
where $K(E,T)$ contains the terms aside from $S(E)$ in Eq.~\eqref{eqn:thermal_average}, the sums run over all polynomial coefficients, and $\mathrm{cov}(a_i,a_j)$ is the covariance between the two polynomial coefficients $a_i$ and $a_j$ from fitting. The reaction rate uncertainty is inflated by $\chi_\nu^2$ (per degree of freedom) as in Refs.~\cite{Serpico2004, Pisanti2021}. We only consider \ddp~and \ddn~individually in the Monte Carlo tests for polynomial fitting; uncertainties in D/H are given by inputting the mean $R(T)+\sqrt{\Delta R^2(T)}$ into LINX. 

\end{document}